\documentclass[12pt]{article}
\usepackage{epsfig,graphics}

\newcommand{\be}{\begin{equation}}
\newcommand{\ee}{\end{equation}}

\newlength{\figsize}
\begin{document}

\begin{titlepage}

\vspace*{0.7in}
 
\begin{center}
{\large\bf Topology of SU(N) gauge theories at 
${\bf T\simeq 0}$ and  ${\bf T\simeq T_c}$ \\ }
\vspace*{0.5in}
{Biagio Lucini$^{a,b}$, Michael Teper$^{a}$ and Urs Wenger$^{a,c}$\\
\vspace*{.2in}
$^{a}$Theoretical Physics, University of Oxford,\\
1 Keble Road, Oxford OX1 3NP, U.K.\\
\vspace*{.1in}
$^{b}$Institute for Theoretical Physics, ETH Z\"urich,\\
CH-8093 Z\"urich, Switzerland\\
\vspace*{.1in}
$^{c}$NIC/DESY Zeuthen, Platanenallee 6, 15738 Zeuthen, Germany
}
\end{center}

\vspace*{0.6in}

\begin{center}
{\bf Abstract}
\end{center}

We calculate the topological charge density of $SU(N)$ lattice
gauge fields for values of $N$ up to $N=8$. Our $T\simeq 0$
topological susceptibility appears to approach a finite non-zero
limit at $N=\infty$ that is consistent with earlier
extrapolations from smaller values of $N$.
Near the deconfining temperature $T_c$, we are able to
investigate separately the confined and deconfined phases,
since the transition is quite strongly first order. We find 
that the topological susceptibility of the confined phase is 
always very similar to that at $T=0$. By contrast, in the deconfined 
vacuum at larger $N$ there are no topological fluctuations 
except for rare, isolated and small instantons.
This shows that as $N \to \infty$ the large-$T$ suppression of 
large instantons and the large-$N$ suppression of small instantons
overlap, even at $T \simeq T_c$, so as to suppress $all$
topological fluctuations  in the deconfined  phase.
In the confined phase by contrast, the size distribution is much
the same at all $T$, becoming more peaked as $N$ grows, 
suggesting that $D(\rho) \propto \delta(\rho-\rho_c)$
at $N=\infty$, with $\rho_c \sim 1/T_c$.

\end{titlepage}

\setcounter{page}{1}
\newpage
\pagestyle{plain}

\section{Introduction}
\label          {intro}

The finite-temperature deconfinement transition of $SU(N)$ 
gauge theories has, recently, been studied numerically for 
gauge groups ranging from $N=2$ to $N=8$
\cite{letterTc,recentTc}.
The transition is first order for $N\geq 3$ and, as with much 
of the other physics of these gauge theories 
\cite{blmt-glue},
the approach to the $N=\infty$ limit appears to be rapid.

As part of this ongoing study
\cite{futureTc}
we have investigated the topological properties of the
gauge fields in the neighbourhood of the transition. Since 
for $N\geq 4$ the transition is quite strongly first order,
it is possible to do this for the confined and deconfined
phases separately, at exactly the same temperature, when
$T\simeq T_c$. What happens as $N\to\infty$ is of
particular interest and what we find is that the topological 
properties in the confined phase are the same for all $T$
as at $T=0$, while in the deconfined phase there are no
topological fluctuations at all, even at $T\simeq T_c$.
Since at $N=\infty$ QCD (with $m_q > 0$) and quenched QCD
are the same, this conclusion also applies to the deconfined
phase of  QCD$_{N=\infty}$. Moreover, as we shall see, the
suppression of topological fluctuations in the deconfined phase,
and the corresponding restoration of the  $U_A(1)$ symmetry,
is very rapid in $N$ (probably exponential).

We have also performed some $SU(6)$ and $SU(8)$ calculations
at $T=0$ in order to obtain the string tension, $\sigma$,
which then provides a scale in which to express the 
calculated values of the deconfining temperature $T_c$.
We have, at the same time, calculated the topological charge 
and this allows us to extend earlier $T=0$ calculations, both
of the topological susceptibility
\cite{blmt-glue,pisaQ}
and of the size distribution of the topological charges
\cite{blmt-glue,gw02}. 
This size distribution becomes narrower as $N$ increases
suggesting that it might be tending
to a simple $\delta$-function.

\section{Some expectations at large N}
\label{section_expect}

We review here some expectations that one has about
topology in the large-$N$ limit, that follow from
general counting arguments
\cite{largeN}.

\subsection{The topological susceptibility}
\label{subsection_khi}

The topological susceptibility is defined as
$\chi_t \equiv \langle Q^2 \rangle/V$,
where $V$ is the volume of space-time and $Q$ is the total
topological charge. Since fluctuations 
are suppressed as $N\to\infty$, so that one can imagine
$N=\infty$ physics being given by a single (gauge-invariant) 
Master Field 
\cite{Master}, and since $\langle Q^2 \rangle$ is a measure of
fluctuations, the leading-order value of $\chi_t$ should
vanish and one needs to ask at what order in $1/N$
one expects it to appear.

We can write
\be
\langle Q^2 \rangle
=
\sum_{x,y} \langle Q(x)Q(y) \rangle
=
V \sum_{x} \langle Q(x)Q(0) \rangle
\label{eqNhiN1}
\ee
and we expect  $\langle Q(x)Q(y) \rangle$ to factorise up
to relative corrections of $O(1/N^2)$
\be
\langle Q(x)Q(0) \rangle
\stackrel{N\to\infty}{=}
\langle Q(x) \rangle\langle Q(0) \rangle
+f(x)O({N^{\zeta - 2}})
=
\langle Q(0) \rangle^2
+f(x)O({N^{\zeta - 2}})
\label{eqNhiN2}
\ee
where we take the $N$-dependence of the leading factorised term
to be $N^{\zeta}$, and we
use the fact that $\langle Q(x) \rangle = \langle Q(0) \rangle$
by translation invariance. Putting this together and using 
$\langle Q(0) \rangle = \langle Q \rangle/V$
we obtain
\be
{{\langle Q^2 \rangle}\over{V}}
=
\sum_x \Biggl\{{{\langle Q \rangle}\over{V}}\Biggr\}^2
+\sum_x f(x)O({N^{\zeta - 2}})
=
{{{\langle Q \rangle}^2}\over{V}} 
+ \sum_x f(x) O({N^{\zeta - 2}}) .
\label{eqNhiN3}
\ee
In order to determine $\zeta$ 
let us suppose for the moment that we have added a $\theta$-term to
the action, i.e. $\delta S = i\theta Q$, and that we are working 
with a generic nonzero value of $\theta$ so that 
$\langle Q \rangle \not= 0$. We have
\be
\langle Q \rangle
=
-i{{d}\over{d\theta}}\ln Z(\theta)
=
i{{d}\over{d\theta}}\epsilon(\theta)V
\label{eqn_QZE}
\ee
where we have used $ Z = \exp -\epsilon V$ where $\epsilon$
is the vacuum energy per unit volume. Now we
expect from the usual  large-$N$ counting arguments
\cite{Wtheta}
that $\epsilon \propto N^2$ and that a smooth large-$N$ 
limit is reached if one keeps $\theta/N$ fixed i.e.
\be
\epsilon(\theta)
=
N^2 h(\theta/N).
\ee
Plugging this into eqn(\ref{eqn_QZE}) and using the
notation $\psi \equiv \theta/N$, we immediately
see that
\be
\langle Q \rangle
=
i{{d}\over{d\theta}}\epsilon(\theta)V
=
N V i{{d}\over{d\psi}} h(\psi)
\label{eqn_QN}
\ee
i.e. $\langle Q \rangle \propto NV$. Thus $\zeta = 2$ and 
we see from eqn(\ref{eqNhiN3}) that for any $\psi\not= 0$
\be
\chi_t
\equiv
{{\langle (Q-\langle Q \rangle)^2 \rangle}\over{V}}
\propto
O(V^0)O(N^0)
\label{eqNhiN4}
\ee
where we have generalised the definition of $\chi_t$ to
allow for $\langle Q \rangle\not= 0$. In obtaining
eqn(\ref{eqNhiN4}) we use the fact that $\sum_x f(x) = O(V^0)$ 
which follows from the fact that
the mass of the lightest glueball with the quantum numbers 
of $Q(x)$, i.e. $J^{PC}=0^{-+}$, is non-zero. By continuity
we expect that at $\theta=0$ eqn(\ref{eqNhiN4}) will be valid
and that $\chi_t\propto O(N^0)$.

We therefore expect that $\chi_t$ should have a finite non-zero limit
as $N\to\infty$, even though this is really an $O(1/N^2)$ correction
in that limit. This is of course nothing but the conventional 
expectation, needed to explain the mass of the $\eta^\prime$
\cite{eta-thooft,WV}.
Lattice calculations 
\cite{blmt-glue,pisaQ}
support this expectation and we shall provide further 
evidence in this paper.

\subsection{The topological charge}
\label{subsection_Q}

We have noted above that for $\theta\not= 0$ we have
\be
{{\langle Q \rangle}\over{V}} \propto N. 
\ee
This is not unexpected: instantons are $SU(2)$ objects and 
(in the semiclassical limit) can be simultaneously 
placed in any of the $\sim N/2$ non-overlapping subgroups 
of $SU(N)$ without interacting -- even if they are centered
at the same point in space-time. At  $\theta = 0$, 
where we work, one continues to expect $O(N)$ charges in
every finite region of space-time, although now they
will largely cancel in $\langle Q \rangle$. Since the general
arguments given in Section~\ref{subsection_khi} imply that 
$\langle Q^2 \rangle\propto N^0$ rather 
than $\propto N^1$, this cancellation is clearly much
stronger than the usual `random walk' of a dilute gas.

There are detailed variational treatments of the
instanton partition function that explicitly show how this 
cancellation can happen
\cite{Diak,TSI}.
It is instructive to sketch the core of the argument.
Since the (anti)instantons can be thrown into $\propto N$
SU(2) subgroups in the space-time volume $V$, we can expect
their numbers to be $n_+,n_- \propto V N$ and equal
on the average. That is to say $n_+ +n_- \propto N$
(we drop the ubiquitous factor of $V$ from now on)
while the distribution in $Q = n_+ -n_-$ is flat
for $Q \ll \surd N$. (It is easy to see this in an
explicit dilute gas calculation.) For the fields not to interact 
they have to be in subgroups that are exactly orthogonal and
this is a set of fields of measure zero. Allowing fluctuations
away from these particular subgroups involves interactions.
Let $\gamma_s,\gamma_a$ label the interactions between
like and unlike charges that are in the same volume of
space-time. The simplest instanton calculation
\cite{Diak}
tells us that the interaction is repulsive with 
$\gamma_{a,s} \propto 1/N$. Thus the extra interaction 
piece in the action will be
\be
\delta S_{int} \propto 
\frac{1}{2}\gamma_sn_+(n_+ - 1)
+
\frac{1}{2}\gamma_sn_-(n_- - 1)
+
\gamma_an_-n_+
\ee
which we can rewrite as
\be
\delta S_{int} \propto 
\frac{1}{2}\frac{1+r}{2}\gamma_s (n_+ + n_-)^2
+
\frac{1}{2}\frac{1-r}{2}\gamma_s (n_+ - n_-)^2
\label{diak_int}
\ee
defining $r=\gamma_a/\gamma_s$ (and suppressing a non-leading
term linear in $n_+ + n_-$). The second term provides
a factor in the partition function of $\propto \exp -cQ^2$
where $c=O(1)$ since the factor of $1/N$ from $\gamma_s$
will combine with the usual $1/g^2$ factor to give the
't Hooft coupling that is kept constant as $N\to\infty$.
Of course, for $c > 0$ we must assume that  $r <1$ i.e.
$\gamma_a < \gamma_s$, since otherwise the favoured
configuration would clearly have charges of the same sign,
so cancellations would be suppressed and we would have
$Q \propto N$. If all this is the case then,
given that the partition function
is otherwise flat in $Q = n_+ -n_-$ for $Q \ll \sqrt{N}$,
this factor of $\exp -cQ^2$ will ensure that
$\langle Q^2 \rangle$ is $O(1)$ rather than $O(N)$.
What happens to the term $\propto (n_+ + n_-)^2$ in
eqn(\ref{diak_int}) is more difficult to see,
and requires the full apparatus of a variational
treatment
\cite{Diak}.
But the suppression of $\langle Q^2 \rangle$ from
$O(N)$ to $O(1)$ is (in this instanton picture) simply 
the result of the fact that it induces a suppression in 
the action that is quadratic in $Q$ because it is quadratic
in the $n_+,n_-$. This might appear to be a severe breakdown 
of the dilute gas approximation, but it occurs within 
the same region of space-time, and still leaves open
the possibility that a dilute gas/liquid picture can 
describe the interactions of the net charges that
are located in different regions of space-time.

It would clearly be interesting to confirm the above 
expectation, and to explore how it works in the full 
non-perturbative vacuum of the gauge theory. 
Unfortunately our lattice calculations were not specifically
designed to address this issue. Nonetheless we do calculate
two potentially relevant quantities: the total action and
$Q_{mod} \equiv \sum_x |Q(x)|$ as a function of cooling.
If there are $O(N)$ topological charges in every region
of space-time, and if there is some scatter in the positions
and sizes of these charges, then this might be revealed in  
a component of these two quantities that grows with $N$.
Unfortunately our calculations of $Q_{mod}$ have shown
nothing significant (which may be due to the coarseness
of our lattice spacing) and while the average plaquette after
a given number of `cooling sweeps' 
(see Section~\ref{section_lattice}) is roughly independent
of $N$, so that the normalised action is $\propto N^2$
(recall the factor $\beta=2N/g^2=2N^2/(g^2N)$ in the 
lattice action -- see Section~\ref{section_lattice} -- 
and that we keep $g^2N$ fixed as $N\to\infty$),
this is difficult to interpret on its own since we of course
expect the full action to show this behaviour irrespective
of topological considerations.

\subsection{Small instantons at all T}
\label{subsection_Ismall}

For large $N$, instantons are exponentially suppressed
in $N$ 
\cite{WI}:
\be
D(\rho)
\propto 
e^{-S_I}
\propto 
e^{-{{8\pi^2}\over{g^2(\rho)}}}
=
e^{-{{8\pi^2}\over{\lambda(\rho)}}N}
\label{eqn_WI}
\ee
where $S_I$ is the instanton action, $\rho$ is the instanton 
size and $\lambda\equiv g^2 N $ is the 't Hooft coupling 
which is kept constant as $N \to \infty$. This simple
semiclassical argument is incomplete (see e.g.
\cite{MTI,HNI,TSI}) 
but is essentially correct for topological charges that 
are small enough and using the one-loop expression for
$g^2(\rho)$ we expect that
\be
D(\rho)
\propto 
e^{-S_I}
\stackrel{\rho \to 0}{\propto}
\rho^{{{11N}\over{3}}-5} .
\label{eqn_WI1loop}
\ee
Thus we expect that there is some critical size $\rho_c$ such that
\be
D(\rho)
\stackrel{N \to \infty}{\longrightarrow}
0
\ \ \ \ \ \ \ ; \ \forall \rho < \rho_c .
\label{eqn_Ismall}
\ee
There is some evidence for this from earlier
lattice calculations
\cite{blmt-glue}.

This has an important practical consequence for lattice
calculations. Since the lattice Monte Carlo is a local
process, in that the elementary step consists of changing 
one link matrix at a time, the value of $Q$ can only
change in a sequence of Monte Carlo generated gauge fields
if an instanton gradually shrinks until it disappears
within a lattice hypercube (or the converse process). For
small lattice spacings this requires the presence of lattice 
fields containing an instanton of size $a \ll \rho \ll \rho_c$.
From eqn(\ref{eqn_Ismall}) the probability of such a lattice
field $\to 0$ exponentially fast in $N$ as $N\to\infty$.
Thus at fixed $a$ topology suffers a critical slowing
down in $N$, i.e. the value of $Q$ ceases to change and
we cannot calculate $\langle Q^2 \rangle$ in the usual
fashion. (Of course, fluctuations corresponding to
topological charges of opposite sign can still appear and
disappear and in principle one could calculate  
$\langle Q^2 \rangle$ from the charge $Q$ within a large 
sub-volume of a very much larger space-time volume -- with 
the latter chosen large enough
that the constraint of a fixed total topological charge 
is negligible.) In practice this means that in our $SU(8)$
calculations, the value of $Q$ ceases to change as
we reduce $a$ and we are unable to perform a continuum
extrapolation for the susceptibility. Instead we will
perform a comparison of the susceptibility at a fixed
not-too-small value of $a$.

\subsection{Large instantons at high T}
\label{subsection_Ilarge}

At high $T$, chromo-electric fields are screened on a scale
characterised by the electric screening mass $m_{el}$.
At high enough $T$ the characteristic expansion parameter, $g^2(T)$, 
is small and it suffices to calculate  $m_{el}$ to leading
order in perturbation theory, whereupon one finds
\cite{GPY}
\be
m^2_{el}(T)
=
{{N}\over{3}} g^2(T) T^2.
\label{eqn_mel}
\ee
Here the power of $T$ follows on simple dimensional grounds,
and the coupling appears in the familiar combination, $g^2N$, 
which is kept constant as $N\to\infty$.
The chromoelectric fields in the core of an instanton
are coherent over its size $\rho$. On the other hand they
will necessarily be screened over a distance $O(1/m_{el})$.
This means that instantons of size 
$\rho \gg 1/m_{el} \propto 1/T$ will be strongly
suppressed, and we expect the suppression to be a function
of $\rho T $. A careful calculation 
\cite{GPY}
shows that the instanton density, $D(\rho)$, is modified  
as follows for $\pi\rho T\gg 1$ and $g^2(T)\ll 1$:
\be
D(\rho,T) 
= 
D(\rho,0) 
\exp(-{{2N}\over{3}}\{\pi\rho T\}^2 - \gamma (\rho T))
\label{eqn_DrhoT}
\ee
where $\gamma (\rho T)$ is a weak function of its
argument, which we can neglect for our purposes.
We note that the dominant suppression in eqn(\ref{eqn_DrhoT}) 
is a function of $\rho T$ as expected. It also contains 
a factor of $N$, which one would expect for the same reason
that it appears in the classical action in eqn(\ref{eqn_WI}). 

When valid, eqn(\ref{eqn_DrhoT}) implies that
topological fluctuations on size scales
$\rho \gg 1/\pi \sqrt{N} T$ will be heavily suppressed,
and that at fixed $\rho$ the suppression will
be exponential in $N$. So what is its domain of validity?
The derivation of
electric screening requires that we be in the
deconfined phase and the leading-order perturbative 
calculation is only guaranteed to be valid for $T\gg T_c$.
So an interesting question is down to what value of $T$ 
does this calculation hold? The most extreme and
yet most elegant scenario is one in
which it is  valid at all $T$ in the deconfined phase
and that as $N\to \infty$ the small instanton suppression
described in the previous section, overlaps with this
larger instanton suppression, so that all topological
fluctuations vanish exponentially with $N$ at any value
of $T$, in the deconfined phase. This, as we shall see 
below, is indeed what appears to happen.

\section{The lattice calculation}
\label{section_lattice}

The lattice calculations have been described in detail
elsewhere
\cite{letterTc,recentTc}
and the calculation of the topological charge density
is precisely as in 
\cite{blmt-glue}.
We shall therefore restrict ourselves to a minimal 
description here and refer the interested reader to
those earlier papers.

We work on periodic hypercubic lattices . We label the
lattice spacing by $a$ and the size, in lattice units, by 
$L^3_s L_t$. For $L_s$ sufficiently large this corresponds 
to the field theory at a finite temperature
\be
T = {1\over{aL_t}}.
\label{eqn_T}
\ee
If $L_s, L_t \gg 1/aT_c$ we refer to the system as being
at $T=0$, although this is of course not exactly the case.
The variables are $SU(N)$ matrices assigned to the links of 
the lattice. We employ the usual plaquette action
\cite{blmt-glue}
which contains the factor
\be
\beta = {{2N}\over{g^2}}.
\label{eqn_beta}
\ee
Here $g^2$ is the bare coupling defined at the cut-off
scale $a$, and by varying the value of $\beta$ we vary
the value of $a$. If we tune $a\to0$ we recover
the continuum field theory in Euclidean space-time,
if $T=0$, or the statistical physics of the theory
at temperature $T$, more generally.

We evaluate the lattice Feynman Path Integral/Partition
Function using standard Monte Carlo techniques.
At a fixed value of $\beta$ we can calculate the $T=0$ 
string tension in lattice units, $a^2(\beta)\sigma$, and 
its value allows us to express the lattice spacing
$a(\beta)$ in physical units. If we increase $\beta$
on a lattice with $L_s \gg L_t$ we will find a deconfining
transition at some $\beta=\beta_c$. The corresponding
deconfining temperature is
\be
a(\beta_c)T_c = {1\over{L_t}}.
\label{eqn_Tc}
\ee
Dividing this by the $T=0$ value of $a(\beta=\beta_c)\surd\sigma$
we obtain a value of the dimensionless ratio $T_c/\surd\sigma$
which differs from the continuum value by
lattice corrections of order $a^2(\beta_c)$.
For larger $N$ we have typically calculated $\beta_c$ for
$L_t=5,6,8$ and this allows us to make the desired $a\to 0$ 
continuum extrapolation.

We calculate the topological charge density using one of
the  standard methods
\cite{blmt-glue,mt_Qrev}. 
For smooth lattice gauge fields one can define a lattice 
topological charge density $Q_L(x)$ as follows:
\be
Q_L(x) = {1\over{32\pi^2}}
\varepsilon_{\mu\nu\rho\sigma}
ReTr\{U_{\mu\nu}(x)U_{\rho\sigma}(x)\}
\stackrel{a \to 0}{\longrightarrow}
a^4 Q(x) + O(a^6)
\label{eqn_Qlattice}
\ee
where $Q(x)$ is the continuum topological charge density.
To obtain a smooth lattice field from a rough Monte
Carlo generated lattice field we apply an iterative
`cooling' procedure
\cite{blmt-glue,mt_Qrev}.
Using eqn(\ref{eqn_Qlattice}) on the cooled lattice field
we obtain its total topological charge $Q_L = \sum_x Q_L(x)$.
This will not be an integer because of the lattice
corrections in  eqn(\ref{eqn_Qlattice}), but can easily
be rounded to the appropriate integer. We can also attempt
to identify individual instantons by looking for peaks
in $Q_L(x)$ and applying the classical continuum relation
\be
Q_{peak} = {6\over{\pi^2\rho^4}}.
\label{eqn_Qpeak}
\ee
to extract the instanton size, $\rho$, from the peak height.
In this way we can calculate an instanton size density,
$D(\rho)$.

If we use a modest number of cooling sweeps (typically
10 or 20) the calculated topological charge is accurate 
except in  the presence of very narrow charges whose
interpretation is naturally ambiguous on a lattice.
One can see this by comparing with a fermionic calculation
of $Q$ using fermions that satisfy the index theorem  
\cite{gw02}.
The topological charge density, on the other hand, is 
steadily distorted by cooling and so the information
one obtains on $D(\rho)$ is less reliable. (Although
here too there is some evidence from using massless fermions 
as a probe, that the distortions are not large for a 
small number of cooling sweeps
\cite{gw02b}.)
Together with the crudeness of eqn(\ref{eqn_Qpeak}) as
a pattern recognition algorithm for a dense `instanton'
ensemble (see e.g.
\cite{dsmt}
for more sophistication) it is clear that our estimate
of $D(\rho)$ must be regarded as semi-quantitative.
In particular, while small instantons are easy to
identify, since the peak is large, very large instantons
correspond to very small and smooth peaks which
might be remnant artifacts of the cooling procedure.
In this paper we will make some progress in dealing
with this latter uncertainty.

\section{Results}
\label{section_results}

\subsection{Topological susceptibility at T=0}
\label{subsectioNhiT0}

In our  `$T=0$' $SU(8)$ calculation, it is only
for the coarsest lattice spacings that $Q$ changes
sufficiently frequently that we obtain a usefully
accurate value for $\langle Q^2 \rangle$. This
means that we cannot reliably extrapolate the 
susceptibility to the continuum limit and
compare its value with continuum values previously
obtained for $N\leq 5$ in
\cite{blmt-glue}
and for $N\leq 6$ in
\cite{pisaQ}
as we would ideally wish to do.
Instead we will calculate the dimensionless ratio of
the susceptibility, $a^4 \chi_t$, and the string
tension, $a^2\sigma$, i.e.
$a\chi^{1/4}_t/a\surd\sigma \equiv \chi^{1/4}_t/\surd\sigma$,
at a fixed value of $a$, chosen so that
we are able to obtain an accurate $SU(8)$ value.
We do so for  $N=2,3,4,6,8$. The large-$N$
counting for the lattice gauge theory is the
same as for the continuum gauge theory
\cite{lattice-thooft}, 
and so one can perform an extrapolation to $N=\infty$, 
using corrections that are powers of $1/N^2$.

If we want to use the `same' value of $a$ in different
$SU(N)$ gauge theories, we need to measure it in
units of some physical quantity that has a smooth
large-$N$ limit. We shall choose to fix the value of 
$a$ in units of the deconfining temperature, $T_c$. 
To be precise, the value of $a$ that we choose corresponds 
to $a=1/5T_c$. That is to say, we perform our calculations
at the value of $\beta$ at which the gauge theory on an
$L^3 5$ lattice (with $L\gg 5$) will undergo a
deconfining transition. We could have used some other
physical quantity, such as the string tension,
to fix $a$ and the resulting values of $\beta$ would
have differed, because $T_c/\surd\sigma$ does vary
with $N$. But because such physical ratios possess a 
smooth large-$N$ limit, any difference will correspond
to an change in the coefficients of terms that are
higher order in $1/N^2$.

In Table~\ref{table_chi5a} we list our results for
$\langle Q^2 \rangle$ for $N=2,...,8$. We show the lattice
sizes, the string tensions, and the values of $\beta$ 
for each value of $N$. We also show for each $N$ the critical 
value of $\beta$ at a lattice spacing $a=1/5T_c$. We see that
the value of $\beta$ we use is always, within errors,
equal to $\beta_c(a=1/5T_c)$, i.e. the lattice spacing is 
constant across all $N$ when expressed in units of $T_c$.
At a finite lattice spacing different methods of
calculating $Q$ will differ. The value we show is obtained 
after 20 cooling sweeps. $Q_r$ is the real number obtained
directly from summing our twisted plaquette operator,
while  $Q_I$ has been rounded to the appropriate
neighbouring integer value. We show in
Table~\ref{table_chi5b} the resulting values
of the dimensionless ratio $\chi^{1/4}_t/\surd\sigma$
that we wish to compare across $N$.

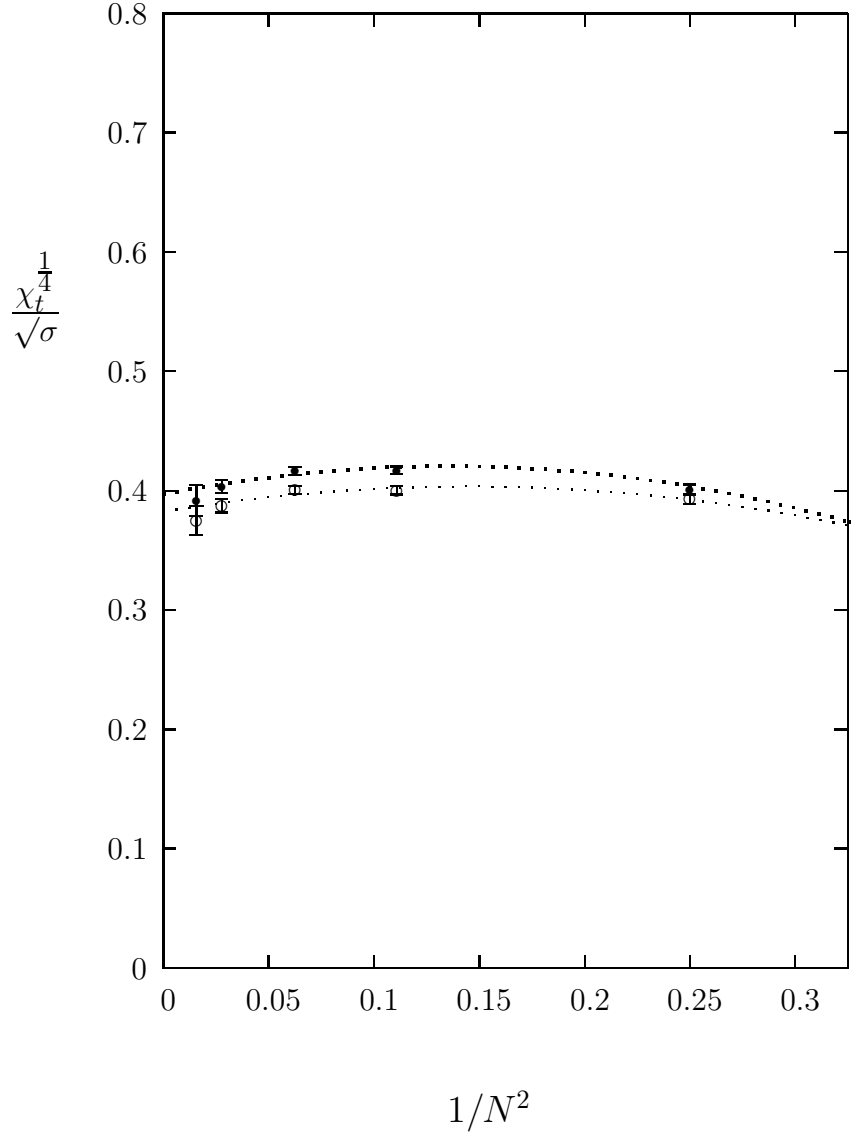
\begin  {figure}[p]
\begin  {center}
\leavevmode
\setlength{\unitlength}{0.240900pt}
\ifx\plotpoint\undefined\newsavebox{\plotpoint}\fi
\sbox{\plotpoint}{\rule[-0.200pt]{0.400pt}{0.400pt}}%
\begin{picture}(1500,1800)(0,0)
\font\gnuplot=cmr10 at 12pt
\gnuplot
\sbox{\plotpoint}{\rule[-0.200pt]{0.400pt}{0.400pt}}%
\put(350.0,250.0){\rule[-0.200pt]{4.818pt}{0.400pt}}
\put(325,250){\makebox(0,0)[r]{\ \ {$0$}}}
\put(1405.0,250.0){\rule[-0.200pt]{4.818pt}{0.400pt}}
\put(350.0,438.0){\rule[-0.200pt]{4.818pt}{0.400pt}}
\put(325,438){\makebox(0,0)[r]{\ \ {$0.1$}}}
\put(1405.0,438.0){\rule[-0.200pt]{4.818pt}{0.400pt}}
\put(350.0,625.0){\rule[-0.200pt]{4.818pt}{0.400pt}}
\put(325,625){\makebox(0,0)[r]{\ \ {$0.2$}}}
\put(1405.0,625.0){\rule[-0.200pt]{4.818pt}{0.400pt}}
\put(350.0,813.0){\rule[-0.200pt]{4.818pt}{0.400pt}}
\put(325,813){\makebox(0,0)[r]{\ \ {$0.3$}}}
\put(1405.0,813.0){\rule[-0.200pt]{4.818pt}{0.400pt}}
\put(350.0,1000.0){\rule[-0.200pt]{4.818pt}{0.400pt}}
\put(325,1000){\makebox(0,0)[r]{\ \ {$0.4$}}}
\put(1405.0,1000.0){\rule[-0.200pt]{4.818pt}{0.400pt}}
\put(350.0,1188.0){\rule[-0.200pt]{4.818pt}{0.400pt}}
\put(325,1188){\makebox(0,0)[r]{\ \ {$0.5$}}}
\put(1405.0,1188.0){\rule[-0.200pt]{4.818pt}{0.400pt}}
\put(350.0,1375.0){\rule[-0.200pt]{4.818pt}{0.400pt}}
\put(325,1375){\makebox(0,0)[r]{\ \ {$0.6$}}}
\put(1405.0,1375.0){\rule[-0.200pt]{4.818pt}{0.400pt}}
\put(350.0,1563.0){\rule[-0.200pt]{4.818pt}{0.400pt}}
\put(325,1563){\makebox(0,0)[r]{\ \ {$0.7$}}}
\put(1405.0,1563.0){\rule[-0.200pt]{4.818pt}{0.400pt}}
\put(350.0,1750.0){\rule[-0.200pt]{4.818pt}{0.400pt}}
\put(325,1750){\makebox(0,0)[r]{\ \ {$0.8$}}}
\put(1405.0,1750.0){\rule[-0.200pt]{4.818pt}{0.400pt}}
\put(350.0,250.0){\rule[-0.200pt]{0.400pt}{4.818pt}}
\put(350,200){\makebox(0,0){\ {$0$}}}
\put(350.0,1730.0){\rule[-0.200pt]{0.400pt}{4.818pt}}
\put(515.0,250.0){\rule[-0.200pt]{0.400pt}{4.818pt}}
\put(515,200){\makebox(0,0){\ {$0.05$}}}
\put(515.0,1730.0){\rule[-0.200pt]{0.400pt}{4.818pt}}
\put(681.0,250.0){\rule[-0.200pt]{0.400pt}{4.818pt}}
\put(681,200){\makebox(0,0){\ {$0.1$}}}
\put(681.0,1730.0){\rule[-0.200pt]{0.400pt}{4.818pt}}
\put(846.0,250.0){\rule[-0.200pt]{0.400pt}{4.818pt}}
\put(846,200){\makebox(0,0){\ {$0.15$}}}
\put(846.0,1730.0){\rule[-0.200pt]{0.400pt}{4.818pt}}
\put(1012.0,250.0){\rule[-0.200pt]{0.400pt}{4.818pt}}
\put(1012,200){\makebox(0,0){\ {$0.2$}}}
\put(1012.0,1730.0){\rule[-0.200pt]{0.400pt}{4.818pt}}
\put(1177.0,250.0){\rule[-0.200pt]{0.400pt}{4.818pt}}
\put(1177,200){\makebox(0,0){\ {$0.25$}}}
\put(1177.0,1730.0){\rule[-0.200pt]{0.400pt}{4.818pt}}
\put(1342.0,250.0){\rule[-0.200pt]{0.400pt}{4.818pt}}
\put(1342,200){\makebox(0,0){\ {$0.3$}}}
\put(1342.0,1730.0){\rule[-0.200pt]{0.400pt}{4.818pt}}
\put(350.0,250.0){\rule[-0.200pt]{258.967pt}{0.400pt}}
\put(1425.0,250.0){\rule[-0.200pt]{0.400pt}{361.350pt}}
\put(350.0,1750.0){\rule[-0.200pt]{258.967pt}{0.400pt}}
\put(150,1300){\makebox(0,0){\Large{${{\chi^{\frac{1}{4}}_t} \over{\surd\sigma} }$}}}
\put(862,25){\makebox(0,0){\large{$1/N^2$}}}
\put(350.0,250.0){\rule[-0.200pt]{0.400pt}{361.350pt}}
\put(1177.0,995.0){\rule[-0.200pt]{0.400pt}{3.613pt}}
\put(1167.0,995.0){\rule[-0.200pt]{4.818pt}{0.400pt}}
\put(1167.0,1010.0){\rule[-0.200pt]{4.818pt}{0.400pt}}
\put(717.0,1026.0){\rule[-0.200pt]{0.400pt}{2.891pt}}
\put(707.0,1026.0){\rule[-0.200pt]{4.818pt}{0.400pt}}
\put(707.0,1038.0){\rule[-0.200pt]{4.818pt}{0.400pt}}
\put(557.0,1025.0){\rule[-0.200pt]{0.400pt}{2.891pt}}
\put(547.0,1025.0){\rule[-0.200pt]{4.818pt}{0.400pt}}
\put(547.0,1037.0){\rule[-0.200pt]{4.818pt}{0.400pt}}
\put(442.0,996.0){\rule[-0.200pt]{0.400pt}{5.059pt}}
\put(432.0,996.0){\rule[-0.200pt]{4.818pt}{0.400pt}}
\put(432.0,1017.0){\rule[-0.200pt]{4.818pt}{0.400pt}}
\put(402.0,960.0){\rule[-0.200pt]{0.400pt}{11.804pt}}
\put(392.0,960.0){\rule[-0.200pt]{4.818pt}{0.400pt}}
\put(1177,1002){\circle*{12}}
\put(717,1032){\circle*{12}}
\put(557,1031){\circle*{12}}
\put(442,1006){\circle*{12}}
\put(402,984){\circle*{12}}
\put(392.0,1009.0){\rule[-0.200pt]{4.818pt}{0.400pt}}
\put(1177.0,979.0){\rule[-0.200pt]{0.400pt}{3.613pt}}
\put(1167.0,979.0){\rule[-0.200pt]{4.818pt}{0.400pt}}
\put(1167.0,994.0){\rule[-0.200pt]{4.818pt}{0.400pt}}
\put(717.0,994.0){\rule[-0.200pt]{0.400pt}{3.132pt}}
\put(707.0,994.0){\rule[-0.200pt]{4.818pt}{0.400pt}}
\put(707.0,1007.0){\rule[-0.200pt]{4.818pt}{0.400pt}}
\put(557.0,995.0){\rule[-0.200pt]{0.400pt}{2.891pt}}
\put(547.0,995.0){\rule[-0.200pt]{4.818pt}{0.400pt}}
\put(547.0,1007.0){\rule[-0.200pt]{4.818pt}{0.400pt}}
\put(442.0,966.0){\rule[-0.200pt]{0.400pt}{5.059pt}}
\put(432.0,966.0){\rule[-0.200pt]{4.818pt}{0.400pt}}
\put(432.0,987.0){\rule[-0.200pt]{4.818pt}{0.400pt}}
\put(402.0,930.0){\rule[-0.200pt]{0.400pt}{11.081pt}}
\put(392.0,930.0){\rule[-0.200pt]{4.818pt}{0.400pt}}
\put(1177,987){\circle{18}}
\put(717,1000){\circle{18}}
\put(557,1001){\circle{18}}
\put(442,977){\circle{18}}
\put(402,953){\circle{18}}
\put(392.0,976.0){\rule[-0.200pt]{4.818pt}{0.400pt}}
\sbox{\plotpoint}{\rule[-0.500pt]{1.000pt}{1.000pt}}%
\put(350,995){\usebox{\plotpoint}}
\put(350.00,995.00){\usebox{\plotpoint}}
\put(370.42,998.71){\usebox{\plotpoint}}
\put(390.82,1002.56){\usebox{\plotpoint}}
\put(411.23,1006.31){\usebox{\plotpoint}}
\put(431.72,1009.52){\usebox{\plotpoint}}
\put(452.20,1012.76){\usebox{\plotpoint}}
\put(472.74,1015.68){\usebox{\plotpoint}}
\put(493.29,1018.42){\usebox{\plotpoint}}
\put(513.84,1021.15){\usebox{\plotpoint}}
\put(534.39,1023.94){\usebox{\plotpoint}}
\put(555.05,1025.91){\usebox{\plotpoint}}
\put(575.72,1027.79){\usebox{\plotpoint}}
\put(596.39,1029.67){\usebox{\plotpoint}}
\put(617.06,1031.61){\usebox{\plotpoint}}
\put(637.72,1033.52){\usebox{\plotpoint}}
\put(658.41,1035.00){\usebox{\plotpoint}}
\put(679.12,1036.00){\usebox{\plotpoint}}
\put(699.83,1037.00){\usebox{\plotpoint}}
\put(720.54,1038.00){\usebox{\plotpoint}}
\put(741.29,1038.03){\usebox{\plotpoint}}
\put(762.00,1039.00){\usebox{\plotpoint}}
\put(782.76,1039.00){\usebox{\plotpoint}}
\put(803.51,1039.00){\usebox{\plotpoint}}
\put(824.27,1039.00){\usebox{\plotpoint}}
\put(845.00,1038.40){\usebox{\plotpoint}}
\put(865.73,1038.00){\usebox{\plotpoint}}
\put(886.44,1037.00){\usebox{\plotpoint}}
\put(907.15,1036.00){\usebox{\plotpoint}}
\put(927.86,1035.00){\usebox{\plotpoint}}
\put(948.56,1033.86){\usebox{\plotpoint}}
\put(969.23,1031.98){\usebox{\plotpoint}}
\put(989.95,1031.00){\usebox{\plotpoint}}
\put(1010.61,1029.14){\usebox{\plotpoint}}
\put(1031.18,1026.51){\usebox{\plotpoint}}
\put(1051.82,1024.38){\usebox{\plotpoint}}
\put(1072.42,1022.01){\usebox{\plotpoint}}
\put(1093.02,1019.54){\usebox{\plotpoint}}
\put(1113.51,1016.36){\usebox{\plotpoint}}
\put(1134.06,1013.63){\usebox{\plotpoint}}
\put(1154.61,1010.88){\usebox{\plotpoint}}
\put(1175.00,1007.00){\usebox{\plotpoint}}
\put(1195.42,1003.29){\usebox{\plotpoint}}
\put(1215.84,999.57){\usebox{\plotpoint}}
\put(1236.24,995.75){\usebox{\plotpoint}}
\put(1256.65,991.97){\usebox{\plotpoint}}
\put(1276.85,987.30){\usebox{\plotpoint}}
\put(1297.23,983.39){\usebox{\plotpoint}}
\put(1317.42,978.61){\usebox{\plotpoint}}
\put(1337.65,974.06){\usebox{\plotpoint}}
\put(1357.68,968.63){\usebox{\plotpoint}}
\put(1377.84,963.76){\usebox{\plotpoint}}
\put(1397.87,958.40){\usebox{\plotpoint}}
\put(1417.89,952.94){\usebox{\plotpoint}}
\put(1425,951){\usebox{\plotpoint}}
\sbox{\plotpoint}{\rule[-0.200pt]{0.400pt}{0.400pt}}%
\put(350,967){\usebox{\plotpoint}}
\put(350.00,967.00){\usebox{\plotpoint}}
\put(370.42,970.71){\usebox{\plotpoint}}
\put(390.93,973.79){\usebox{\plotpoint}}
\put(411.46,976.68){\usebox{\plotpoint}}
\put(432.00,979.55){\usebox{\plotpoint}}
\put(452.54,982.41){\usebox{\plotpoint}}
\put(473.05,985.37){\usebox{\plotpoint}}
\put(493.72,987.25){\usebox{\plotpoint}}
\put(514.26,990.11){\usebox{\plotpoint}}
\put(534.93,991.99){\usebox{\plotpoint}}
\put(555.59,993.96){\usebox{\plotpoint}}
\put(576.26,995.84){\usebox{\plotpoint}}
\put(596.93,997.72){\usebox{\plotpoint}}
\put(617.63,999.00){\usebox{\plotpoint}}
\put(638.31,1000.57){\usebox{\plotpoint}}
\put(659.03,1001.46){\usebox{\plotpoint}}
\put(679.71,1003.00){\usebox{\plotpoint}}
\put(700.42,1004.00){\usebox{\plotpoint}}
\put(721.13,1005.00){\usebox{\plotpoint}}
\put(741.88,1005.08){\usebox{\plotpoint}}
\put(762.59,1006.00){\usebox{\plotpoint}}
\put(783.35,1006.00){\usebox{\plotpoint}}
\put(804.07,1006.82){\usebox{\plotpoint}}
\put(824.81,1007.00){\usebox{\plotpoint}}
\put(845.57,1007.00){\usebox{\plotpoint}}
\put(866.30,1006.43){\usebox{\plotpoint}}
\put(887.04,1006.00){\usebox{\plotpoint}}
\put(907.78,1005.66){\usebox{\plotpoint}}
\put(928.50,1005.00){\usebox{\plotpoint}}
\put(949.21,1004.00){\usebox{\plotpoint}}
\put(969.92,1003.00){\usebox{\plotpoint}}
\put(990.63,1002.03){\usebox{\plotpoint}}
\put(1011.34,1001.07){\usebox{\plotpoint}}
\put(1032.01,999.18){\usebox{\plotpoint}}
\put(1052.73,998.30){\usebox{\plotpoint}}
\put(1073.40,996.42){\usebox{\plotpoint}}
\put(1094.06,994.45){\usebox{\plotpoint}}
\put(1114.73,992.57){\usebox{\plotpoint}}
\put(1135.40,990.69){\usebox{\plotpoint}}
\put(1156.04,988.59){\usebox{\plotpoint}}
\put(1176.58,985.86){\usebox{\plotpoint}}
\put(1197.12,982.99){\usebox{\plotpoint}}
\put(1217.79,981.11){\usebox{\plotpoint}}
\put(1238.32,978.17){\usebox{\plotpoint}}
\put(1258.85,975.29){\usebox{\plotpoint}}
\put(1279.31,971.85){\usebox{\plotpoint}}
\put(1299.86,969.12){\usebox{\plotpoint}}
\put(1320.25,965.23){\usebox{\plotpoint}}
\put(1340.67,961.51){\usebox{\plotpoint}}
\put(1361.09,957.80){\usebox{\plotpoint}}
\put(1381.51,954.09){\usebox{\plotpoint}}
\put(1401.90,950.20){\usebox{\plotpoint}}
\put(1422.32,946.49){\usebox{\plotpoint}}
\put(1425,946){\usebox{\plotpoint}}
\end{picture}
\end    {center}
\vskip 0.15in
\caption{The topological susceptibility, $\chi_t$, in units of the
string tension, $\sigma$, plotted versus $1/N^2$. The calculations
are performed at a fixed lattice spacing $a \simeq 1/5 T_c$. The
line is a large-$N$ extrapolation that includes $O(1/N^2)$ and
$O(1/N^4)$ corrections.}
\label{fig_chiN}
\end    {figure}

In Fig.~\ref{fig_chiN} we plot the values of 
$\chi^{1/4}_t/\surd\sigma$ against the expected
expansion parameter $1/N^2$. To fit the $N$-dependence
we  use as few correction terms as possible. However 
a simple $1/N^2$ correction does not provide an acceptable
fit, if we insist on fitting all the way down to $N=2$
(or even only down to $N=3$). We therefore introduce an  
additional $1/N^4$ correction and this enables us to obtain 
an acceptable fit. In that case our best fits are
\be
{{\chi^{\frac{1}{4}}_t}\over{\surd\sigma}}
= \left \{ \begin{array}{ll}
0.397(7) + {{\displaystyle 0.35(13)}\over{\displaystyle N^2}} -
{{\displaystyle 1.32(41)}\over{\displaystyle N^4}} 
& \ \ \ \ \mbox{using $Q_I$,} \\
& \\
0.382(7) + {{\displaystyle 0.30(13)}\over{\displaystyle N^2}} -
{{\displaystyle 1.02(42)}\over{\displaystyle N^4}} 
&  \ \ \ \ \mbox{using $Q_r$,}
\end{array}
\right.
\label{eqNhi5}
\ee
with a goodness of fit $\chi^2 \simeq 1.0, 1.3$ per degree 
of freedom respectively. (The difference between the 
two fits is an $O(a^2)$ lattice correction and should
disappear in the continuum limit.)
As we see from Fig.~\ref{fig_chiN} the convergence to
the $N=\infty$ limit is rapid, and our values are
accurate enough, and extend over a large enough range of 
$N$, to make this a reasonably compelling conclusion.

As an aside we note that the coefficient of the $1/N^4$ term 
in eqn(\ref{eqNhi5}) is quite large compared
to the other coefficients, and one might be tempted
to ascribe this to the fact that at this lattice spacing
we are very close to the bulk transition
\cite{blmt-glue,recentTc,futureTc}.
However, although the original $a=0$ large-$N$ calculations of 
$\chi^{1/4}_t/\surd\sigma$ 
\cite{blmt-glue}
were able to interpolate within
errors down to $N=2$ with just a simple $O(1/N^2)$ correction,
and therefore the coefficient of the $O(1/N^4)$ term could
not be usefully estimated, there are now much more 
accurate $a=0$ calculations
\cite{pisaQ}
from which we can extract the coefficient of the $1/N^4$ term
in the continuum limit and we find that it has a similar size 
to the one we have found at $a=1/5T_c$, although with the
opposite sign.

\subsection{Topological fluctuations at ${\bf T\simeq T_c}$}
\label{subsection_topTc}

When the deconfining transition is first order, as it is 
for $N \geq 3$, and when $T$ is sufficiently close to $T_c$,
the system will repeatedly tunnel between confined and deconfined
phases. At $T\simeq T_c$ the tunneling probability $\to 0$ as 
$V\to\infty$, and what that means for our Monte Carlo
calculation is that for large $V$ we have very long sequences
of lattice gauge fields that remain in a single phase.
Thus we can calculate the topological charge separately in
the confined and deconfined phases when $T$ is near $T_c$. 
Indeed there exist calculations of this kind in $SU(3)$ (see
\cite{mt-TcQsu3}
for early examples)
which show that at $T\simeq T_c$ the topological susceptibility 
is markedly suppressed in the deconfined phase as compared to 
the confined phase. In this Section we extend these
calculations to $N > 3$ so as to learn what happens as
$N \to \infty$.

In Table~\ref{table_chiTc} we present some results on
the value that $\chi_t$ takes in the confined and deconfined
phases at  $T\simeq T_c$, for various values of $N$. The 
calculations are at a finite lattice spacing and the value of
$a$ is the same for all $N$ when expressed in units of $T_c$:
specifically $aT_c \simeq 5$. We present the susceptibility
in the form of the dimensionless ratio ${{\chi_t}/{\sigma^2}}$,
which possesses a simple interpretation
in the limit of a dilute gas of topological charges:
\be
{{\chi_t}\over{\sigma^2}}
\equiv
{{\langle Q^2 \rangle}\over{\{aL_s\}^3{aL_t}\sigma^2}} 
=
{{\langle N_I \rangle}\over{\{aL_s\}^3{aL_t}\sigma^2}} 
\label{eqNhiTc}
\ee
where $N_I$ is the total number of topological charges in the field on 
the  $L^3_sL_t$ lattice. Thus the quantity ${{\chi_t}/{\sigma^2}}$
measures the total number of topological charges per
unit (four)volume, where the volume is measured in units
of the confining string tension. (Recall that in QCD$_{N=3}$, 
one has $\surd\sigma \simeq 440 {\mathrm{MeV}} \simeq 
0.45 {\mathrm{fm}}$.) This interpretation is of course only 
exact in the dilute gas limit, which is certainly not valid
in the confining phase of the $SU(N)$ gauge theory, although,
as we shall see below, it does appear to be the case in the deconfined 
phase.

Since we are at a finite $a$, different ways of calculating
$Q$ will differ -- typically by $O(a^2)$ lattice artifacts.
The results we present in the remainder of this sub-section are
based on the real-valued topological charge, $Q_r$, calculated
after 20 cooling sweeps. The  results one would obtain 
using the integer value topological charge and/or fewer, say 10, 
cooling sweeps lead to the same conclusions.

For $SU(8)$ we generated two sequences of 50,000 fields on
$12^3 5$ lattices with one starting in the confined phase and one
in the deconfined phase, and with the inverse coupling $\beta$ 
chosen so that $T = 1/5a(\beta) \simeq T_c$. The spatial volume
is large enough that not only are there no phase transitions 
but there is no sign of even a partial tunneling. (We use the
time-like Polyakov loop averaged over the lattice as an order
parameter to discriminate phases.) Thus there is no ambiguity 
in calculating quantities separately in the confined and
deconfined phases.
For $SU(6)$ we do the same (with sequences of 100,000 fields) 
on $16^3 5$ lattices, for $SU(4)$ (with sequences of 25,000 
fields) on $32^3 5$ lattices and for $SU(3)$ on $32^3 5$ lattices. 
In the last case the volume is small enough that we observe
several tunnelings between the phases, but these are
sufficiently well-defined that there is no significant ambiguity 
in separating the phases from each other. That one can
work on smaller volumes as $N\uparrow$ is mainly due to 
the fact that the surface tension of the bubble separating
the confined and deconfined vacua increases with $N$
\cite{recentTc}
so that for a given volume $V$ the tunneling probability
$\to 0$ as $N\to\infty$ at $T=T_c$, and so we have longer sequences
of fields in which the system remains in a single phase. 
Finally we remark that 
while we have not included $SU(2)$ here, since
its transition is second order, one also finds in that case 
a suppression of $\chi_t$ for $T > T_c$
\cite{mt-TcQsu2}.

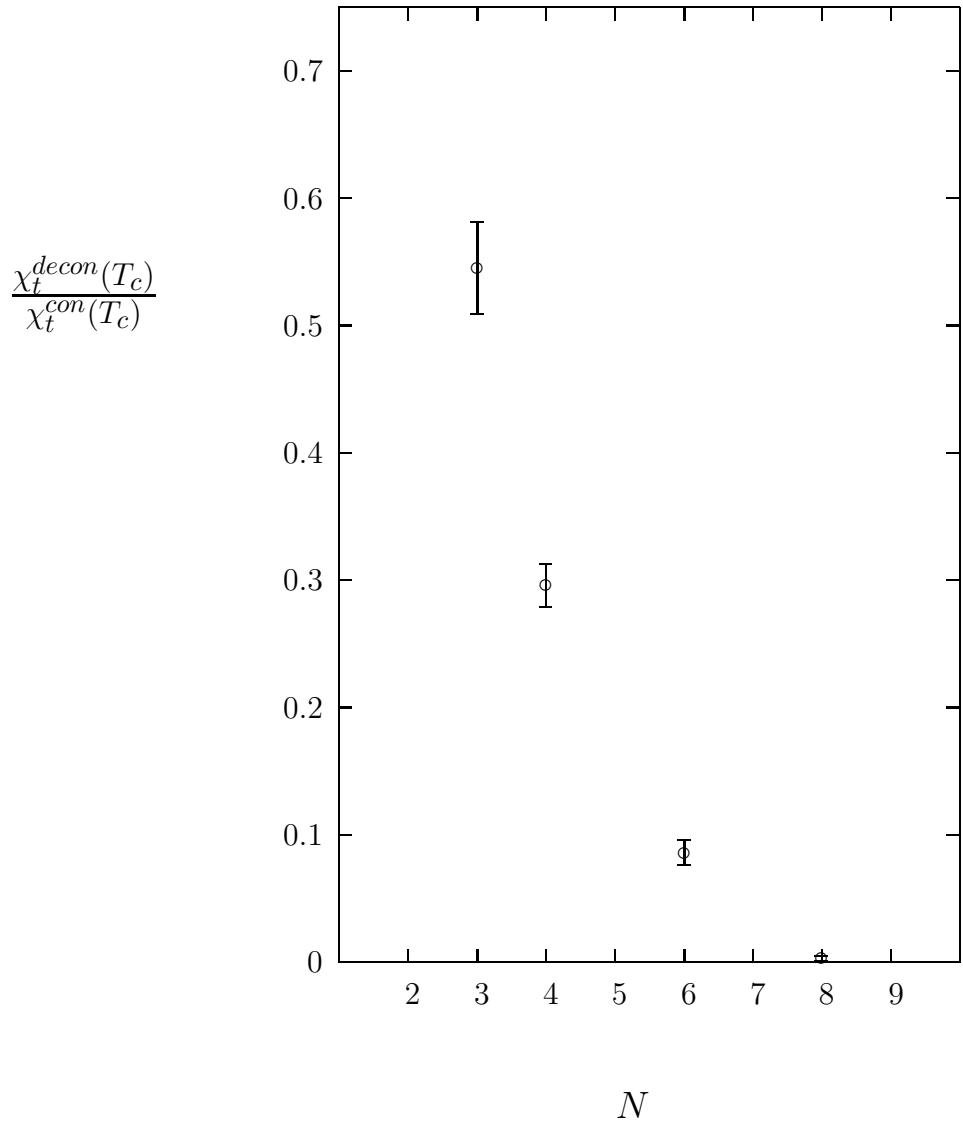
\begin  {figure}[p]
\begin  {center}
\leavevmode
\setlength{\unitlength}{0.240900pt}
\ifx\plotpoint\undefined\newsavebox{\plotpoint}\fi
\sbox{\plotpoint}{\rule[-0.200pt]{0.400pt}{0.400pt}}%
\begin{picture}(1500,1800)(0,0)
\font\gnuplot=cmr10 at 12pt
\gnuplot
\sbox{\plotpoint}{\rule[-0.200pt]{0.400pt}{0.400pt}}%
\put(450.0,250.0){\rule[-0.200pt]{4.818pt}{0.400pt}}
\put(425,250){\makebox(0,0)[r]{\ \ {$0$}}}
\put(1405.0,250.0){\rule[-0.200pt]{4.818pt}{0.400pt}}
\put(450.0,450.0){\rule[-0.200pt]{4.818pt}{0.400pt}}
\put(425,450){\makebox(0,0)[r]{\ \ {$0.1$}}}
\put(1405.0,450.0){\rule[-0.200pt]{4.818pt}{0.400pt}}
\put(450.0,650.0){\rule[-0.200pt]{4.818pt}{0.400pt}}
\put(425,650){\makebox(0,0)[r]{\ \ {$0.2$}}}
\put(1405.0,650.0){\rule[-0.200pt]{4.818pt}{0.400pt}}
\put(450.0,850.0){\rule[-0.200pt]{4.818pt}{0.400pt}}
\put(425,850){\makebox(0,0)[r]{\ \ {$0.3$}}}
\put(1405.0,850.0){\rule[-0.200pt]{4.818pt}{0.400pt}}
\put(450.0,1050.0){\rule[-0.200pt]{4.818pt}{0.400pt}}
\put(425,1050){\makebox(0,0)[r]{\ \ {$0.4$}}}
\put(1405.0,1050.0){\rule[-0.200pt]{4.818pt}{0.400pt}}
\put(450.0,1250.0){\rule[-0.200pt]{4.818pt}{0.400pt}}
\put(425,1250){\makebox(0,0)[r]{\ \ {$0.5$}}}
\put(1405.0,1250.0){\rule[-0.200pt]{4.818pt}{0.400pt}}
\put(450.0,1450.0){\rule[-0.200pt]{4.818pt}{0.400pt}}
\put(425,1450){\makebox(0,0)[r]{\ \ {$0.6$}}}
\put(1405.0,1450.0){\rule[-0.200pt]{4.818pt}{0.400pt}}
\put(450.0,1650.0){\rule[-0.200pt]{4.818pt}{0.400pt}}
\put(425,1650){\makebox(0,0)[r]{\ \ {$0.7$}}}
\put(1405.0,1650.0){\rule[-0.200pt]{4.818pt}{0.400pt}}
\put(558.0,250.0){\rule[-0.200pt]{0.400pt}{4.818pt}}
\put(558,200){\makebox(0,0){\ {$2$}}}
\put(558.0,1730.0){\rule[-0.200pt]{0.400pt}{4.818pt}}
\put(667.0,250.0){\rule[-0.200pt]{0.400pt}{4.818pt}}
\put(667,200){\makebox(0,0){\ {$3$}}}
\put(667.0,1730.0){\rule[-0.200pt]{0.400pt}{4.818pt}}
\put(775.0,250.0){\rule[-0.200pt]{0.400pt}{4.818pt}}
\put(775,200){\makebox(0,0){\ {$4$}}}
\put(775.0,1730.0){\rule[-0.200pt]{0.400pt}{4.818pt}}
\put(883.0,250.0){\rule[-0.200pt]{0.400pt}{4.818pt}}
\put(883,200){\makebox(0,0){\ {$5$}}}
\put(883.0,1730.0){\rule[-0.200pt]{0.400pt}{4.818pt}}
\put(992.0,250.0){\rule[-0.200pt]{0.400pt}{4.818pt}}
\put(992,200){\makebox(0,0){\ {$6$}}}
\put(992.0,1730.0){\rule[-0.200pt]{0.400pt}{4.818pt}}
\put(1100.0,250.0){\rule[-0.200pt]{0.400pt}{4.818pt}}
\put(1100,200){\makebox(0,0){\ {$7$}}}
\put(1100.0,1730.0){\rule[-0.200pt]{0.400pt}{4.818pt}}
\put(1208.0,250.0){\rule[-0.200pt]{0.400pt}{4.818pt}}
\put(1208,200){\makebox(0,0){\ {$8$}}}
\put(1208.0,1730.0){\rule[-0.200pt]{0.400pt}{4.818pt}}
\put(1317.0,250.0){\rule[-0.200pt]{0.400pt}{4.818pt}}
\put(1317,200){\makebox(0,0){\ {$9$}}}
\put(1317.0,1730.0){\rule[-0.200pt]{0.400pt}{4.818pt}}
\put(450.0,250.0){\rule[-0.200pt]{234.877pt}{0.400pt}}
\put(1425.0,250.0){\rule[-0.200pt]{0.400pt}{361.350pt}}
\put(450.0,1750.0){\rule[-0.200pt]{234.877pt}{0.400pt}}
\put(50,1300){\makebox(0,0){\Large{${{\chi_t^{decon}(T_c)}\over{\chi_t^{con}(T_c)} }$}}}
\put(912,25){\makebox(0,0){\large{$N$}}}
\put(450.0,250.0){\rule[-0.200pt]{0.400pt}{361.350pt}}
\put(667.0,1268.0){\rule[-0.200pt]{0.400pt}{34.690pt}}
\put(657.0,1268.0){\rule[-0.200pt]{4.818pt}{0.400pt}}
\put(657.0,1412.0){\rule[-0.200pt]{4.818pt}{0.400pt}}
\put(775.0,808.0){\rule[-0.200pt]{0.400pt}{16.381pt}}
\put(765.0,808.0){\rule[-0.200pt]{4.818pt}{0.400pt}}
\put(765.0,876.0){\rule[-0.200pt]{4.818pt}{0.400pt}}
\put(992.0,402.0){\rule[-0.200pt]{0.400pt}{9.636pt}}
\put(982.0,402.0){\rule[-0.200pt]{4.818pt}{0.400pt}}
\put(982.0,442.0){\rule[-0.200pt]{4.818pt}{0.400pt}}
\put(1208.0,252.0){\rule[-0.200pt]{0.400pt}{1.927pt}}
\put(1198.0,252.0){\rule[-0.200pt]{4.818pt}{0.400pt}}
\put(667,1340){\circle{18}}
\put(775,842){\circle{18}}
\put(992,422){\circle{18}}
\put(1208,256){\circle{18}}
\put(1198.0,260.0){\rule[-0.200pt]{4.818pt}{0.400pt}}
\end{picture}
\end    {center}
\vskip 0.15in
\caption{The ratio of the topological susceptibility, $\chi_t$, 
in the deconfined and confined phases at $T\simeq T_c$.}
\label{fig_chiNTc}
\end    {figure}

For comparison we  also show in Table~\ref{table_chiTc} 
the `T=0' value of ${{\chi_t}/{\sigma^2}}$, as obtained from 
Table~\ref{table_chi5b} (or a slight extrapolation
thereof). Comparing the various values in that Table, we see  
that ${{\chi_t}/{\sigma^2}}$ remains approximately
constant in the confining phase at all $T$, but that it is
strongly suppressed in the deconfined phase. Moreover
the suppression becomes rapidly more severe as $N$ grows.
To illustrate this we show in Fig.~\ref{fig_chiNTc} how the 
ratio of $\chi_t$ in the deconfined phase to its value in
the confined phase varies with $N$ at $T=T_c$ . It seems
quite clear that in the deconfined phase $\chi_t(T=T_c) \to 0$ 
as $N \to \infty$.

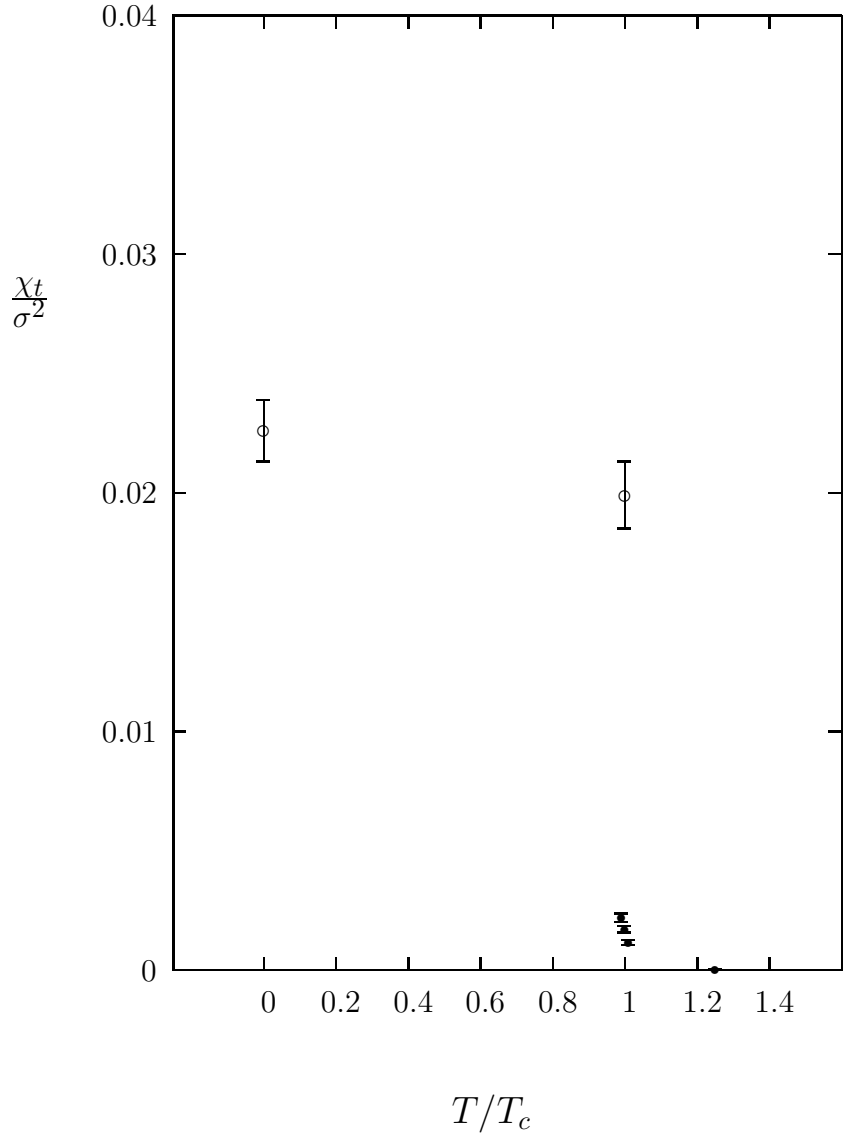
\begin  {figure}[p]
\begin  {center}
\leavevmode
\setlength{\unitlength}{0.240900pt}
\ifx\plotpoint\undefined\newsavebox{\plotpoint}\fi
\sbox{\plotpoint}{\rule[-0.200pt]{0.400pt}{0.400pt}}%
\begin{picture}(1500,1800)(0,0)
\font\gnuplot=cmr10 at 12pt
\gnuplot
\sbox{\plotpoint}{\rule[-0.200pt]{0.400pt}{0.400pt}}%
\put(375.0,250.0){\rule[-0.200pt]{4.818pt}{0.400pt}}
\put(350,250){\makebox(0,0)[r]{\ \ {$0$}}}
\put(1405.0,250.0){\rule[-0.200pt]{4.818pt}{0.400pt}}
\put(375.0,625.0){\rule[-0.200pt]{4.818pt}{0.400pt}}
\put(350,625){\makebox(0,0)[r]{\ \ {$0.01$}}}
\put(1405.0,625.0){\rule[-0.200pt]{4.818pt}{0.400pt}}
\put(375.0,1000.0){\rule[-0.200pt]{4.818pt}{0.400pt}}
\put(350,1000){\makebox(0,0)[r]{\ \ {$0.02$}}}
\put(1405.0,1000.0){\rule[-0.200pt]{4.818pt}{0.400pt}}
\put(375.0,1375.0){\rule[-0.200pt]{4.818pt}{0.400pt}}
\put(350,1375){\makebox(0,0)[r]{\ \ {$0.03$}}}
\put(1405.0,1375.0){\rule[-0.200pt]{4.818pt}{0.400pt}}
\put(375.0,1750.0){\rule[-0.200pt]{4.818pt}{0.400pt}}
\put(350,1750){\makebox(0,0)[r]{\ \ {$0.04$}}}
\put(1405.0,1750.0){\rule[-0.200pt]{4.818pt}{0.400pt}}
\put(517.0,250.0){\rule[-0.200pt]{0.400pt}{4.818pt}}
\put(517,200){\makebox(0,0){\ {$0$}}}
\put(517.0,1730.0){\rule[-0.200pt]{0.400pt}{4.818pt}}
\put(630.0,250.0){\rule[-0.200pt]{0.400pt}{4.818pt}}
\put(630,200){\makebox(0,0){\ {$0.2$}}}
\put(630.0,1730.0){\rule[-0.200pt]{0.400pt}{4.818pt}}
\put(744.0,250.0){\rule[-0.200pt]{0.400pt}{4.818pt}}
\put(744,200){\makebox(0,0){\ {$0.4$}}}
\put(744.0,1730.0){\rule[-0.200pt]{0.400pt}{4.818pt}}
\put(857.0,250.0){\rule[-0.200pt]{0.400pt}{4.818pt}}
\put(857,200){\makebox(0,0){\ {$0.6$}}}
\put(857.0,1730.0){\rule[-0.200pt]{0.400pt}{4.818pt}}
\put(971.0,250.0){\rule[-0.200pt]{0.400pt}{4.818pt}}
\put(971,200){\makebox(0,0){\ {$0.8$}}}
\put(971.0,1730.0){\rule[-0.200pt]{0.400pt}{4.818pt}}
\put(1084.0,250.0){\rule[-0.200pt]{0.400pt}{4.818pt}}
\put(1084,200){\makebox(0,0){\ {$1$}}}
\put(1084.0,1730.0){\rule[-0.200pt]{0.400pt}{4.818pt}}
\put(1198.0,250.0){\rule[-0.200pt]{0.400pt}{4.818pt}}
\put(1198,200){\makebox(0,0){\ {$1.2$}}}
\put(1198.0,1730.0){\rule[-0.200pt]{0.400pt}{4.818pt}}
\put(1311.0,250.0){\rule[-0.200pt]{0.400pt}{4.818pt}}
\put(1311,200){\makebox(0,0){\ {$1.4$}}}
\put(1311.0,1730.0){\rule[-0.200pt]{0.400pt}{4.818pt}}
\put(375.0,250.0){\rule[-0.200pt]{252.945pt}{0.400pt}}
\put(1425.0,250.0){\rule[-0.200pt]{0.400pt}{361.350pt}}
\put(375.0,1750.0){\rule[-0.200pt]{252.945pt}{0.400pt}}
\put(150,1300){\makebox(0,0){\Large{${{\chi_t}\over{\sigma^2}}$}}}
\put(875,25){\makebox(0,0){\large{$T/T_c$}}}
\put(375.0,250.0){\rule[-0.200pt]{0.400pt}{361.350pt}}
\put(517.0,1049.0){\rule[-0.200pt]{0.400pt}{23.367pt}}
\put(507.0,1049.0){\rule[-0.200pt]{4.818pt}{0.400pt}}
\put(507.0,1146.0){\rule[-0.200pt]{4.818pt}{0.400pt}}
\put(1084.0,944.0){\rule[-0.200pt]{0.400pt}{25.294pt}}
\put(1074.0,944.0){\rule[-0.200pt]{4.818pt}{0.400pt}}
\put(517,1098){\circle{18}}
\put(1084,996){\circle{18}}
\put(1074.0,1049.0){\rule[-0.200pt]{4.818pt}{0.400pt}}
\put(1079.0,325.0){\rule[-0.200pt]{0.400pt}{3.373pt}}
\put(1069.0,325.0){\rule[-0.200pt]{4.818pt}{0.400pt}}
\put(1069.0,339.0){\rule[-0.200pt]{4.818pt}{0.400pt}}
\put(1084.0,309.0){\rule[-0.200pt]{0.400pt}{2.409pt}}
\put(1074.0,309.0){\rule[-0.200pt]{4.818pt}{0.400pt}}
\put(1074.0,319.0){\rule[-0.200pt]{4.818pt}{0.400pt}}
\put(1090.0,289.0){\rule[-0.200pt]{0.400pt}{1.927pt}}
\put(1080.0,289.0){\rule[-0.200pt]{4.818pt}{0.400pt}}
\put(1080.0,297.0){\rule[-0.200pt]{4.818pt}{0.400pt}}
\put(1226.0,250.0){\rule[-0.200pt]{0.400pt}{0.482pt}}
\put(1216.0,250.0){\rule[-0.200pt]{4.818pt}{0.400pt}}
\put(1079,332){\circle*{12}}
\put(1084,314){\circle*{12}}
\put(1090,293){\circle*{12}}
\put(1226,250){\circle*{12}}
\put(1216.0,252.0){\rule[-0.200pt]{4.818pt}{0.400pt}}
\end{picture}
\end    {center}
\vskip 0.15in
\caption{The topological susceptibility, $\chi_t$, in units of the
string tension, $\sigma$, plotted versus $T/T_c$ for $SU(6)$, plotted
separately for the confined, $\circ$, and deconfined, $\bullet$, phases.   
The calculations are performed at a lattice spacing $a \simeq 1/5 T_c$.}
\label{fig_chi6T}
\end    {figure}

In Tables~\ref{table_chiTN8} to ~\ref{table_chiTN3} we list
some further calculations, some at higher $T$. (The values
marked by a $\star$ in Tables~\ref{table_chiTc} to 
~\ref{table_chiTN3} have been obtained from sequences of
fields containing sufficient tunnelings for there to
be some danger that the quoted error is underestimated.)
Together with our other calculations, these provide more 
information on the $T$ dependence of $\chi_t$. As an 
example, we plot in Fig.~\ref{fig_chi6T} the $T$
dependence of ${{\chi_t}/{\sigma^2}}$ for $SU(6)$. We
see that the value of  $\chi_t$ in the deconfined phase
shows a strong temperature dependence, vanishing rapidly
as $T$ grows.

We conclude that at $N =\infty$ a rather simple picture
emerges, 
\be
\lim_{N\to\infty}{{\chi^{\frac{1}{4}}_t}\over{\surd\sigma}}
=\left \{ \begin{array}{ll}
0
& \ \ \ \ \mbox{deconfined $\forall T$,} \\
& \\
const (\not= 0)
& \ \ \ \  \ \mbox{confined $\forall T$,}
\end{array}
\right.
\label{eqNhiNinf}
\ee
with the suppression in the deconfined phase being very
rapid -- probably exponential in a power of $N$. 
This shows that the the rapid  instanton suppression, 
for which we have reliable semiclassical arguments only 
at high-$T$ where $g^2(T)$ is small, in fact occurs at all 
values of $T$, even down to $T=T_c$, and indeed for all values of
the instanton size $\rho$.
The confined phase, on the other hand, appears
to be impervious to the large-$\rho$ suppression, even
at $T\simeq T_c$. To make this interpretation convincing, 
we need to estimate the instanton size distribution,
and this we shall now do.

\subsection{Instanton size distributions}
\label{subsection_size}

To obtain information on the instanton size density, $D(\rho)$, 
we smoothen (`cool') the lattice fields, so that our lattice
topological charge density operator should approximate the
continuum one, and we then  locate the  peaks in that
density, $Q(x)$. We identify each of these with an instanton, 
whose size $\rho$ is given by  eqn(\ref{eqn_Qpeak}). (As described 
in Section~\ref{section_lattice}.) This procedure is clearly
unambiguous for small instantons, which correspond to large peaks 
in $Q(x)$, but becomes increasingly more  ambiguous for larger 
instantons that correspond to very low and smooth bumps in  $Q(x)$.
To help avoid mere fluctuations in the instanton profile being 
misinterpreted as extra instantons, we do not count peaks that are 
within a distance of $2a$ of a larger peak. This algorithm is
of course very crude (as emphasised in Section~\ref{section_lattice})
but unfortunately we are not aware of a really reliable alternative.
We shall therefore focus on the qualitative features of our
calculated size densities.

\subsubsection{T=0}
\label{subsubsection_sizeT0}

The $T=0$ size density was calculated for $2\leq N \leq 5$ in 
\cite{blmt-glue}.
Here we shall extend the comparison to $N=8$. However none
of our lattice spacings is as fine as that used in Fig.~13 of 
\cite{blmt-glue}
and this means that our comparison will be more affected
by lattice spacing artifacts.

\begin  {figure}[p]
\begin  {center}
\leavevmode
\setlength{\unitlength}{0.240900pt}
\ifx\plotpoint\undefined\newsavebox{\plotpoint}\fi
\sbox{\plotpoint}{\rule[-0.200pt]{0.400pt}{0.400pt}}%
\begin{picture}(1500,1800)(0,0)
\font\gnuplot=cmr10 at 12pt
\gnuplot
\sbox{\plotpoint}{\rule[-0.200pt]{0.400pt}{0.400pt}}%
\put(375.0,250.0){\rule[-0.200pt]{4.818pt}{0.400pt}}
\put(350,250){\makebox(0,0)[r]{\ \ {$0$}}}
\put(1405.0,250.0){\rule[-0.200pt]{4.818pt}{0.400pt}}
\put(375.0,420.0){\rule[-0.200pt]{4.818pt}{0.400pt}}
\put(350,420){\makebox(0,0)[r]{\ \ {$0.25$}}}
\put(1405.0,420.0){\rule[-0.200pt]{4.818pt}{0.400pt}}
\put(375.0,591.0){\rule[-0.200pt]{4.818pt}{0.400pt}}
\put(350,591){\makebox(0,0)[r]{\ \ {$0.5$}}}
\put(1405.0,591.0){\rule[-0.200pt]{4.818pt}{0.400pt}}
\put(375.0,761.0){\rule[-0.200pt]{4.818pt}{0.400pt}}
\put(350,761){\makebox(0,0)[r]{\ \ {$0.75$}}}
\put(1405.0,761.0){\rule[-0.200pt]{4.818pt}{0.400pt}}
\put(375.0,932.0){\rule[-0.200pt]{4.818pt}{0.400pt}}
\put(350,932){\makebox(0,0)[r]{\ \ {$1$}}}
\put(1405.0,932.0){\rule[-0.200pt]{4.818pt}{0.400pt}}
\put(375.0,1102.0){\rule[-0.200pt]{4.818pt}{0.400pt}}
\put(350,1102){\makebox(0,0)[r]{\ \ {$1.25$}}}
\put(1405.0,1102.0){\rule[-0.200pt]{4.818pt}{0.400pt}}
\put(375.0,1273.0){\rule[-0.200pt]{4.818pt}{0.400pt}}
\put(350,1273){\makebox(0,0)[r]{\ \ {$1.5$}}}
\put(1405.0,1273.0){\rule[-0.200pt]{4.818pt}{0.400pt}}
\put(375.0,1443.0){\rule[-0.200pt]{4.818pt}{0.400pt}}
\put(350,1443){\makebox(0,0)[r]{\ \ {$1.75$}}}
\put(1405.0,1443.0){\rule[-0.200pt]{4.818pt}{0.400pt}}
\put(375.0,1614.0){\rule[-0.200pt]{4.818pt}{0.400pt}}
\put(350,1614){\makebox(0,0)[r]{\ \ {$2$}}}
\put(1405.0,1614.0){\rule[-0.200pt]{4.818pt}{0.400pt}}
\put(375.0,250.0){\rule[-0.200pt]{0.400pt}{4.818pt}}
\put(375,200){\makebox(0,0){\ {$0$}}}
\put(375.0,1730.0){\rule[-0.200pt]{0.400pt}{4.818pt}}
\put(463.0,250.0){\rule[-0.200pt]{0.400pt}{4.818pt}}
\put(463,200){\makebox(0,0){\ {$1$}}}
\put(463.0,1730.0){\rule[-0.200pt]{0.400pt}{4.818pt}}
\put(550.0,250.0){\rule[-0.200pt]{0.400pt}{4.818pt}}
\put(550,200){\makebox(0,0){\ {$2$}}}
\put(550.0,1730.0){\rule[-0.200pt]{0.400pt}{4.818pt}}
\put(638.0,250.0){\rule[-0.200pt]{0.400pt}{4.818pt}}
\put(638,200){\makebox(0,0){\ {$3$}}}
\put(638.0,1730.0){\rule[-0.200pt]{0.400pt}{4.818pt}}
\put(725.0,250.0){\rule[-0.200pt]{0.400pt}{4.818pt}}
\put(725,200){\makebox(0,0){\ {$4$}}}
\put(725.0,1730.0){\rule[-0.200pt]{0.400pt}{4.818pt}}
\put(813.0,250.0){\rule[-0.200pt]{0.400pt}{4.818pt}}
\put(813,200){\makebox(0,0){\ {$5$}}}
\put(813.0,1730.0){\rule[-0.200pt]{0.400pt}{4.818pt}}
\put(900.0,250.0){\rule[-0.200pt]{0.400pt}{4.818pt}}
\put(900,200){\makebox(0,0){\ {$6$}}}
\put(900.0,1730.0){\rule[-0.200pt]{0.400pt}{4.818pt}}
\put(988.0,250.0){\rule[-0.200pt]{0.400pt}{4.818pt}}
\put(988,200){\makebox(0,0){\ {$7$}}}
\put(988.0,1730.0){\rule[-0.200pt]{0.400pt}{4.818pt}}
\put(1075.0,250.0){\rule[-0.200pt]{0.400pt}{4.818pt}}
\put(1075,200){\makebox(0,0){\ {$8$}}}
\put(1075.0,1730.0){\rule[-0.200pt]{0.400pt}{4.818pt}}
\put(1163.0,250.0){\rule[-0.200pt]{0.400pt}{4.818pt}}
\put(1163,200){\makebox(0,0){\ {$9$}}}
\put(1163.0,1730.0){\rule[-0.200pt]{0.400pt}{4.818pt}}
\put(1250.0,250.0){\rule[-0.200pt]{0.400pt}{4.818pt}}
\put(1250,200){\makebox(0,0){\ {$10$}}}
\put(1250.0,1730.0){\rule[-0.200pt]{0.400pt}{4.818pt}}
\put(1338.0,250.0){\rule[-0.200pt]{0.400pt}{4.818pt}}
\put(1338,200){\makebox(0,0){\ {$11$}}}
\put(1338.0,1730.0){\rule[-0.200pt]{0.400pt}{4.818pt}}
\put(1425.0,250.0){\rule[-0.200pt]{0.400pt}{4.818pt}}
\put(1425,200){\makebox(0,0){\ {$12$}}}
\put(1425.0,1730.0){\rule[-0.200pt]{0.400pt}{4.818pt}}
\put(375.0,250.0){\rule[-0.200pt]{252.945pt}{0.400pt}}
\put(1425.0,250.0){\rule[-0.200pt]{0.400pt}{361.350pt}}
\put(375.0,1750.0){\rule[-0.200pt]{252.945pt}{0.400pt}}
\put(150,1300){\makebox(0,0){\Large{$D(\rho)$}}}
\put(875,25){\makebox(0,0){\large{$\rho$}}}
\put(375.0,250.0){\rule[-0.200pt]{0.400pt}{361.350pt}}
\put(564,251){\circle*{12}}
\put(587,252){\circle*{12}}
\put(607,253){\circle*{12}}
\put(630,256){\circle*{12}}
\put(651,278){\circle*{12}}
\put(672,309){\circle*{12}}
\put(694,398){\circle*{12}}
\put(715,552){\circle*{12}}
\put(736,759){\circle*{12}}
\put(758,1030){\circle*{12}}
\put(780,1205){\circle*{12}}
\put(801,1305){\circle*{12}}
\put(824,1333){\circle*{12}}
\put(845,1248){\circle*{12}}
\put(867,1098){\circle*{12}}
\put(889,909){\circle*{12}}
\put(910,787){\circle*{12}}
\put(932,679){\circle*{12}}
\put(954,527){\circle*{12}}
\put(976,491){\circle*{12}}
\put(998,408){\circle*{12}}
\put(1021,394){\circle*{12}}
\put(1041,356){\circle*{12}}
\put(1064,341){\circle*{12}}
\put(1086,307){\circle*{12}}
\put(1109,292){\circle*{12}}
\put(1131,285){\circle*{12}}
\put(1152,278){\circle*{12}}
\put(1171,270){\circle*{12}}
\put(1195,275){\circle*{12}}
\put(1216,264){\circle*{12}}
\put(1233,260){\circle*{12}}
\put(547,251){\makebox(0,0){$\times$}}
\put(567,253){\makebox(0,0){$\times$}}
\put(584,258){\makebox(0,0){$\times$}}
\put(604,275){\makebox(0,0){$\times$}}
\put(628,299){\makebox(0,0){$\times$}}
\put(649,343){\makebox(0,0){$\times$}}
\put(672,428){\makebox(0,0){$\times$}}
\put(693,522){\makebox(0,0){$\times$}}
\put(715,666){\makebox(0,0){$\times$}}
\put(736,812){\makebox(0,0){$\times$}}
\put(758,954){\makebox(0,0){$\times$}}
\put(780,1046){\makebox(0,0){$\times$}}
\put(801,1055){\makebox(0,0){$\times$}}
\put(823,1034){\makebox(0,0){$\times$}}
\put(846,962){\makebox(0,0){$\times$}}
\put(867,831){\makebox(0,0){$\times$}}
\put(889,741){\makebox(0,0){$\times$}}
\put(911,688){\makebox(0,0){$\times$}}
\put(932,587){\makebox(0,0){$\times$}}
\put(954,513){\makebox(0,0){$\times$}}
\put(977,469){\makebox(0,0){$\times$}}
\put(999,418){\makebox(0,0){$\times$}}
\put(1020,379){\makebox(0,0){$\times$}}
\put(1041,356){\makebox(0,0){$\times$}}
\put(1064,348){\makebox(0,0){$\times$}}
\put(1086,307){\makebox(0,0){$\times$}}
\put(1109,312){\makebox(0,0){$\times$}}
\put(1130,293){\makebox(0,0){$\times$}}
\put(1151,287){\makebox(0,0){$\times$}}
\put(1170,277){\makebox(0,0){$\times$}}
\put(1193,282){\makebox(0,0){$\times$}}
\put(1217,269){\makebox(0,0){$\times$}}
\put(1233,262){\makebox(0,0){$\times$}}
\put(546,256){\circle{18}}
\put(563,291){\circle{18}}
\put(584,317){\circle{18}}
\put(606,361){\circle{18}}
\put(627,413){\circle{18}}
\put(649,480){\circle{18}}
\put(671,543){\circle{18}}
\put(693,608){\circle{18}}
\put(715,663){\circle{18}}
\put(736,730){\circle{18}}
\put(758,728){\circle{18}}
\put(779,750){\circle{18}}
\put(802,732){\circle{18}}
\put(823,740){\circle{18}}
\put(845,682){\circle{18}}
\put(867,625){\circle{18}}
\put(889,546){\circle{18}}
\put(911,543){\circle{18}}
\put(932,469){\circle{18}}
\put(954,442){\circle{18}}
\put(976,421){\circle{18}}
\put(999,378){\circle{18}}
\put(1021,370){\circle{18}}
\put(1042,346){\circle{18}}
\put(1064,338){\circle{18}}
\put(1086,321){\circle{18}}
\put(1107,317){\circle{18}}
\put(1131,300){\circle{18}}
\put(1150,282){\circle{18}}
\put(1171,282){\circle{18}}
\put(1194,289){\circle{18}}
\put(1217,270){\circle{18}}
\put(1233,266){\circle{18}}
\put(545,260){\makebox(0,0){$+$}}
\put(563,320){\makebox(0,0){$+$}}
\put(583,370){\makebox(0,0){$+$}}
\put(606,436){\makebox(0,0){$+$}}
\put(627,458){\makebox(0,0){$+$}}
\put(649,486){\makebox(0,0){$+$}}
\put(671,543){\makebox(0,0){$+$}}
\put(692,583){\makebox(0,0){$+$}}
\put(715,596){\makebox(0,0){$+$}}
\put(736,612){\makebox(0,0){$+$}}
\put(758,616){\makebox(0,0){$+$}}
\put(779,593){\makebox(0,0){$+$}}
\put(801,587){\makebox(0,0){$+$}}
\put(823,554){\makebox(0,0){$+$}}
\put(845,544){\makebox(0,0){$+$}}
\put(868,510){\makebox(0,0){$+$}}
\put(889,473){\makebox(0,0){$+$}}
\put(911,455){\makebox(0,0){$+$}}
\put(933,426){\makebox(0,0){$+$}}
\put(955,412){\makebox(0,0){$+$}}
\put(976,395){\makebox(0,0){$+$}}
\put(999,355){\makebox(0,0){$+$}}
\put(1019,353){\makebox(0,0){$+$}}
\put(1042,342){\makebox(0,0){$+$}}
\put(1065,346){\makebox(0,0){$+$}}
\put(1086,318){\makebox(0,0){$+$}}
\put(1107,314){\makebox(0,0){$+$}}
\put(1131,308){\makebox(0,0){$+$}}
\put(1151,296){\makebox(0,0){$+$}}
\put(1170,294){\makebox(0,0){$+$}}
\put(1195,301){\makebox(0,0){$+$}}
\put(1215,275){\makebox(0,0){$+$}}
\put(1233,274){\makebox(0,0){$+$}}
\put(526,250){\makebox(0,0){$\star$}}
\put(544,271){\makebox(0,0){$\star$}}
\put(562,375){\makebox(0,0){$\star$}}
\put(583,413){\makebox(0,0){$\star$}}
\put(605,440){\makebox(0,0){$\star$}}
\put(627,444){\makebox(0,0){$\star$}}
\put(649,464){\makebox(0,0){$\star$}}
\put(670,467){\makebox(0,0){$\star$}}
\put(692,484){\makebox(0,0){$\star$}}
\put(715,462){\makebox(0,0){$\star$}}
\put(736,459){\makebox(0,0){$\star$}}
\put(757,444){\makebox(0,0){$\star$}}
\put(780,448){\makebox(0,0){$\star$}}
\put(801,442){\makebox(0,0){$\star$}}
\put(823,428){\makebox(0,0){$\star$}}
\put(845,410){\makebox(0,0){$\star$}}
\put(867,417){\makebox(0,0){$\star$}}
\put(889,391){\makebox(0,0){$\star$}}
\put(911,389){\makebox(0,0){$\star$}}
\put(932,367){\makebox(0,0){$\star$}}
\put(954,358){\makebox(0,0){$\star$}}
\put(976,350){\makebox(0,0){$\star$}}
\put(999,333){\makebox(0,0){$\star$}}
\put(1021,339){\makebox(0,0){$\star$}}
\put(1043,319){\makebox(0,0){$\star$}}
\put(1064,327){\makebox(0,0){$\star$}}
\put(1087,314){\makebox(0,0){$\star$}}
\put(1108,312){\makebox(0,0){$\star$}}
\put(1131,309){\makebox(0,0){$\star$}}
\put(1151,291){\makebox(0,0){$\star$}}
\put(1172,297){\makebox(0,0){$\star$}}
\put(1195,296){\makebox(0,0){$\star$}}
\put(1216,286){\makebox(0,0){$\star$}}
\put(1234,278){\makebox(0,0){$\star$}}
\end{picture}
\end    {center}
\vskip 0.15in
\caption{The instanton size density, $D(\rho)$, for
$N=2(\star),3(+),4(\circ),6(\times),8(\bullet)$
on (mostly) $10^4$ lattices with $a= 1/5T_c$.}
\label{fig_drho10}
\end    {figure}
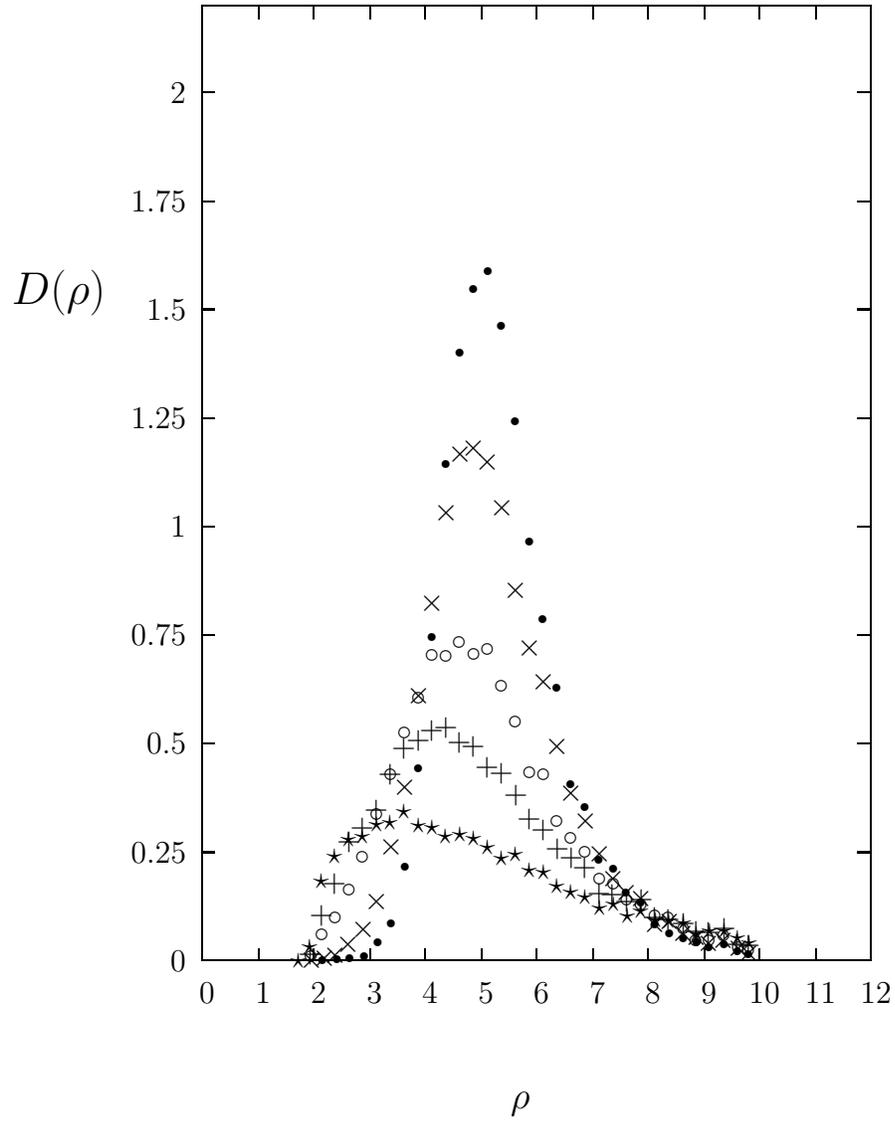

We start with the calculations listed in Table~\ref{table_chi5a}.
In each case we calculated the topological charge density
after 20 cooling sweeps and the resulting number densities are 
shown in Fig.~\ref{fig_drho10}. What we plot is the number of 
charges per $10^4$ lattice fields with $a=1/5T_c$. 
(If one wishes one can translate the volume
into more `physical' units using the listed values of the
string tension, $a\surd\sigma$, and using the real world
value of $\surd\sigma \simeq 440 {\mathrm{MeV}}$.)
The histograms are for numbers of instantons in 
bins of $\rho$ that are of size $\delta\rho = 0.25a$.
Each point is placed at the average value of $\rho$
within that bin, rather than in the middle of the bin,
which means that the points for different $N$ are
slightly shifted (in $\rho$) with respect to each other.

The value of $a$ is very coarse and this means that the 
effective cut-off induced by the cooling, which here we
can estimate to be at about $\rho \sim 2-2.5a$ from 
the $SU(2)$ density, is uncomfortably close to the typical instanton 
size. Nonetheless some qualitative features are clear.
First is the steepening of the distribution at small $\rho$.
This is expected theoretically and was already observed 
up to $N=5$ in
\cite{blmt-glue}.
More interesting, and only becoming apparent from the greater range 
of $N$  available to us in the present calculation, is that the 
whole distribution, at both small and large $\rho$, is becoming 
much more peaked as $N$ grows. Indeed, if we take the distributions
in Fig.~\ref{fig_drho10} seriously, we infer that
\be
\lim_{N\to\infty} D(\rho) = c(N)\delta(\rho-\rho_c)
 \ \ \ \  \ ; \  \rho_c \sim {1\over{T_c}}.
\label{eqn_drhoNinf}
\ee
The main uncertainty is to do with the reliability of our
identification of the larger topological charges. However since the 
size distribution narrows with increasing $N$, 
the relevant range of $\rho$ comes closer to the average, and
so our calculation should be more reliable at larger $N$. Some explicit
checks showing that our conclusions are robust against the
detailed cuts of our algorithm would be useful, but
have not been performed in the present calculation. Equally,
a systematic finite volume study would be useful. For
the case of $SU(3)$ there exist such finite volume studies (e.g.
\cite{dsmt}) 
and for higher $N$ our present study has larger
spatial volumes at $T\simeq T_c$, which suggest that finite
volume corrections are unimportant.

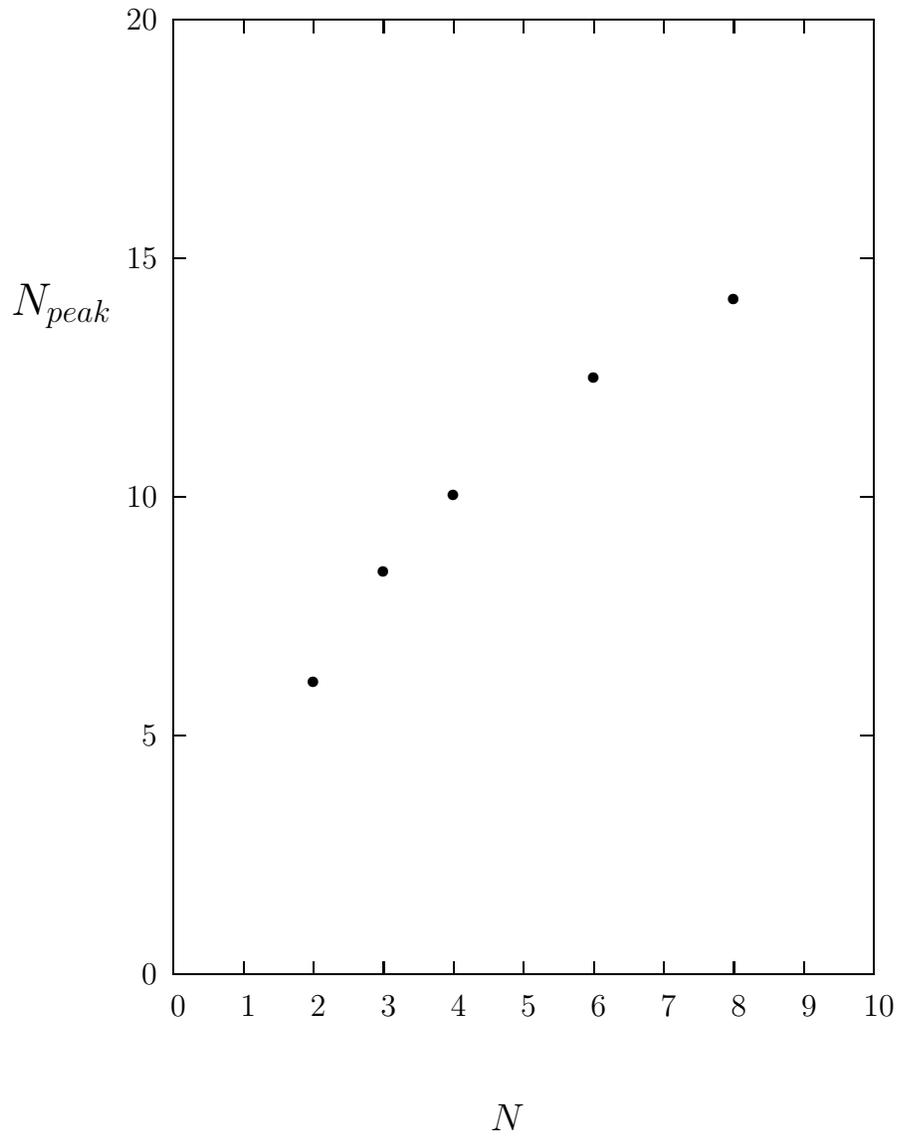
\begin  {figure}[p]
\begin  {center}
\leavevmode
\setlength{\unitlength}{0.240900pt}
\ifx\plotpoint\undefined\newsavebox{\plotpoint}\fi
\begin{picture}(1500,1800)(0,0)
\font\gnuplot=cmr10 at 12pt
\gnuplot
\sbox{\plotpoint}{\rule[-0.200pt]{0.400pt}{0.400pt}}%
\put(325.0,250.0){\rule[-0.200pt]{4.818pt}{0.400pt}}
\put(300,250){\makebox(0,0)[r]{\ \ {$0$}}}
\put(1405.0,250.0){\rule[-0.200pt]{4.818pt}{0.400pt}}
\put(325.0,625.0){\rule[-0.200pt]{4.818pt}{0.400pt}}
\put(300,625){\makebox(0,0)[r]{\ \ {$5$}}}
\put(1405.0,625.0){\rule[-0.200pt]{4.818pt}{0.400pt}}
\put(325.0,1000.0){\rule[-0.200pt]{4.818pt}{0.400pt}}
\put(300,1000){\makebox(0,0)[r]{\ \ {$10$}}}
\put(1405.0,1000.0){\rule[-0.200pt]{4.818pt}{0.400pt}}
\put(325.0,1375.0){\rule[-0.200pt]{4.818pt}{0.400pt}}
\put(300,1375){\makebox(0,0)[r]{\ \ {$15$}}}
\put(1405.0,1375.0){\rule[-0.200pt]{4.818pt}{0.400pt}}
\put(325.0,1750.0){\rule[-0.200pt]{4.818pt}{0.400pt}}
\put(300,1750){\makebox(0,0)[r]{\ \ {$20$}}}
\put(1405.0,1750.0){\rule[-0.200pt]{4.818pt}{0.400pt}}
\put(325.0,250.0){\rule[-0.200pt]{0.400pt}{4.818pt}}
\put(325,200){\makebox(0,0){\ {$0$}}}
\put(325.0,1730.0){\rule[-0.200pt]{0.400pt}{4.818pt}}
\put(435.0,250.0){\rule[-0.200pt]{0.400pt}{4.818pt}}
\put(435,200){\makebox(0,0){\ {$1$}}}
\put(435.0,1730.0){\rule[-0.200pt]{0.400pt}{4.818pt}}
\put(545.0,250.0){\rule[-0.200pt]{0.400pt}{4.818pt}}
\put(545,200){\makebox(0,0){\ {$2$}}}
\put(545.0,1730.0){\rule[-0.200pt]{0.400pt}{4.818pt}}
\put(655.0,250.0){\rule[-0.200pt]{0.400pt}{4.818pt}}
\put(655,200){\makebox(0,0){\ {$3$}}}
\put(655.0,1730.0){\rule[-0.200pt]{0.400pt}{4.818pt}}
\put(765.0,250.0){\rule[-0.200pt]{0.400pt}{4.818pt}}
\put(765,200){\makebox(0,0){\ {$4$}}}
\put(765.0,1730.0){\rule[-0.200pt]{0.400pt}{4.818pt}}
\put(875.0,250.0){\rule[-0.200pt]{0.400pt}{4.818pt}}
\put(875,200){\makebox(0,0){\ {$5$}}}
\put(875.0,1730.0){\rule[-0.200pt]{0.400pt}{4.818pt}}
\put(985.0,250.0){\rule[-0.200pt]{0.400pt}{4.818pt}}
\put(985,200){\makebox(0,0){\ {$6$}}}
\put(985.0,1730.0){\rule[-0.200pt]{0.400pt}{4.818pt}}
\put(1095.0,250.0){\rule[-0.200pt]{0.400pt}{4.818pt}}
\put(1095,200){\makebox(0,0){\ {$7$}}}
\put(1095.0,1730.0){\rule[-0.200pt]{0.400pt}{4.818pt}}
\put(1205.0,250.0){\rule[-0.200pt]{0.400pt}{4.818pt}}
\put(1205,200){\makebox(0,0){\ {$8$}}}
\put(1205.0,1730.0){\rule[-0.200pt]{0.400pt}{4.818pt}}
\put(1315.0,250.0){\rule[-0.200pt]{0.400pt}{4.818pt}}
\put(1315,200){\makebox(0,0){\ {$9$}}}
\put(1315.0,1730.0){\rule[-0.200pt]{0.400pt}{4.818pt}}
\put(1425.0,250.0){\rule[-0.200pt]{0.400pt}{4.818pt}}
\put(1425,200){\makebox(0,0){\ {$10$}}}
\put(1425.0,1730.0){\rule[-0.200pt]{0.400pt}{4.818pt}}
\put(325.0,250.0){\rule[-0.200pt]{264.990pt}{0.400pt}}
\put(1425.0,250.0){\rule[-0.200pt]{0.400pt}{361.350pt}}
\put(325.0,1750.0){\rule[-0.200pt]{264.990pt}{0.400pt}}
\put(150,1300){\makebox(0,0){\Large{$N_{peak}$}}}
\put(850,25){\makebox(0,0){\large{$N$}}}
\put(325.0,250.0){\rule[-0.200pt]{0.400pt}{361.350pt}}
\put(545,709){\circle*{18}}
\put(655,883){\circle*{18}}
\put(765,1003){\circle*{18}}
\put(985,1188){\circle*{18}}
\put(1205,1311){\circle*{18}}
\end{picture}
\end    {center}
\vskip 0.15in
\caption{The average number of instantons versus $N$ 
at $a=1/5T_c$ on a $10^4$ volume, from the calculations
listed in Table~\ref{table_chi5a}}
\label{fig_numpk}
\end    {figure}

The integrated number of topological charges also appears to 
increase with $N$, as shown in Fig.~\ref{fig_numpk}. One might
imagine that one is seeing the theoretically expected growth 
that was discussed in Section~\ref{subsection_Q}. However one 
needs to be cautious. The volumes at different $N$ that are 
used here are only constant if expressed in units of $T_c$.
If instead they are expressed in units of, say, $\surd\sigma$
most of the variation that we see in Fig.~\ref{fig_numpk} 
for $N \leq 4$ disappears. This is because the volume goes
as the 4'th power of the scale. It is only for $N \geq 6$
that we are close enough to $N=\infty$ for this ambiguity
to become negligible. 

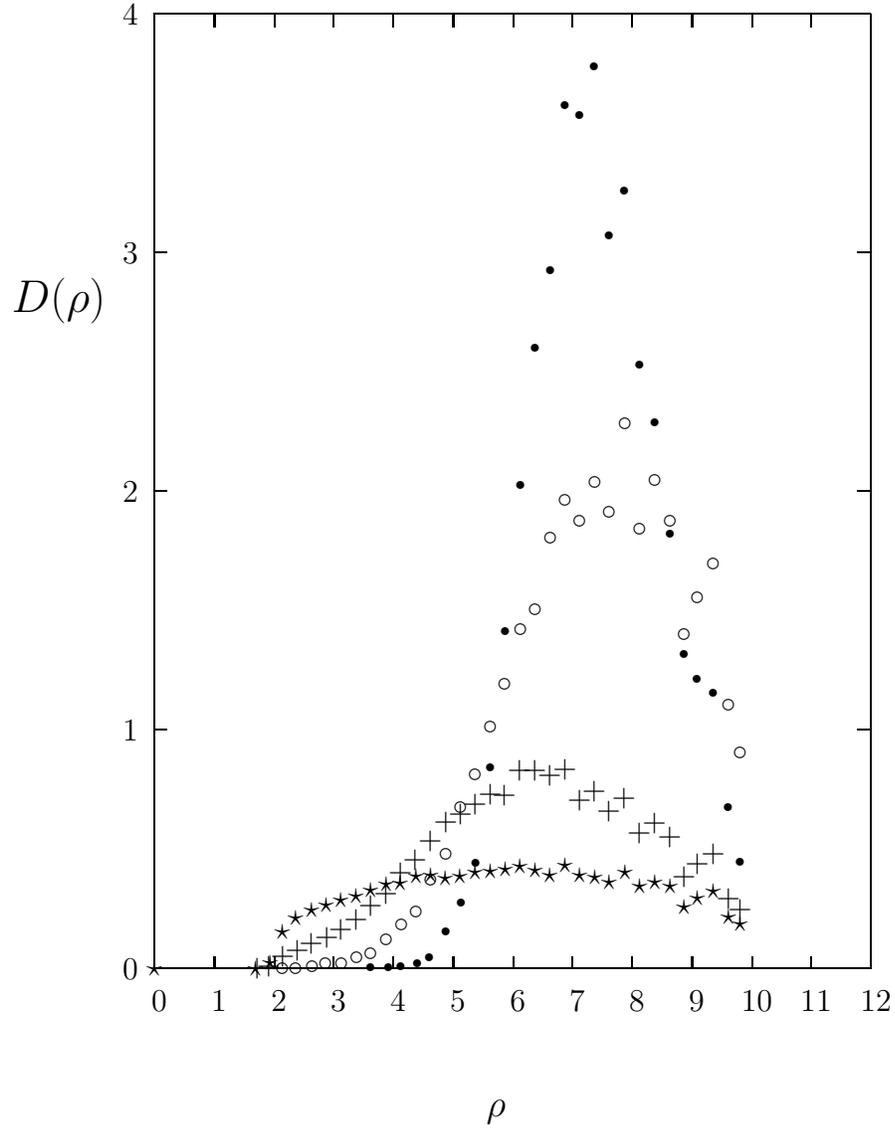
\begin  {figure}[p]
\begin  {center}
\leavevmode
\setlength{\unitlength}{0.240900pt}
\ifx\plotpoint\undefined\newsavebox{\plotpoint}\fi
\begin{picture}(1500,1800)(0,0)
\font\gnuplot=cmr10 at 12pt
\gnuplot
\sbox{\plotpoint}{\rule[-0.200pt]{0.400pt}{0.400pt}}%
\put(300.0,250.0){\rule[-0.200pt]{4.818pt}{0.400pt}}
\put(275,250){\makebox(0,0)[r]{\ \ {$0$}}}
\put(1405.0,250.0){\rule[-0.200pt]{4.818pt}{0.400pt}}
\put(300.0,625.0){\rule[-0.200pt]{4.818pt}{0.400pt}}
\put(275,625){\makebox(0,0)[r]{\ \ {$1$}}}
\put(1405.0,625.0){\rule[-0.200pt]{4.818pt}{0.400pt}}
\put(300.0,1000.0){\rule[-0.200pt]{4.818pt}{0.400pt}}
\put(275,1000){\makebox(0,0)[r]{\ \ {$2$}}}
\put(1405.0,1000.0){\rule[-0.200pt]{4.818pt}{0.400pt}}
\put(300.0,1375.0){\rule[-0.200pt]{4.818pt}{0.400pt}}
\put(275,1375){\makebox(0,0)[r]{\ \ {$3$}}}
\put(1405.0,1375.0){\rule[-0.200pt]{4.818pt}{0.400pt}}
\put(300.0,1750.0){\rule[-0.200pt]{4.818pt}{0.400pt}}
\put(275,1750){\makebox(0,0)[r]{\ \ {$4$}}}
\put(1405.0,1750.0){\rule[-0.200pt]{4.818pt}{0.400pt}}
\put(300.0,250.0){\rule[-0.200pt]{0.400pt}{4.818pt}}
\put(300,200){\makebox(0,0){\ {$0$}}}
\put(300.0,1730.0){\rule[-0.200pt]{0.400pt}{4.818pt}}
\put(394.0,250.0){\rule[-0.200pt]{0.400pt}{4.818pt}}
\put(394,200){\makebox(0,0){\ {$1$}}}
\put(394.0,1730.0){\rule[-0.200pt]{0.400pt}{4.818pt}}
\put(488.0,250.0){\rule[-0.200pt]{0.400pt}{4.818pt}}
\put(488,200){\makebox(0,0){\ {$2$}}}
\put(488.0,1730.0){\rule[-0.200pt]{0.400pt}{4.818pt}}
\put(581.0,250.0){\rule[-0.200pt]{0.400pt}{4.818pt}}
\put(581,200){\makebox(0,0){\ {$3$}}}
\put(581.0,1730.0){\rule[-0.200pt]{0.400pt}{4.818pt}}
\put(675.0,250.0){\rule[-0.200pt]{0.400pt}{4.818pt}}
\put(675,200){\makebox(0,0){\ {$4$}}}
\put(675.0,1730.0){\rule[-0.200pt]{0.400pt}{4.818pt}}
\put(769.0,250.0){\rule[-0.200pt]{0.400pt}{4.818pt}}
\put(769,200){\makebox(0,0){\ {$5$}}}
\put(769.0,1730.0){\rule[-0.200pt]{0.400pt}{4.818pt}}
\put(863.0,250.0){\rule[-0.200pt]{0.400pt}{4.818pt}}
\put(863,200){\makebox(0,0){\ {$6$}}}
\put(863.0,1730.0){\rule[-0.200pt]{0.400pt}{4.818pt}}
\put(956.0,250.0){\rule[-0.200pt]{0.400pt}{4.818pt}}
\put(956,200){\makebox(0,0){\ {$7$}}}
\put(956.0,1730.0){\rule[-0.200pt]{0.400pt}{4.818pt}}
\put(1050.0,250.0){\rule[-0.200pt]{0.400pt}{4.818pt}}
\put(1050,200){\makebox(0,0){\ {$8$}}}
\put(1050.0,1730.0){\rule[-0.200pt]{0.400pt}{4.818pt}}
\put(1144.0,250.0){\rule[-0.200pt]{0.400pt}{4.818pt}}
\put(1144,200){\makebox(0,0){\ {$9$}}}
\put(1144.0,1730.0){\rule[-0.200pt]{0.400pt}{4.818pt}}
\put(1238.0,250.0){\rule[-0.200pt]{0.400pt}{4.818pt}}
\put(1238,200){\makebox(0,0){\ {$10$}}}
\put(1238.0,1730.0){\rule[-0.200pt]{0.400pt}{4.818pt}}
\put(1331.0,250.0){\rule[-0.200pt]{0.400pt}{4.818pt}}
\put(1331,200){\makebox(0,0){\ {$11$}}}
\put(1331.0,1730.0){\rule[-0.200pt]{0.400pt}{4.818pt}}
\put(1425.0,250.0){\rule[-0.200pt]{0.400pt}{4.818pt}}
\put(1425,200){\makebox(0,0){\ {$12$}}}
\put(1425.0,1730.0){\rule[-0.200pt]{0.400pt}{4.818pt}}
\put(300.0,250.0){\rule[-0.200pt]{271.012pt}{0.400pt}}
\put(1425.0,250.0){\rule[-0.200pt]{0.400pt}{361.350pt}}
\put(300.0,1750.0){\rule[-0.200pt]{271.012pt}{0.400pt}}
\put(150,1300){\makebox(0,0){\Large{$D(\rho)$}}}
\put(837,25){\makebox(0,0){\large{$\rho$}}}
\put(300.0,250.0){\rule[-0.200pt]{0.400pt}{361.350pt}}
\put(640,252){\circle*{12}}
\put(668,252){\circle*{12}}
\put(687,254){\circle*{12}}
\put(713,258){\circle*{12}}
\put(732,268){\circle*{12}}
\put(758,309){\circle*{12}}
\put(782,353){\circle*{12}}
\put(805,416){\circle*{12}}
\put(828,566){\circle*{12}}
\put(851,780){\circle*{12}}
\put(875,1010){\circle*{12}}
\put(898,1226){\circle*{12}}
\put(922,1347){\circle*{12}}
\put(945,1607){\circle*{12}}
\put(968,1591){\circle*{12}}
\put(991,1668){\circle*{12}}
\put(1014,1402){\circle*{12}}
\put(1038,1473){\circle*{12}}
\put(1062,1199){\circle*{12}}
\put(1086,1108){\circle*{12}}
\put(1110,934){\circle*{12}}
\put(1132,744){\circle*{12}}
\put(1152,705){\circle*{12}}
\put(1178,683){\circle*{12}}
\put(1201,503){\circle*{12}}
\put(1220,418){\circle*{12}}
\put(501,251){\circle{18}}
\put(522,251){\circle{18}}
\put(548,253){\circle{18}}
\put(569,259){\circle{18}}
\put(594,258){\circle{18}}
\put(618,267){\circle{18}}
\put(640,274){\circle{18}}
\put(664,296){\circle{18}}
\put(688,319){\circle{18}}
\put(711,339){\circle{18}}
\put(734,389){\circle{18}}
\put(758,430){\circle{18}}
\put(781,503){\circle{18}}
\put(804,555){\circle{18}}
\put(828,630){\circle{18}}
\put(850,698){\circle{18}}
\put(875,784){\circle{18}}
\put(898,815){\circle{18}}
\put(922,927){\circle{18}}
\put(945,986){\circle{18}}
\put(968,954){\circle{18}}
\put(992,1015){\circle{18}}
\put(1014,968){\circle{18}}
\put(1039,1106){\circle{18}}
\put(1062,941){\circle{18}}
\put(1086,1017){\circle{18}}
\put(1110,954){\circle{18}}
\put(1132,775){\circle{18}}
\put(1153,833){\circle{18}}
\put(1178,887){\circle{18}}
\put(1202,665){\circle{18}}
\put(1220,590){\circle{18}}
\put(462,250){\makebox(0,0){$+$}}
\put(480,254){\makebox(0,0){$+$}}
\put(502,269){\makebox(0,0){$+$}}
\put(525,278){\makebox(0,0){$+$}}
\put(547,289){\makebox(0,0){$+$}}
\put(571,299){\makebox(0,0){$+$}}
\put(593,312){\makebox(0,0){$+$}}
\put(617,327){\makebox(0,0){$+$}}
\put(640,349){\makebox(0,0){$+$}}
\put(664,368){\makebox(0,0){$+$}}
\put(687,400){\makebox(0,0){$+$}}
\put(710,421){\makebox(0,0){$+$}}
\put(734,451){\makebox(0,0){$+$}}
\put(758,480){\makebox(0,0){$+$}}
\put(781,492){\makebox(0,0){$+$}}
\put(804,508){\makebox(0,0){$+$}}
\put(828,524){\makebox(0,0){$+$}}
\put(850,523){\makebox(0,0){$+$}}
\put(874,562){\makebox(0,0){$+$}}
\put(898,562){\makebox(0,0){$+$}}
\put(921,554){\makebox(0,0){$+$}}
\put(945,563){\makebox(0,0){$+$}}
\put(968,515){\makebox(0,0){$+$}}
\put(991,528){\makebox(0,0){$+$}}
\put(1014,498){\makebox(0,0){$+$}}
\put(1038,518){\makebox(0,0){$+$}}
\put(1062,463){\makebox(0,0){$+$}}
\put(1086,478){\makebox(0,0){$+$}}
\put(1110,457){\makebox(0,0){$+$}}
\put(1132,394){\makebox(0,0){$+$}}
\put(1153,414){\makebox(0,0){$+$}}
\put(1178,430){\makebox(0,0){$+$}}
\put(1202,360){\makebox(0,0){$+$}}
\put(1220,343){\makebox(0,0){$+$}}
\put(300,250){\makebox(0,0){$\star$}}
\put(459,250){\makebox(0,0){$\star$}}
\put(482,259){\makebox(0,0){$\star$}}
\put(501,308){\makebox(0,0){$\star$}}
\put(522,329){\makebox(0,0){$\star$}}
\put(547,342){\makebox(0,0){$\star$}}
\put(570,350){\makebox(0,0){$\star$}}
\put(593,358){\makebox(0,0){$\star$}}
\put(617,364){\makebox(0,0){$\star$}}
\put(640,373){\makebox(0,0){$\star$}}
\put(664,382){\makebox(0,0){$\star$}}
\put(686,384){\makebox(0,0){$\star$}}
\put(711,394){\makebox(0,0){$\star$}}
\put(734,397){\makebox(0,0){$\star$}}
\put(757,392){\makebox(0,0){$\star$}}
\put(780,395){\makebox(0,0){$\star$}}
\put(804,401){\makebox(0,0){$\star$}}
\put(828,402){\makebox(0,0){$\star$}}
\put(851,405){\makebox(0,0){$\star$}}
\put(874,411){\makebox(0,0){$\star$}}
\put(898,404){\makebox(0,0){$\star$}}
\put(921,397){\makebox(0,0){$\star$}}
\put(945,412){\makebox(0,0){$\star$}}
\put(968,396){\makebox(0,0){$\star$}}
\put(992,393){\makebox(0,0){$\star$}}
\put(1014,385){\makebox(0,0){$\star$}}
\put(1039,401){\makebox(0,0){$\star$}}
\put(1062,379){\makebox(0,0){$\star$}}
\put(1086,386){\makebox(0,0){$\star$}}
\put(1110,379){\makebox(0,0){$\star$}}
\put(1132,346){\makebox(0,0){$\star$}}
\put(1153,361){\makebox(0,0){$\star$}}
\put(1178,372){\makebox(0,0){$\star$}}
\put(1202,331){\makebox(0,0){$\star$}}
\put(1220,320){\makebox(0,0){$\star$}}
\end{picture}
\end    {center}
\vskip 0.15in
\caption{The instanton size density, $D(\rho)$, for
$N=2(\star),3(+),4(\circ),8(\bullet)$
on $16^4$ lattices with $a \simeq 1/8T_c$.}
\label{fig_drho16}
\end    {figure}

A study similar to that in Fig.~\ref{fig_drho10} but for
a smaller value of $a$ would be very useful. Unfortunately we
have only performed such a calculation for $SU(8)$.
In Fig.~\ref{fig_drho16} we show the resulting $SU(8)$ instanton
size distribution obtained on a $16^4$ lattice with
$a\simeq 1/8T_c$. To compare with distributions at smaller $N$ 
we use data on $16^4$ lattices obtained in
\cite{blmt-glue}.
These are mostly at rather different values of $a$ and we
rescale the sizes and volumes to $a\simeq 1/8T_c$ in each
case. The resulting distributions are plotted in 
Fig.~\ref{fig_drho16}. The cooling/lattice cut-offs are
now far from the typical instanton size, and we have an
extended region of $\rho$ in which we can observe the
dramatic and expected suppression of small instantons.
Again we see a rapid narrowing of the distribution
as $N$ grows, consistent with eqn(\ref{eqn_drhoNinf}).
The total number of charges appears to grow with $N$, but 
once again there are various reasons for considering this
to be a much less robust observation.

\subsubsection{${\bf T\geq T_c}$}
\label{subsubsection_sizeTc}

We saw in Section~\ref{subsection_topTc} how, in the deconfined
phase, the fluctuations of $Q$ are rapidly suppressed as
$N$ grows. In this section we shall look at the size distribution, 
$D(\rho)$, in that phase. In doing so we shall continue to use the 
relation eqn(\ref{eqn_Qpeak})
to extract $\rho$, although for larger instantons at high $T$ the 
solution changes (becoming periodic), as described in
\cite{GPY}.
The reason is that both in theory and in practice such periodic 
instantons are highly suppressed at larger $N$.

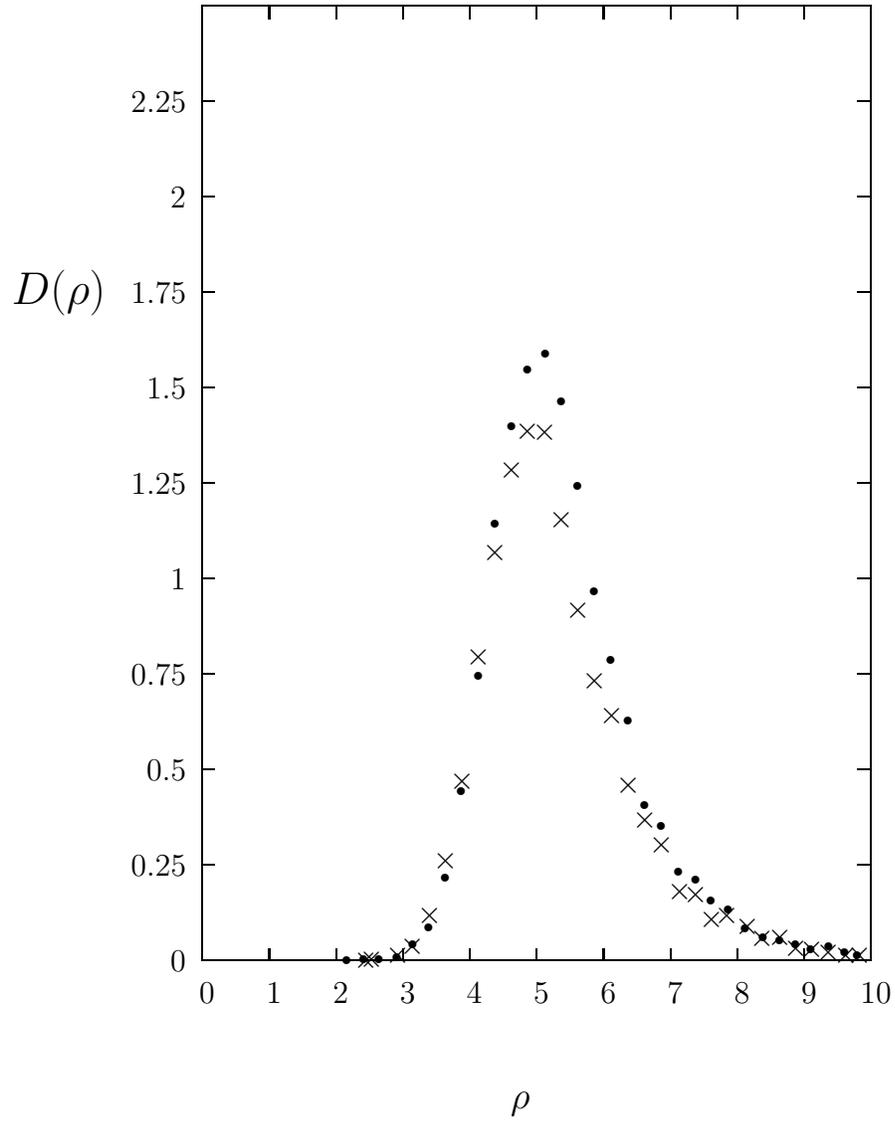
\begin  {figure}[p]
\begin  {center}
\leavevmode
\setlength{\unitlength}{0.240900pt}
\ifx\plotpoint\undefined\newsavebox{\plotpoint}\fi
\begin{picture}(1500,1800)(0,0)
\font\gnuplot=cmr10 at 12pt
\gnuplot
\sbox{\plotpoint}{\rule[-0.200pt]{0.400pt}{0.400pt}}%
\put(375.0,250.0){\rule[-0.200pt]{4.818pt}{0.400pt}}
\put(350,250){\makebox(0,0)[r]{\ \ {$0$}}}
\put(1405.0,250.0){\rule[-0.200pt]{4.818pt}{0.400pt}}
\put(375.0,400.0){\rule[-0.200pt]{4.818pt}{0.400pt}}
\put(350,400){\makebox(0,0)[r]{\ \ {$0.25$}}}
\put(1405.0,400.0){\rule[-0.200pt]{4.818pt}{0.400pt}}
\put(375.0,550.0){\rule[-0.200pt]{4.818pt}{0.400pt}}
\put(350,550){\makebox(0,0)[r]{\ \ {$0.5$}}}
\put(1405.0,550.0){\rule[-0.200pt]{4.818pt}{0.400pt}}
\put(375.0,700.0){\rule[-0.200pt]{4.818pt}{0.400pt}}
\put(350,700){\makebox(0,0)[r]{\ \ {$0.75$}}}
\put(1405.0,700.0){\rule[-0.200pt]{4.818pt}{0.400pt}}
\put(375.0,850.0){\rule[-0.200pt]{4.818pt}{0.400pt}}
\put(350,850){\makebox(0,0)[r]{\ \ {$1$}}}
\put(1405.0,850.0){\rule[-0.200pt]{4.818pt}{0.400pt}}
\put(375.0,1000.0){\rule[-0.200pt]{4.818pt}{0.400pt}}
\put(350,1000){\makebox(0,0)[r]{\ \ {$1.25$}}}
\put(1405.0,1000.0){\rule[-0.200pt]{4.818pt}{0.400pt}}
\put(375.0,1150.0){\rule[-0.200pt]{4.818pt}{0.400pt}}
\put(350,1150){\makebox(0,0)[r]{\ \ {$1.5$}}}
\put(1405.0,1150.0){\rule[-0.200pt]{4.818pt}{0.400pt}}
\put(375.0,1300.0){\rule[-0.200pt]{4.818pt}{0.400pt}}
\put(350,1300){\makebox(0,0)[r]{\ \ {$1.75$}}}
\put(1405.0,1300.0){\rule[-0.200pt]{4.818pt}{0.400pt}}
\put(375.0,1450.0){\rule[-0.200pt]{4.818pt}{0.400pt}}
\put(350,1450){\makebox(0,0)[r]{\ \ {$2$}}}
\put(1405.0,1450.0){\rule[-0.200pt]{4.818pt}{0.400pt}}
\put(375.0,1600.0){\rule[-0.200pt]{4.818pt}{0.400pt}}
\put(350,1600){\makebox(0,0)[r]{\ \ {$2.25$}}}
\put(1405.0,1600.0){\rule[-0.200pt]{4.818pt}{0.400pt}}
\put(375.0,250.0){\rule[-0.200pt]{0.400pt}{4.818pt}}
\put(375,200){\makebox(0,0){\ {$0$}}}
\put(375.0,1730.0){\rule[-0.200pt]{0.400pt}{4.818pt}}
\put(480.0,250.0){\rule[-0.200pt]{0.400pt}{4.818pt}}
\put(480,200){\makebox(0,0){\ {$1$}}}
\put(480.0,1730.0){\rule[-0.200pt]{0.400pt}{4.818pt}}
\put(585.0,250.0){\rule[-0.200pt]{0.400pt}{4.818pt}}
\put(585,200){\makebox(0,0){\ {$2$}}}
\put(585.0,1730.0){\rule[-0.200pt]{0.400pt}{4.818pt}}
\put(690.0,250.0){\rule[-0.200pt]{0.400pt}{4.818pt}}
\put(690,200){\makebox(0,0){\ {$3$}}}
\put(690.0,1730.0){\rule[-0.200pt]{0.400pt}{4.818pt}}
\put(795.0,250.0){\rule[-0.200pt]{0.400pt}{4.818pt}}
\put(795,200){\makebox(0,0){\ {$4$}}}
\put(795.0,1730.0){\rule[-0.200pt]{0.400pt}{4.818pt}}
\put(900.0,250.0){\rule[-0.200pt]{0.400pt}{4.818pt}}
\put(900,200){\makebox(0,0){\ {$5$}}}
\put(900.0,1730.0){\rule[-0.200pt]{0.400pt}{4.818pt}}
\put(1005.0,250.0){\rule[-0.200pt]{0.400pt}{4.818pt}}
\put(1005,200){\makebox(0,0){\ {$6$}}}
\put(1005.0,1730.0){\rule[-0.200pt]{0.400pt}{4.818pt}}
\put(1110.0,250.0){\rule[-0.200pt]{0.400pt}{4.818pt}}
\put(1110,200){\makebox(0,0){\ {$7$}}}
\put(1110.0,1730.0){\rule[-0.200pt]{0.400pt}{4.818pt}}
\put(1215.0,250.0){\rule[-0.200pt]{0.400pt}{4.818pt}}
\put(1215,200){\makebox(0,0){\ {$8$}}}
\put(1215.0,1730.0){\rule[-0.200pt]{0.400pt}{4.818pt}}
\put(1320.0,250.0){\rule[-0.200pt]{0.400pt}{4.818pt}}
\put(1320,200){\makebox(0,0){\ {$9$}}}
\put(1320.0,1730.0){\rule[-0.200pt]{0.400pt}{4.818pt}}
\put(1425.0,250.0){\rule[-0.200pt]{0.400pt}{4.818pt}}
\put(1425,200){\makebox(0,0){\ {$10$}}}
\put(1425.0,1730.0){\rule[-0.200pt]{0.400pt}{4.818pt}}
\put(375.0,250.0){\rule[-0.200pt]{252.945pt}{0.400pt}}
\put(1425.0,250.0){\rule[-0.200pt]{0.400pt}{361.350pt}}
\put(375.0,1750.0){\rule[-0.200pt]{252.945pt}{0.400pt}}
\put(150,1300){\makebox(0,0){\Large{$D(\rho)$}}}
\put(875,25){\makebox(0,0){\large{$\rho$}}}
\put(375.0,250.0){\rule[-0.200pt]{0.400pt}{361.350pt}}
\put(602,251){\circle*{12}}
\put(629,252){\circle*{12}}
\put(653,252){\circle*{12}}
\put(681,255){\circle*{12}}
\put(706,275){\circle*{12}}
\put(731,302){\circle*{12}}
\put(757,380){\circle*{12}}
\put(782,516){\circle*{12}}
\put(809,698){\circle*{12}}
\put(835,936){\circle*{12}}
\put(861,1090){\circle*{12}}
\put(886,1178){\circle*{12}}
\put(914,1203){\circle*{12}}
\put(939,1128){\circle*{12}}
\put(965,996){\circle*{12}}
\put(991,830){\circle*{12}}
\put(1017,722){\circle*{12}}
\put(1044,627){\circle*{12}}
\put(1070,494){\circle*{12}}
\put(1096,462){\circle*{12}}
\put(1123,389){\circle*{12}}
\put(1150,377){\circle*{12}}
\put(1174,344){\circle*{12}}
\put(1201,330){\circle*{12}}
\put(1228,300){\circle*{12}}
\put(1256,287){\circle*{12}}
\put(1282,281){\circle*{12}}
\put(1307,275){\circle*{12}}
\put(1331,268){\circle*{12}}
\put(1359,272){\circle*{12}}
\put(1384,263){\circle*{12}}
\put(1404,259){\circle*{12}}
\put(632,251){\makebox(0,0){$\times$}}
\put(641,252){\makebox(0,0){$\times$}}
\put(682,259){\makebox(0,0){$\times$}}
\put(705,272){\makebox(0,0){$\times$}}
\put(732,320){\makebox(0,0){$\times$}}
\put(757,407){\makebox(0,0){$\times$}}
\put(783,531){\makebox(0,0){$\times$}}
\put(809,727){\makebox(0,0){$\times$}}
\put(835,891){\makebox(0,0){$\times$}}
\put(861,1021){\makebox(0,0){$\times$}}
\put(886,1081){\makebox(0,0){$\times$}}
\put(913,1080){\makebox(0,0){$\times$}}
\put(939,942){\makebox(0,0){$\times$}}
\put(965,800){\makebox(0,0){$\times$}}
\put(991,689){\makebox(0,0){$\times$}}
\put(1018,635){\makebox(0,0){$\times$}}
\put(1044,526){\makebox(0,0){$\times$}}
\put(1070,470){\makebox(0,0){$\times$}}
\put(1096,431){\makebox(0,0){$\times$}}
\put(1125,359){\makebox(0,0){$\times$}}
\put(1150,354){\makebox(0,0){$\times$}}
\put(1175,314){\makebox(0,0){$\times$}}
\put(1199,321){\makebox(0,0){$\times$}}
\put(1231,304){\makebox(0,0){$\times$}}
\put(1254,285){\makebox(0,0){$\times$}}
\put(1282,286){\makebox(0,0){$\times$}}
\put(1307,269){\makebox(0,0){$\times$}}
\put(1332,267){\makebox(0,0){$\times$}}
\put(1358,263){\makebox(0,0){$\times$}}
\put(1386,258){\makebox(0,0){$\times$}}
\put(1407,258){\makebox(0,0){$\times$}}
\end{picture}
\end    {center}
\vskip 0.15in
\caption{The $SU(8)$ instanton size density, $D(\rho)$, at 
$T\simeq 0$ ($\bullet$) and at $T\simeq T_c$ ($\times$) in 
the confined phase, with $a\simeq 1/5T_c$ in both
cases. Normalised to a $10^4$ volume.}
\label{fig_drhoTconf}
\end    {figure}

Before looking at the deconfined phase we briefly look at the 
instanton density in the confined phase, and compare densities
at $T\simeq 0$ and $T=T_c$. We focus on $SU(8)$. The  $T=T_c$
calculation has been performed on a $12^3 5$ lattice at
$\beta=43.965$ and is compared in Fig.~\ref{fig_drhoTconf} to the 
$T\simeq 0$ calculation on a $10^4$ lattice at $\beta=44.0$. In the
Figure sizes have been rescaled so that the numbers are per
$10^4$ volume at  $\beta=44.0$. What we see is that there is
no significant difference between the densities at $T\simeq 0$
and at $T\simeq T_c$, just as we saw in Table~\ref{table_chiTc}
no significant difference in the susceptibilities. The same pattern
appears at other values of $N$. We conclude that the topological 
content of the confined phase is very much the same at any value of
$T$.

\begin  {figure}[p]
\begin  {center}
\leavevmode
\setlength{\unitlength}{0.240900pt}
\ifx\plotpoint\undefined\newsavebox{\plotpoint}\fi
\begin{picture}(1500,1800)(0,0)
\font\gnuplot=cmr10 at 12pt
\gnuplot
\sbox{\plotpoint}{\rule[-0.200pt]{0.400pt}{0.400pt}}%
\put(375.0,250.0){\rule[-0.200pt]{4.818pt}{0.400pt}}
\put(350,250){\makebox(0,0)[r]{\ \ {$0$}}}
\put(1405.0,250.0){\rule[-0.200pt]{4.818pt}{0.400pt}}
\put(375.0,400.0){\rule[-0.200pt]{4.818pt}{0.400pt}}
\put(350,400){\makebox(0,0)[r]{\ \ {$0.25$}}}
\put(1405.0,400.0){\rule[-0.200pt]{4.818pt}{0.400pt}}
\put(375.0,550.0){\rule[-0.200pt]{4.818pt}{0.400pt}}
\put(350,550){\makebox(0,0)[r]{\ \ {$0.5$}}}
\put(1405.0,550.0){\rule[-0.200pt]{4.818pt}{0.400pt}}
\put(375.0,700.0){\rule[-0.200pt]{4.818pt}{0.400pt}}
\put(350,700){\makebox(0,0)[r]{\ \ {$0.75$}}}
\put(1405.0,700.0){\rule[-0.200pt]{4.818pt}{0.400pt}}
\put(375.0,850.0){\rule[-0.200pt]{4.818pt}{0.400pt}}
\put(350,850){\makebox(0,0)[r]{\ \ {$1$}}}
\put(1405.0,850.0){\rule[-0.200pt]{4.818pt}{0.400pt}}
\put(375.0,1000.0){\rule[-0.200pt]{4.818pt}{0.400pt}}
\put(350,1000){\makebox(0,0)[r]{\ \ {$1.25$}}}
\put(1405.0,1000.0){\rule[-0.200pt]{4.818pt}{0.400pt}}
\put(375.0,1150.0){\rule[-0.200pt]{4.818pt}{0.400pt}}
\put(350,1150){\makebox(0,0)[r]{\ \ {$1.5$}}}
\put(1405.0,1150.0){\rule[-0.200pt]{4.818pt}{0.400pt}}
\put(375.0,1300.0){\rule[-0.200pt]{4.818pt}{0.400pt}}
\put(350,1300){\makebox(0,0)[r]{\ \ {$1.75$}}}
\put(1405.0,1300.0){\rule[-0.200pt]{4.818pt}{0.400pt}}
\put(375.0,1450.0){\rule[-0.200pt]{4.818pt}{0.400pt}}
\put(350,1450){\makebox(0,0)[r]{\ \ {$2$}}}
\put(1405.0,1450.0){\rule[-0.200pt]{4.818pt}{0.400pt}}
\put(375.0,1600.0){\rule[-0.200pt]{4.818pt}{0.400pt}}
\put(350,1600){\makebox(0,0)[r]{\ \ {$2.25$}}}
\put(1405.0,1600.0){\rule[-0.200pt]{4.818pt}{0.400pt}}
\put(375.0,250.0){\rule[-0.200pt]{0.400pt}{4.818pt}}
\put(375,200){\makebox(0,0){\ {$0$}}}
\put(375.0,1730.0){\rule[-0.200pt]{0.400pt}{4.818pt}}
\put(480.0,250.0){\rule[-0.200pt]{0.400pt}{4.818pt}}
\put(480,200){\makebox(0,0){\ {$1$}}}
\put(480.0,1730.0){\rule[-0.200pt]{0.400pt}{4.818pt}}
\put(585.0,250.0){\rule[-0.200pt]{0.400pt}{4.818pt}}
\put(585,200){\makebox(0,0){\ {$2$}}}
\put(585.0,1730.0){\rule[-0.200pt]{0.400pt}{4.818pt}}
\put(690.0,250.0){\rule[-0.200pt]{0.400pt}{4.818pt}}
\put(690,200){\makebox(0,0){\ {$3$}}}
\put(690.0,1730.0){\rule[-0.200pt]{0.400pt}{4.818pt}}
\put(795.0,250.0){\rule[-0.200pt]{0.400pt}{4.818pt}}
\put(795,200){\makebox(0,0){\ {$4$}}}
\put(795.0,1730.0){\rule[-0.200pt]{0.400pt}{4.818pt}}
\put(900.0,250.0){\rule[-0.200pt]{0.400pt}{4.818pt}}
\put(900,200){\makebox(0,0){\ {$5$}}}
\put(900.0,1730.0){\rule[-0.200pt]{0.400pt}{4.818pt}}
\put(1005.0,250.0){\rule[-0.200pt]{0.400pt}{4.818pt}}
\put(1005,200){\makebox(0,0){\ {$6$}}}
\put(1005.0,1730.0){\rule[-0.200pt]{0.400pt}{4.818pt}}
\put(1110.0,250.0){\rule[-0.200pt]{0.400pt}{4.818pt}}
\put(1110,200){\makebox(0,0){\ {$7$}}}
\put(1110.0,1730.0){\rule[-0.200pt]{0.400pt}{4.818pt}}
\put(1215.0,250.0){\rule[-0.200pt]{0.400pt}{4.818pt}}
\put(1215,200){\makebox(0,0){\ {$8$}}}
\put(1215.0,1730.0){\rule[-0.200pt]{0.400pt}{4.818pt}}
\put(1320.0,250.0){\rule[-0.200pt]{0.400pt}{4.818pt}}
\put(1320,200){\makebox(0,0){\ {$9$}}}
\put(1320.0,1730.0){\rule[-0.200pt]{0.400pt}{4.818pt}}
\put(1425.0,250.0){\rule[-0.200pt]{0.400pt}{4.818pt}}
\put(1425,200){\makebox(0,0){\ {$10$}}}
\put(1425.0,1730.0){\rule[-0.200pt]{0.400pt}{4.818pt}}
\put(375.0,250.0){\rule[-0.200pt]{252.945pt}{0.400pt}}
\put(1425.0,250.0){\rule[-0.200pt]{0.400pt}{361.350pt}}
\put(375.0,1750.0){\rule[-0.200pt]{252.945pt}{0.400pt}}
\put(150,1300){\makebox(0,0){\Large{$D(\rho)$}}}
\put(875,25){\makebox(0,0){\large{$\rho$}}}
\put(375.0,250.0){\rule[-0.200pt]{0.400pt}{361.350pt}}
\put(629,252){\circle*{12}}
\put(684,252){\circle*{12}}
\put(710,251){\circle*{12}}
\put(1005,251){\circle*{12}}
\put(1045,256){\circle*{12}}
\put(1074,285){\circle*{12}}
\put(1098,344){\circle*{12}}
\put(1126,432){\circle*{12}}
\put(1150,533){\circle*{12}}
\put(1175,682){\circle*{12}}
\put(1202,903){\circle*{12}}
\put(1229,948){\circle*{12}}
\put(1255,1016){\circle*{12}}
\put(1282,1079){\circle*{12}}
\put(1306,865){\circle*{12}}
\put(1331,890){\circle*{12}}
\put(1358,918){\circle*{12}}
\put(1385,645){\circle*{12}}
\put(1405,547){\circle*{12}}
\end{picture}
\end    {center}
\vskip 0.15in
\caption{The $SU(8)$ instanton size density, $D(\rho)$, at 
$T\simeq T_c$ in the deconfined phase, with $a\simeq 1/5T_c$. 
Normalised to a $10^4$ volume.}
\label{fig_drhoTdeconf}
\end    {figure}
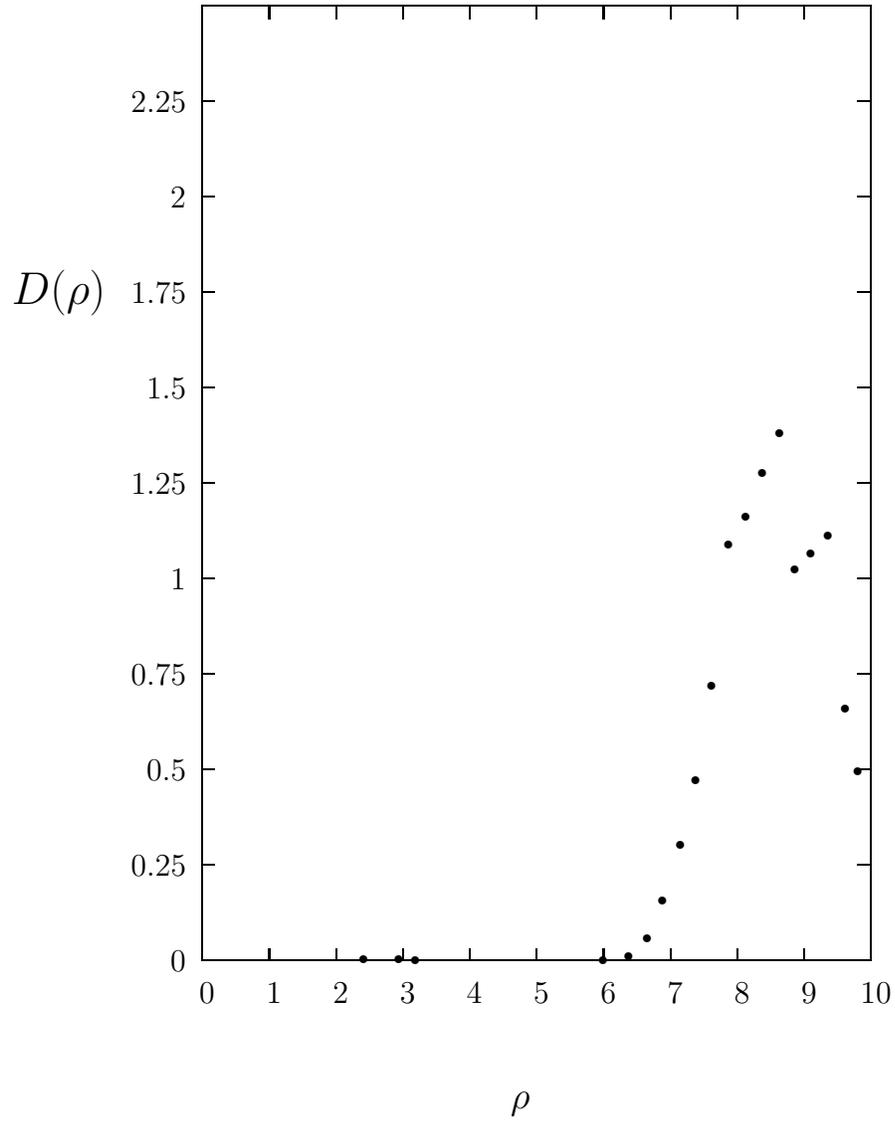

Turning now to the deconfined phase of the $SU(8)$ theory, we plot in 
Fig.~\ref{fig_drhoTdeconf} the size distribution obtained on 500
$12^3 5$ lattice fields that are in the deconfined phase 
and generated at $\beta=43.965$, i.e. at  $T\simeq T_c$.
We see that the distribution is entirely located at very large 
values of $\rho$ (compare Fig.~\ref{fig_drhoTconf} which is
in the confined phase) except for a very few peaks (4 out of
about 6000) that lie in the range $\rho\in [2,3.5]$. 

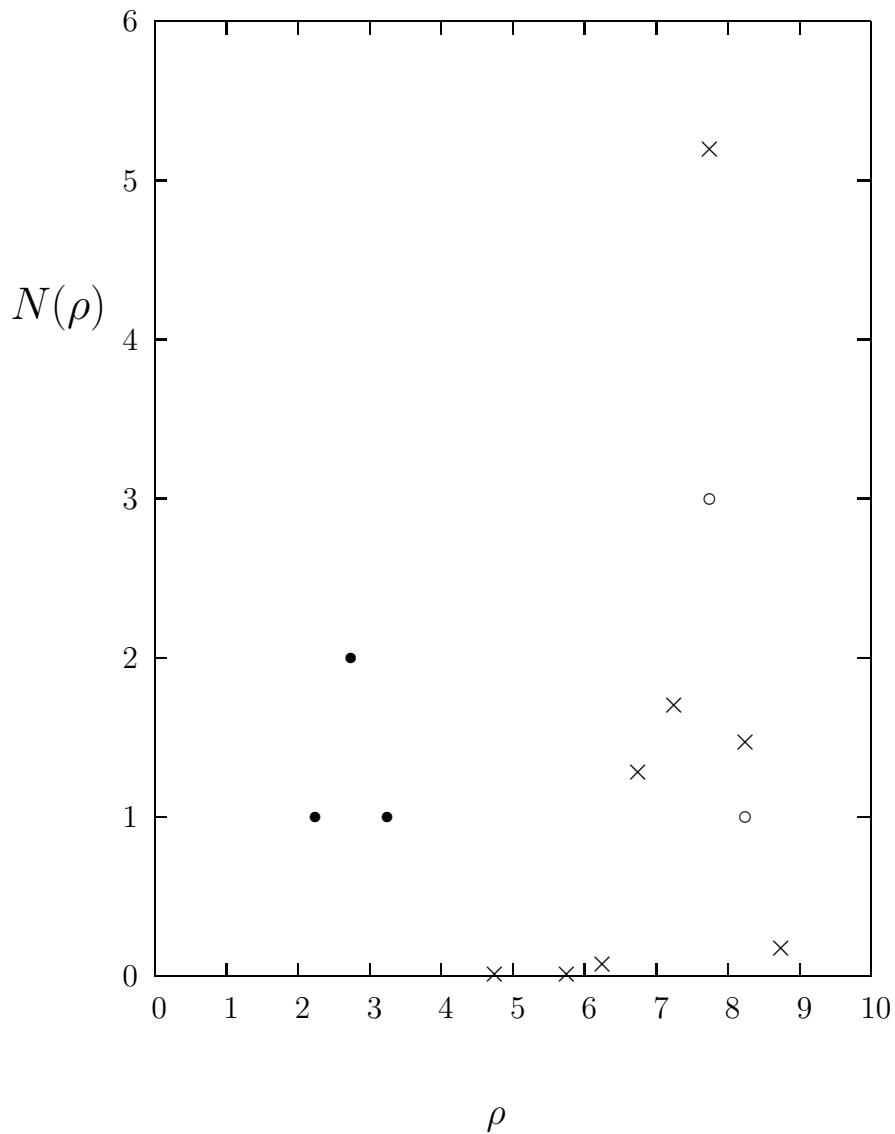
\begin  {figure}[p]
\begin  {center}
\leavevmode
\setlength{\unitlength}{0.240900pt}
\ifx\plotpoint\undefined\newsavebox{\plotpoint}\fi
\sbox{\plotpoint}{\rule[-0.200pt]{0.400pt}{0.400pt}}%
\begin{picture}(1500,1800)(0,0)
\font\gnuplot=cmr10 at 12pt
\gnuplot
\sbox{\plotpoint}{\rule[-0.200pt]{0.400pt}{0.400pt}}%
\put(300.0,250.0){\rule[-0.200pt]{4.818pt}{0.400pt}}
\put(275,250){\makebox(0,0)[r]{\ \ {$0$}}}
\put(1405.0,250.0){\rule[-0.200pt]{4.818pt}{0.400pt}}
\put(300.0,500.0){\rule[-0.200pt]{4.818pt}{0.400pt}}
\put(275,500){\makebox(0,0)[r]{\ \ {$1$}}}
\put(1405.0,500.0){\rule[-0.200pt]{4.818pt}{0.400pt}}
\put(300.0,750.0){\rule[-0.200pt]{4.818pt}{0.400pt}}
\put(275,750){\makebox(0,0)[r]{\ \ {$2$}}}
\put(1405.0,750.0){\rule[-0.200pt]{4.818pt}{0.400pt}}
\put(300.0,1000.0){\rule[-0.200pt]{4.818pt}{0.400pt}}
\put(275,1000){\makebox(0,0)[r]{\ \ {$3$}}}
\put(1405.0,1000.0){\rule[-0.200pt]{4.818pt}{0.400pt}}
\put(300.0,1250.0){\rule[-0.200pt]{4.818pt}{0.400pt}}
\put(275,1250){\makebox(0,0)[r]{\ \ {$4$}}}
\put(1405.0,1250.0){\rule[-0.200pt]{4.818pt}{0.400pt}}
\put(300.0,1500.0){\rule[-0.200pt]{4.818pt}{0.400pt}}
\put(275,1500){\makebox(0,0)[r]{\ \ {$5$}}}
\put(1405.0,1500.0){\rule[-0.200pt]{4.818pt}{0.400pt}}
\put(300.0,1750.0){\rule[-0.200pt]{4.818pt}{0.400pt}}
\put(275,1750){\makebox(0,0)[r]{\ \ {$6$}}}
\put(1405.0,1750.0){\rule[-0.200pt]{4.818pt}{0.400pt}}
\put(300.0,250.0){\rule[-0.200pt]{0.400pt}{4.818pt}}
\put(300,200){\makebox(0,0){\ {$0$}}}
\put(300.0,1730.0){\rule[-0.200pt]{0.400pt}{4.818pt}}
\put(413.0,250.0){\rule[-0.200pt]{0.400pt}{4.818pt}}
\put(413,200){\makebox(0,0){\ {$1$}}}
\put(413.0,1730.0){\rule[-0.200pt]{0.400pt}{4.818pt}}
\put(525.0,250.0){\rule[-0.200pt]{0.400pt}{4.818pt}}
\put(525,200){\makebox(0,0){\ {$2$}}}
\put(525.0,1730.0){\rule[-0.200pt]{0.400pt}{4.818pt}}
\put(638.0,250.0){\rule[-0.200pt]{0.400pt}{4.818pt}}
\put(638,200){\makebox(0,0){\ {$3$}}}
\put(638.0,1730.0){\rule[-0.200pt]{0.400pt}{4.818pt}}
\put(750.0,250.0){\rule[-0.200pt]{0.400pt}{4.818pt}}
\put(750,200){\makebox(0,0){\ {$4$}}}
\put(750.0,1730.0){\rule[-0.200pt]{0.400pt}{4.818pt}}
\put(863.0,250.0){\rule[-0.200pt]{0.400pt}{4.818pt}}
\put(863,200){\makebox(0,0){\ {$5$}}}
\put(863.0,1730.0){\rule[-0.200pt]{0.400pt}{4.818pt}}
\put(975.0,250.0){\rule[-0.200pt]{0.400pt}{4.818pt}}
\put(975,200){\makebox(0,0){\ {$6$}}}
\put(975.0,1730.0){\rule[-0.200pt]{0.400pt}{4.818pt}}
\put(1088.0,250.0){\rule[-0.200pt]{0.400pt}{4.818pt}}
\put(1088,200){\makebox(0,0){\ {$7$}}}
\put(1088.0,1730.0){\rule[-0.200pt]{0.400pt}{4.818pt}}
\put(1200.0,250.0){\rule[-0.200pt]{0.400pt}{4.818pt}}
\put(1200,200){\makebox(0,0){\ {$8$}}}
\put(1200.0,1730.0){\rule[-0.200pt]{0.400pt}{4.818pt}}
\put(1313.0,250.0){\rule[-0.200pt]{0.400pt}{4.818pt}}
\put(1313,200){\makebox(0,0){\ {$9$}}}
\put(1313.0,1730.0){\rule[-0.200pt]{0.400pt}{4.818pt}}
\put(1425.0,250.0){\rule[-0.200pt]{0.400pt}{4.818pt}}
\put(1425,200){\makebox(0,0){\ {$10$}}}
\put(1425.0,1730.0){\rule[-0.200pt]{0.400pt}{4.818pt}}
\put(300.0,250.0){\rule[-0.200pt]{271.012pt}{0.400pt}}
\put(1425.0,250.0){\rule[-0.200pt]{0.400pt}{361.350pt}}
\put(300.0,1750.0){\rule[-0.200pt]{271.012pt}{0.400pt}}
\put(150,1300){\makebox(0,0){\Large{$N(\rho)$}}}
\put(837,25){\makebox(0,0){\large{$\rho$}}}
\put(300.0,250.0){\rule[-0.200pt]{0.400pt}{361.350pt}}
\put(834,253){\makebox(0,0){$\times$}}
\put(947,253){\makebox(0,0){$\times$}}
\put(1003,268){\makebox(0,0){$\times$}}
\put(1059,570){\makebox(0,0){$\times$}}
\put(1116,675){\makebox(0,0){$\times$}}
\put(1172,1550){\makebox(0,0){$\times$}}
\put(1228,618){\makebox(0,0){$\times$}}
\put(1284,293){\makebox(0,0){$\times$}}
\put(553,500){\circle*{18}}
\put(609,750){\circle*{18}}
\put(666,500){\circle*{18}}
\put(1172,1000){\circle{18}}
\put(1228,500){\circle{18}}
\end{picture}
\end    {center}
\vskip 0.15in
\caption{The number of instantons in 500 lattice fields 
using only the largest positive and negative peaks in each 
field. All in the deconfined phase, at $T=T_c$, on a $12^3 5$ 
lattice in $SU(8)$. Separately for the 4 configurations with
$Q\not= 0$ and for the 496 lattice fields with  $Q=0$.
The latter ($\times$) have been divided by 100 so as to
fit on the same plot. For the former we show separately 
sizes from peaks with the same sign as $Q$ ($\bullet$) and
with the opposite sign ($\circ$).}
\label{fig_rhodeconf}
\end    {figure}

To interpret all this we
take these  500 fields and use only the largest positive and 
negative peaks in each field. We then take separately the 
4 configurations with $Q\not= 0$ and the 496 lattice fields with  
$Q=0$, and we look in Fig.~\ref{fig_rhodeconf}
at the size distributions of the two sets 
separately. Moreover in the former case we show separately the size
distribution from peaks with the same sign as $Q$ and from those
with the opposite sign. We see that the latter have large 
values of $\rho$, just like the $Q=0$ fields. The former, by 
contrast, provide the four values $\in [2,3.5]$ that we
noted in the previous paragraph. The interpretation
is clear: real topological charges are narrow while the
other very broad charges are nothing but artifacts of the
partial cooling and do not correspond to real topological
charges at all. Given that the high-$T$ suppression we expect  
theoretically, $\propto \exp (-\frac{2}{3}N\{\pi\rho T\}^2)$, 
is more severe at larger $\rho$, the observed gap in the size 
distribution makes the interpretation of the very low, broad peaks 
as artifacts quite compelling; as does the fact that these
`large charges' do not contribute anything to the net topological
charge $Q$. 

It is interesting to note that for the confined phase
on the same lattice at the same $T$ we find 7 charges 
$\in [2,3]$ (with none narrower). This represents the tail of
the highly suppressed small instantons. What we see in 
the deconfined phase is consistent with being the same
tail. In contrast to the confining phase however,
where the suppression rapidly relaxes as $\rho$ increases,
here there is a cut-off immediately above $\rho\sim 3.5$. 
Thus it would appear that the the high-$T$ exponential suppression
is already effective down to small sizes $\rho\sim 3.5$ and
down to modest temperatures $T\simeq T_c$. 

Our results at other $N$ show the same features, although
the interpretation is simplest in the case of $SU(8)$ because
here the deconfined configurations with $Q\not= 0$ are so rare,
so that the $Q=0$ fields are highly unlikely to contain any
instanton-antiinstanton pairs that would complicate the
argument and the size distributions. To illustrate this we
show in Fig.~\ref{fig_drhoTdeconfn6} the $SU(6)$ deconfined
instanton size density. It show the same features as
we saw in Fig.~\ref{fig_drhoTdeconf} except that the low-$\rho$
peak is much larger -- as we would expect because the
small-$\rho$ suppression is more severe for $SU(8)$ than
for $SU(6)$. For comparison we show the confined density
at $T_c$ in Fig.~\ref{fig_drhoTconfn6}. The total number of
peaks for $\rho\leq 3.0$ is roughly the same in both cases
showing that what is involved is the suppression of charges
with larger $\rho$. (From the ratios of the densities in
the two Figures one can map out the suppression as
a function of $\rho$.) In Fig.~\ref{fig_rhodeconfn6} we
show the $SU(6)$ version of Fig.~\ref{fig_rhodeconf}. Unlike
the latter we now have a small low-$\rho$ peak from
the $Q=0$ lattice fields. This is because, as we can see,
the number of lattice fields with $Q\not= 0$ is much
larger and so there is
now a finite probability to have lattice fields with 
well-separated instanton-antiinstanton pairs. For
comparison, we show in  Fig.~\ref{fig_rhoconfn6} the
same plot for the confined phase.

We conclude therefore that in the deconfined phase the severe 
small-$\rho$ suppression is matched by an even more severe
larger $\rho$ suppression which, already for $N=8$ and
even at $T\simeq T_c$, extends down to very small $\rho$.
It seems clear that as $N\to\infty$ the deconfined phase
will lose its topological charges exponentially fast
at any given $\rho$; and that this will be so for all $T$
in this phase. By contrast the confined phase appears
to show no variation with $T$, even at $T\simeq T_c$. Clearly 
there the semiclassical prediction is obstructed by the 
non-perturbative fluctuations. We thus have a rather simple
picture of the topological properties of the theory as
$N\to\infty$.

\section{Conclusions}
\label{sectioNonclusions}

As we outlined above, there are various theoretical arguments
for the large-$N$  behaviour of topological fluctuations in gauge 
theories, which it is interesting to test in lattice
calculations. One expects the topological susceptibility $\chi_t$
to have a finite and non-zero $N\to\infty$ limit 
(see Section~\ref{subsection_khi}), despite the fact that 
simple arguments (see Section~\ref{subsection_Q})
suggest that the number of topological charges grows $\propto N$. 
One expects smaller instantons, up to some critical size, to be 
completely suppressed as $N\to \infty$ 
(see Section~\ref{subsection_Ismall}) and, since this is
a short distance argument, it should occur at all $T$.
Once the temperature $T$ is high enough for the perturbative 
evaluation of fluctuations around instantons to be reliable, one 
also expects instantons to be exponentially suppressed in
$N\{\pi\rho T\}^2$ (see Section~\ref{subsection_Ilarge}).

Compared to earlier calculations, our range of $N$
is larger, and we explore both the $T$ and the $N$ dependence 
of the topological fluctuations. Since the deconfining phase 
transition is robustly first order for larger $N$, we are
able to investigate the confined and deconfined phases
separately at the same value of $T\simeq T_c$. 

Our results on the $T\simeq 0$ topological susceptibility
(see Fig.~\ref{fig_chiN}) show that the  topological susceptibility
(albeit at a fixed non-zero value of the lattice spacing) 
does indeed have a finite and non-zero $N\to\infty$ limit,
as suggested by earlier work. The size distribution shows the 
expected suppression of small topological charges, not only
at  $T\simeq 0$ (see Fig.~\ref{fig_drho10} and Fig.~\ref{fig_drho16}) 
but also at non-zero $T$, in both the confined
(see Fig.~\ref{fig_drhoTconf}) and the deconfined
(see Fig.~\ref{fig_drhoTdeconf}) phases. Within the confined phase
there appears to be little if any variation with $T$, at any $T$,
in the character or density of the topological fluctuations.
In addition we find that in this phase the size distribution 
appears to become
more peaked as $N$ grows, suggesting that it becomes
$\propto \delta(\rho-\rho_c)$ at $N = \infty$.

In the deconfined phase we find a strong suppression of
all topological fluctuations which becomes rapidly more severe
with increasing $N$. (See Fig.~\ref{fig_chiNTc} and
Fig.~\ref{fig_drhoTdeconf}.) We infer that the large-$T$
exponential suppression occurs at all $T > T_c$ at large $N$,
and that it overlaps with the small-$\rho$ suppression so that
all topological fluctuations are suppressed. 

The low, broad peaks in the deconfined phase (see 
Fig.~\ref{fig_rhodeconf}) are remnant artifacts of the
incomplete cooling. They provide us, for the first time,
with a measure of the `background' that such artifacts
will contribute to all our other size distributions. 
Of course, these artifacts will depend to some extent
on the content of the fields, which is not the
same for the confined and deconfined phases. Nonetheless
the fact that these artifacts are at such large values
of $\rho$ reassures us that the part of the size
distributions that is important to our conclusions 
about the confined phase (in particular the peaking)
is robust.

Our calculations thus support a very simple picture of
topological fluctuations in the $N=\infty$ limit.
In the deconfined phase these disappear completely for
all $T$ and for all $\rho$. In the confined phase 
all fluctuations with $\rho$ less than some $\rho_c\sim 1/T_c$ 
disappear, and there is evidence 
that the same is true for all $\rho > \rho_c$, so
that the size distribution becomes a simple $\delta$-function.

\section*{Acknowledgements}

Our lattice calculations were carried out on PPARC 
and EPSRC funded Alpha Compaq workstations in Oxford Theoretical 
Physics, and on a desktop funded by All Souls College.
During the course of this research, UW was supported 
by a PPARC SPG fellowship, and BL by a
EU Marie Sk{\l}odowska-Curie postdoctoral fellowship.

\vfill\eject

\begin{table}
\begin{center}
\begin{tabular}{|c|c|c|c|c|l|l|}\hline
$N$ & $\beta$ & $\beta_c(aT_c=0.2)$ & lattice  & $a\surd\sigma$ &
$\langle Q^2_I \rangle$ & $\langle Q^2_r \rangle$  \\ \hline
2 & 2.3715 & 2.3714(6) & $12^4$ & 0.2879(13) & 3.69(13)   & 3.39(12) \\
3 & 5.8000 & 5.8000(5) & $10^4$ & 0.3133(13) & 2.917(79)  & 2.469(74) \\ 
4 & 10.637 & 10.637(1) & $10^4$ & 0.3254(14) & 3.375(90)  & 2.881(81) \\ 
6 & 24.515 & 24.514(3) & $10^4$ & 0.3385(15) & 3.48(18)   & 2.96(17)  \\ 
8 & 44.000 & 43.98(3)  & $10^4$ & 0.3413(13) & 3.20(42)   & 2.68(35) \\  \hline
\end{tabular}
\caption{\label{table_chi5a}
Fluctuation of the topological charge in various $SU(N)$ lattice gauge 
theories for a fixed value of the lattice spacing $a = 1/5T_c$. 
$Q_r$ is the raw real-valued lattice charge and $Q_I$ is our best 
estimate of the integer that corresponds to it. Also
shown is the string tension, $\sigma$, the critical value of
$\beta$, and the lattice parameters.}
\end{center}
\end{table}

\begin{table}
\begin{center}
\begin{tabular}{|c|c|l|l|}\hline
$N$ & $\beta$ & $\chi_{I,t}^{1/4}/\surd\sigma$ & 
$\chi_{r,t}^{1/4}/\surd\sigma$ \\ \hline
2 & 2.3715 & 0.4013(40)  & 0.3928(40) \\
3 & 5.8000 & 0.4171(34)  & 0.4001(35) \\ 
4 & 10.637 & 0.4165(33)  & 0.4004(33) \\ 
6 & 24.515 & 0.4034(56)  & 0.3875(58) \\ 
8 & 44.000 & 0.392(13)   & 0.375(13) \\  \hline
\end{tabular}
\caption{\label{table_chi5b}
The topological suceptibility in units of the string tension
for the calculations in Table~\ref{table_chi5a}.} 
\end{center}
\end{table}

\begin{table}
\begin{center}
\begin{tabular}{|c|c|c|c||l|l|l||l|} \hline
\multicolumn{5}{|c|}{} 
& \multicolumn{3}{|c|}{ $\chi_{r,t}/\sigma^2$ } \\  \hline
$N$ & $\beta$ & lattice  & $a\surd\sigma$ & $T/T_c$ &
confined  & deconfined & $T=0$ \\ \hline
3 & 5.8000 & $32^3 5$ & 0.3133(13) & 1.00 & 0.0270(13)$^\star$ & 0.0141(13)$^\star$  
& 0.0256(9) \\ 
4 & 10.635 & $32^3 5$ & 0.3262(15) & 1.00 & 0.0260(19) & 0.0070(7) 
& 0.0257(9) \\ 
6 & 24.515 & $16^3 5$ & 0.3385(15) & 1.00 & 0.0199(14) & 0.00171(14)  
& 0.0226(13) \\ 
8 & 43.965 & $12^3 5$ & 0.3462(15) & 0.995 & 0.0161(28) & 0.00005(3) 
& 0.0198(26) \\  \hline
\end{tabular}
\caption{\label{table_chiTc}
Topological susceptibility in units of the string tension calculated
separately in the confined and deconfined phases, for the
values of $N$, on the lattice volumes and at the values of 
$\beta$ shown. }
\end{center}
\end{table}

\begin{table}
\begin{center}
\begin{tabular}{|c|l|c||l|l|l|}\hline
\multicolumn{6}{|c|}{ $SU(8)$ } \\ \hline
$\beta$ & lattice  &  $a\surd\sigma$ & $T/T_c$ & 
$\chi_{r,t}/\sigma^2$:conf  & $\chi_{r,t}/\sigma^2$:deconf
\\ \hline
44.00 & $8^3 4$  & 0.3413(13) & 1.26 &            & 0.00000(4) \\
43.85 & $8^4  $  & 0.3630(21) & 0    & 0.0191(24) &    \\  \hline
\end{tabular}
\caption{\label{table_chiTN8}
Some additional $SU(8)$ calculations of the
topological susceptibility in units of the string tension.}
\end{center}
\end{table}

\begin{table}
\begin{center}
\begin{tabular}{|c|c|c||c|l|l|}\hline
\multicolumn{6}{|c|}{ $SU(6)$ } \\ \hline
$\beta$ & lattice  & $a\surd\sigma$ & $T/T_c$ & 
$\chi_{r,t}/\sigma^2$:conf  & $\chi_{r,t}/\sigma^2$:deconf
\\ \hline
24.500  & $16^3 5$ & 0.3414(20) &  0.99 &  & 0.00219(18) \\
24.530  & $16^3 5$ & 0.3356(20) &  1.01 &  & 0.00115(10) \\
24.515  & $10^3 4$ & 0.3385(15) &  1.25 &  & 0.000012(5) \\  \hline
\end{tabular}
\caption{\label{table_chiTN6}
Some additional $SU(6)$ calculations of the
topological susceptibility in units of the string tension.}
\end{center}
\end{table}

\begin{table}
\begin{center}
\begin{tabular}{|c|c|c||c|l|l|}\hline
\multicolumn{6}{|c|}{ $SU(4)$ } \\ \hline
$\beta$ & lattice  &  $a\surd\sigma$ & $T/T_c$ & 
$\chi_{r,t}/\sigma^2$:conf  & $\chi_{r,t}/\sigma^2$:deconf
\\ \hline
10.645 & $12^3 5$ & 0.3222(16) & 1.01 &   & 0.0058(3)$^\star$ \\
10.642 & $20^3 5$ & 0.3235(16) & 1.01 &   & 0.0060(3) \\
10.637 & $20^3 5$ & 0.3254(14) & 1.00 & 0.0246(10)$^\star$ & 0.0071(4)$^\star$ \\
10.635 & $20^3 5$ & 0.3262(16) & 1.00 & 0.0247(11)$^\star$ & 0.0076(4)$^\star$ \\
10.633 & $20^3 5$ & 0.3270(16) & 1.00 & 0.0257(8)$^\star$ & 0.0074(5)$^\star$ \\
10.637 & $12^3 4$ & 0.3254(14) & 1.00 &  & 0.000115(22)  \\  \hline
\end{tabular}
\caption{\label{table_chiTN4}
Some additional $SU(4)$ calculations of the
topological susceptibility in units of the string tension.}
\end{center}
\end{table}

\begin{table}
\begin{center}
\begin{tabular}{|c|c|c||c|l|l|}\hline
\multicolumn{6}{|c|}{ $SU(3)$ } \\ \hline
$\beta$ & lattice  &  $a\surd\sigma$ & $T/T_c$ & 
$\chi_{r,t}/\sigma^2$:conf  & $\chi_{r,t}/\sigma^2$:deconf
\\ \hline
5.7950 & $20^3 5$ & 0.3163(20) & 0.99  & 0.0243(10) &  \\
5.7975 & $20^3 5$ & 0.3148(16) & 1.00  & 0.0253(8)$^\star$ & 0.0144(10)$^\star$ \\
5.8000 & $20^3 5$ & 0.3133(13) & 1.00  &  & 0.0132(5) \\
5.8025 & $20^3 5$ & 0.3118(16) & 1.00  &  & 0.0124(5) \\  
5.8000 & $12^3 4$ & 0.3133(13) & 1.25  &    & 0.00042(4) \\  \hline
\end{tabular}
\caption{\label{table_chiTN3}
Some additional $SU(3)$ calculations of the
topological susceptibility in units of the string tension.}
\end{center}
\end{table}

\clearpage

\begin  {figure}[p]
\begin  {center}
\leavevmode
\setlength{\unitlength}{0.240900pt}
\ifx\plotpoint\undefined\newsavebox{\plotpoint}\fi
\begin{picture}(1500,1800)(0,0)
\font\gnuplot=cmr10 at 12pt
\gnuplot
\sbox{\plotpoint}{\rule[-0.200pt]{0.400pt}{0.400pt}}%
\put(375.0,250.0){\rule[-0.200pt]{4.818pt}{0.400pt}}
\put(350,250){\makebox(0,0)[r]{\ \ {$0$}}}
\put(1405.0,250.0){\rule[-0.200pt]{4.818pt}{0.400pt}}
\put(375.0,375.0){\rule[-0.200pt]{4.818pt}{0.400pt}}
\put(350,375){\makebox(0,0)[r]{\ \ {$0.25$}}}
\put(1405.0,375.0){\rule[-0.200pt]{4.818pt}{0.400pt}}
\put(375.0,500.0){\rule[-0.200pt]{4.818pt}{0.400pt}}
\put(350,500){\makebox(0,0)[r]{\ \ {$0.5$}}}
\put(1405.0,500.0){\rule[-0.200pt]{4.818pt}{0.400pt}}
\put(375.0,625.0){\rule[-0.200pt]{4.818pt}{0.400pt}}
\put(350,625){\makebox(0,0)[r]{\ \ {$0.75$}}}
\put(1405.0,625.0){\rule[-0.200pt]{4.818pt}{0.400pt}}
\put(375.0,750.0){\rule[-0.200pt]{4.818pt}{0.400pt}}
\put(350,750){\makebox(0,0)[r]{\ \ {$1$}}}
\put(1405.0,750.0){\rule[-0.200pt]{4.818pt}{0.400pt}}
\put(375.0,875.0){\rule[-0.200pt]{4.818pt}{0.400pt}}
\put(350,875){\makebox(0,0)[r]{\ \ {$1.25$}}}
\put(1405.0,875.0){\rule[-0.200pt]{4.818pt}{0.400pt}}
\put(375.0,1000.0){\rule[-0.200pt]{4.818pt}{0.400pt}}
\put(350,1000){\makebox(0,0)[r]{\ \ {$1.5$}}}
\put(1405.0,1000.0){\rule[-0.200pt]{4.818pt}{0.400pt}}
\put(375.0,1125.0){\rule[-0.200pt]{4.818pt}{0.400pt}}
\put(350,1125){\makebox(0,0)[r]{\ \ {$1.75$}}}
\put(1405.0,1125.0){\rule[-0.200pt]{4.818pt}{0.400pt}}
\put(375.0,1250.0){\rule[-0.200pt]{4.818pt}{0.400pt}}
\put(350,1250){\makebox(0,0)[r]{\ \ {$2$}}}
\put(1405.0,1250.0){\rule[-0.200pt]{4.818pt}{0.400pt}}
\put(375.0,1375.0){\rule[-0.200pt]{4.818pt}{0.400pt}}
\put(350,1375){\makebox(0,0)[r]{\ \ {$2.25$}}}
\put(1405.0,1375.0){\rule[-0.200pt]{4.818pt}{0.400pt}}
\put(375.0,1500.0){\rule[-0.200pt]{4.818pt}{0.400pt}}
\put(350,1500){\makebox(0,0)[r]{\ \ {$2.5$}}}
\put(1405.0,1500.0){\rule[-0.200pt]{4.818pt}{0.400pt}}
\put(375.0,1625.0){\rule[-0.200pt]{4.818pt}{0.400pt}}
\put(350,1625){\makebox(0,0)[r]{\ \ {$2.75$}}}
\put(1405.0,1625.0){\rule[-0.200pt]{4.818pt}{0.400pt}}
\put(375.0,250.0){\rule[-0.200pt]{0.400pt}{4.818pt}}
\put(375,200){\makebox(0,0){\ {$0$}}}
\put(375.0,1730.0){\rule[-0.200pt]{0.400pt}{4.818pt}}
\put(480.0,250.0){\rule[-0.200pt]{0.400pt}{4.818pt}}
\put(480,200){\makebox(0,0){\ {$1$}}}
\put(480.0,1730.0){\rule[-0.200pt]{0.400pt}{4.818pt}}
\put(585.0,250.0){\rule[-0.200pt]{0.400pt}{4.818pt}}
\put(585,200){\makebox(0,0){\ {$2$}}}
\put(585.0,1730.0){\rule[-0.200pt]{0.400pt}{4.818pt}}
\put(690.0,250.0){\rule[-0.200pt]{0.400pt}{4.818pt}}
\put(690,200){\makebox(0,0){\ {$3$}}}
\put(690.0,1730.0){\rule[-0.200pt]{0.400pt}{4.818pt}}
\put(795.0,250.0){\rule[-0.200pt]{0.400pt}{4.818pt}}
\put(795,200){\makebox(0,0){\ {$4$}}}
\put(795.0,1730.0){\rule[-0.200pt]{0.400pt}{4.818pt}}
\put(900.0,250.0){\rule[-0.200pt]{0.400pt}{4.818pt}}
\put(900,200){\makebox(0,0){\ {$5$}}}
\put(900.0,1730.0){\rule[-0.200pt]{0.400pt}{4.818pt}}
\put(1005.0,250.0){\rule[-0.200pt]{0.400pt}{4.818pt}}
\put(1005,200){\makebox(0,0){\ {$6$}}}
\put(1005.0,1730.0){\rule[-0.200pt]{0.400pt}{4.818pt}}
\put(1110.0,250.0){\rule[-0.200pt]{0.400pt}{4.818pt}}
\put(1110,200){\makebox(0,0){\ {$7$}}}
\put(1110.0,1730.0){\rule[-0.200pt]{0.400pt}{4.818pt}}
\put(1215.0,250.0){\rule[-0.200pt]{0.400pt}{4.818pt}}
\put(1215,200){\makebox(0,0){\ {$8$}}}
\put(1215.0,1730.0){\rule[-0.200pt]{0.400pt}{4.818pt}}
\put(1320.0,250.0){\rule[-0.200pt]{0.400pt}{4.818pt}}
\put(1320,200){\makebox(0,0){\ {$9$}}}
\put(1320.0,1730.0){\rule[-0.200pt]{0.400pt}{4.818pt}}
\put(1425.0,250.0){\rule[-0.200pt]{0.400pt}{4.818pt}}
\put(1425,200){\makebox(0,0){\ {$10$}}}
\put(1425.0,1730.0){\rule[-0.200pt]{0.400pt}{4.818pt}}
\put(375.0,250.0){\rule[-0.200pt]{252.945pt}{0.400pt}}
\put(1425.0,250.0){\rule[-0.200pt]{0.400pt}{361.350pt}}
\put(375.0,1750.0){\rule[-0.200pt]{252.945pt}{0.400pt}}
\put(150,1300){\makebox(0,0){\Large{$D(\rho)$}}}
\put(875,25){\makebox(0,0){\large{$\rho$}}}
\put(375.0,250.0){\rule[-0.200pt]{0.400pt}{361.350pt}}
\put(581,252){\circle*{12}}
\put(601,273){\circle*{12}}
\put(627,302){\circle*{12}}
\put(651,348){\circle*{12}}
\put(675,342){\circle*{12}}
\put(701,289){\circle*{12}}
\put(727,262){\circle*{12}}
\put(754,254){\circle*{12}}
\put(781,253){\circle*{12}}
\put(808,252){\circle*{12}}
\put(836,253){\circle*{12}}
\put(860,252){\circle*{12}}
\put(887,253){\circle*{12}}
\put(915,253){\circle*{12}}
\put(936,253){\circle*{12}}
\put(965,253){\circle*{12}}
\put(995,256){\circle*{12}}
\put(1020,269){\circle*{12}}
\put(1047,291){\circle*{12}}
\put(1072,359){\circle*{12}}
\put(1100,487){\circle*{12}}
\put(1125,621){\circle*{12}}
\put(1150,869){\circle*{12}}
\put(1176,1029){\circle*{12}}
\put(1202,1363){\circle*{12}}
\put(1229,1367){\circle*{12}}
\put(1255,1576){\circle*{12}}
\put(1282,1455){\circle*{12}}
\put(1306,1190){\circle*{12}}
\put(1331,1268){\circle*{12}}
\put(1358,1297){\circle*{12}}
\put(1385,907){\circle*{12}}
\put(1405,756){\circle*{12}}
\end{picture}
\end    {center}
\vskip 0.15in
\caption{The $SU(6)$ instanton size density, $D(\rho)$, at 
$T\simeq T_c$ in the deconfined phase, with $a\simeq 1/5T_c$. 
Normalised to a $16^3 5$ volume.}
\label{fig_drhoTdeconfn6}
\end    {figure}

\begin  {figure}[p]
\begin  {center}
\leavevmode
\setlength{\unitlength}{0.240900pt}
\ifx\plotpoint\undefined\newsavebox{\plotpoint}\fi
\begin{picture}(1500,1800)(0,0)
\font\gnuplot=cmr10 at 12pt
\gnuplot
\sbox{\plotpoint}{\rule[-0.200pt]{0.400pt}{0.400pt}}%
\put(375.0,250.0){\rule[-0.200pt]{4.818pt}{0.400pt}}
\put(350,250){\makebox(0,0)[r]{\ \ {$0$}}}
\put(1405.0,250.0){\rule[-0.200pt]{4.818pt}{0.400pt}}
\put(375.0,375.0){\rule[-0.200pt]{4.818pt}{0.400pt}}
\put(350,375){\makebox(0,0)[r]{\ \ {$0.25$}}}
\put(1405.0,375.0){\rule[-0.200pt]{4.818pt}{0.400pt}}
\put(375.0,500.0){\rule[-0.200pt]{4.818pt}{0.400pt}}
\put(350,500){\makebox(0,0)[r]{\ \ {$0.5$}}}
\put(1405.0,500.0){\rule[-0.200pt]{4.818pt}{0.400pt}}
\put(375.0,625.0){\rule[-0.200pt]{4.818pt}{0.400pt}}
\put(350,625){\makebox(0,0)[r]{\ \ {$0.75$}}}
\put(1405.0,625.0){\rule[-0.200pt]{4.818pt}{0.400pt}}
\put(375.0,750.0){\rule[-0.200pt]{4.818pt}{0.400pt}}
\put(350,750){\makebox(0,0)[r]{\ \ {$1$}}}
\put(1405.0,750.0){\rule[-0.200pt]{4.818pt}{0.400pt}}
\put(375.0,875.0){\rule[-0.200pt]{4.818pt}{0.400pt}}
\put(350,875){\makebox(0,0)[r]{\ \ {$1.25$}}}
\put(1405.0,875.0){\rule[-0.200pt]{4.818pt}{0.400pt}}
\put(375.0,1000.0){\rule[-0.200pt]{4.818pt}{0.400pt}}
\put(350,1000){\makebox(0,0)[r]{\ \ {$1.5$}}}
\put(1405.0,1000.0){\rule[-0.200pt]{4.818pt}{0.400pt}}
\put(375.0,1125.0){\rule[-0.200pt]{4.818pt}{0.400pt}}
\put(350,1125){\makebox(0,0)[r]{\ \ {$1.75$}}}
\put(1405.0,1125.0){\rule[-0.200pt]{4.818pt}{0.400pt}}
\put(375.0,1250.0){\rule[-0.200pt]{4.818pt}{0.400pt}}
\put(350,1250){\makebox(0,0)[r]{\ \ {$2$}}}
\put(1405.0,1250.0){\rule[-0.200pt]{4.818pt}{0.400pt}}
\put(375.0,1375.0){\rule[-0.200pt]{4.818pt}{0.400pt}}
\put(350,1375){\makebox(0,0)[r]{\ \ {$2.25$}}}
\put(1405.0,1375.0){\rule[-0.200pt]{4.818pt}{0.400pt}}
\put(375.0,1500.0){\rule[-0.200pt]{4.818pt}{0.400pt}}
\put(350,1500){\makebox(0,0)[r]{\ \ {$2.5$}}}
\put(1405.0,1500.0){\rule[-0.200pt]{4.818pt}{0.400pt}}
\put(375.0,1625.0){\rule[-0.200pt]{4.818pt}{0.400pt}}
\put(350,1625){\makebox(0,0)[r]{\ \ {$2.75$}}}
\put(1405.0,1625.0){\rule[-0.200pt]{4.818pt}{0.400pt}}
\put(375.0,250.0){\rule[-0.200pt]{0.400pt}{4.818pt}}
\put(375,200){\makebox(0,0){\ {$0$}}}
\put(375.0,1730.0){\rule[-0.200pt]{0.400pt}{4.818pt}}
\put(480.0,250.0){\rule[-0.200pt]{0.400pt}{4.818pt}}
\put(480,200){\makebox(0,0){\ {$1$}}}
\put(480.0,1730.0){\rule[-0.200pt]{0.400pt}{4.818pt}}
\put(585.0,250.0){\rule[-0.200pt]{0.400pt}{4.818pt}}
\put(585,200){\makebox(0,0){\ {$2$}}}
\put(585.0,1730.0){\rule[-0.200pt]{0.400pt}{4.818pt}}
\put(690.0,250.0){\rule[-0.200pt]{0.400pt}{4.818pt}}
\put(690,200){\makebox(0,0){\ {$3$}}}
\put(690.0,1730.0){\rule[-0.200pt]{0.400pt}{4.818pt}}
\put(795.0,250.0){\rule[-0.200pt]{0.400pt}{4.818pt}}
\put(795,200){\makebox(0,0){\ {$4$}}}
\put(795.0,1730.0){\rule[-0.200pt]{0.400pt}{4.818pt}}
\put(900.0,250.0){\rule[-0.200pt]{0.400pt}{4.818pt}}
\put(900,200){\makebox(0,0){\ {$5$}}}
\put(900.0,1730.0){\rule[-0.200pt]{0.400pt}{4.818pt}}
\put(1005.0,250.0){\rule[-0.200pt]{0.400pt}{4.818pt}}
\put(1005,200){\makebox(0,0){\ {$6$}}}
\put(1005.0,1730.0){\rule[-0.200pt]{0.400pt}{4.818pt}}
\put(1110.0,250.0){\rule[-0.200pt]{0.400pt}{4.818pt}}
\put(1110,200){\makebox(0,0){\ {$7$}}}
\put(1110.0,1730.0){\rule[-0.200pt]{0.400pt}{4.818pt}}
\put(1215.0,250.0){\rule[-0.200pt]{0.400pt}{4.818pt}}
\put(1215,200){\makebox(0,0){\ {$8$}}}
\put(1215.0,1730.0){\rule[-0.200pt]{0.400pt}{4.818pt}}
\put(1320.0,250.0){\rule[-0.200pt]{0.400pt}{4.818pt}}
\put(1320,200){\makebox(0,0){\ {$9$}}}
\put(1320.0,1730.0){\rule[-0.200pt]{0.400pt}{4.818pt}}
\put(1425.0,250.0){\rule[-0.200pt]{0.400pt}{4.818pt}}
\put(1425,200){\makebox(0,0){\ {$10$}}}
\put(1425.0,1730.0){\rule[-0.200pt]{0.400pt}{4.818pt}}
\put(375.0,250.0){\rule[-0.200pt]{252.945pt}{0.400pt}}
\put(1425.0,250.0){\rule[-0.200pt]{0.400pt}{361.350pt}}
\put(375.0,1750.0){\rule[-0.200pt]{252.945pt}{0.400pt}}
\put(150,1300){\makebox(0,0){\Large{$D(\rho)$}}}
\put(875,25){\makebox(0,0){\large{$\rho$}}}
\put(375.0,250.0){\rule[-0.200pt]{0.400pt}{361.350pt}}
\put(600,258){\circle*{12}}
\put(627,270){\circle*{12}}
\put(652,303){\circle*{12}}
\put(678,340){\circle*{12}}
\put(705,418){\circle*{12}}
\put(730,562){\circle*{12}}
\put(756,739){\circle*{12}}
\put(782,973){\circle*{12}}
\put(809,1204){\circle*{12}}
\put(835,1388){\circle*{12}}
\put(860,1457){\circle*{12}}
\put(886,1499){\circle*{12}}
\put(913,1428){\circle*{12}}
\put(939,1302){\circle*{12}}
\put(965,1159){\circle*{12}}
\put(991,994){\circle*{12}}
\put(1018,912){\circle*{12}}
\put(1044,741){\circle*{12}}
\put(1070,669){\circle*{12}}
\put(1096,593){\circle*{12}}
\put(1124,484){\circle*{12}}
\put(1149,454){\circle*{12}}
\put(1175,398){\circle*{12}}
\put(1201,383){\circle*{12}}
\put(1228,338){\circle*{12}}
\put(1254,348){\circle*{12}}
\put(1282,318){\circle*{12}}
\put(1306,308){\circle*{12}}
\put(1329,301){\circle*{12}}
\put(1359,296){\circle*{12}}
\put(1386,277){\circle*{12}}
\put(1406,270){\circle*{12}}
\end{picture}
\end    {center}
\vskip 0.15in
\caption{The $SU(6)$ instanton size density, $D(\rho)$, at 
$T\simeq T_c$ in the confined phase, with $a\simeq 1/5T_c$. 
Normalised to a  $16^3 5$ volume.}
\label{fig_drhoTconfn6}
\end    {figure}

\begin  {figure}[p]
\begin  {center}
\leavevmode
\setlength{\unitlength}{0.240900pt}
\ifx\plotpoint\undefined\newsavebox{\plotpoint}\fi
\sbox{\plotpoint}{\rule[-0.200pt]{0.400pt}{0.400pt}}%
\begin{picture}(1500,1800)(0,0)
\font\gnuplot=cmr10 at 12pt
\gnuplot
\sbox{\plotpoint}{\rule[-0.200pt]{0.400pt}{0.400pt}}%
\put(350.0,250.0){\rule[-0.200pt]{4.818pt}{0.400pt}}
\put(325,250){\makebox(0,0)[r]{\ \ {$0$}}}
\put(1405.0,250.0){\rule[-0.200pt]{4.818pt}{0.400pt}}
\put(350.0,426.0){\rule[-0.200pt]{4.818pt}{0.400pt}}
\put(325,426){\makebox(0,0)[r]{\ \ {$100$}}}
\put(1405.0,426.0){\rule[-0.200pt]{4.818pt}{0.400pt}}
\put(350.0,603.0){\rule[-0.200pt]{4.818pt}{0.400pt}}
\put(325,603){\makebox(0,0)[r]{\ \ {$200$}}}
\put(1405.0,603.0){\rule[-0.200pt]{4.818pt}{0.400pt}}
\put(350.0,779.0){\rule[-0.200pt]{4.818pt}{0.400pt}}
\put(325,779){\makebox(0,0)[r]{\ \ {$300$}}}
\put(1405.0,779.0){\rule[-0.200pt]{4.818pt}{0.400pt}}
\put(350.0,956.0){\rule[-0.200pt]{4.818pt}{0.400pt}}
\put(325,956){\makebox(0,0)[r]{\ \ {$400$}}}
\put(1405.0,956.0){\rule[-0.200pt]{4.818pt}{0.400pt}}
\put(350.0,1132.0){\rule[-0.200pt]{4.818pt}{0.400pt}}
\put(325,1132){\makebox(0,0)[r]{\ \ {$500$}}}
\put(1405.0,1132.0){\rule[-0.200pt]{4.818pt}{0.400pt}}
\put(350.0,1309.0){\rule[-0.200pt]{4.818pt}{0.400pt}}
\put(325,1309){\makebox(0,0)[r]{\ \ {$600$}}}
\put(1405.0,1309.0){\rule[-0.200pt]{4.818pt}{0.400pt}}
\put(350.0,1485.0){\rule[-0.200pt]{4.818pt}{0.400pt}}
\put(325,1485){\makebox(0,0)[r]{\ \ {$700$}}}
\put(1405.0,1485.0){\rule[-0.200pt]{4.818pt}{0.400pt}}
\put(350.0,1662.0){\rule[-0.200pt]{4.818pt}{0.400pt}}
\put(325,1662){\makebox(0,0)[r]{\ \ {$800$}}}
\put(1405.0,1662.0){\rule[-0.200pt]{4.818pt}{0.400pt}}
\put(350.0,250.0){\rule[-0.200pt]{0.400pt}{4.818pt}}
\put(350,200){\makebox(0,0){\ {$0$}}}
\put(350.0,1730.0){\rule[-0.200pt]{0.400pt}{4.818pt}}
\put(458.0,250.0){\rule[-0.200pt]{0.400pt}{4.818pt}}
\put(458,200){\makebox(0,0){\ {$1$}}}
\put(458.0,1730.0){\rule[-0.200pt]{0.400pt}{4.818pt}}
\put(565.0,250.0){\rule[-0.200pt]{0.400pt}{4.818pt}}
\put(565,200){\makebox(0,0){\ {$2$}}}
\put(565.0,1730.0){\rule[-0.200pt]{0.400pt}{4.818pt}}
\put(673.0,250.0){\rule[-0.200pt]{0.400pt}{4.818pt}}
\put(673,200){\makebox(0,0){\ {$3$}}}
\put(673.0,1730.0){\rule[-0.200pt]{0.400pt}{4.818pt}}
\put(780.0,250.0){\rule[-0.200pt]{0.400pt}{4.818pt}}
\put(780,200){\makebox(0,0){\ {$4$}}}
\put(780.0,1730.0){\rule[-0.200pt]{0.400pt}{4.818pt}}
\put(888.0,250.0){\rule[-0.200pt]{0.400pt}{4.818pt}}
\put(888,200){\makebox(0,0){\ {$5$}}}
\put(888.0,1730.0){\rule[-0.200pt]{0.400pt}{4.818pt}}
\put(995.0,250.0){\rule[-0.200pt]{0.400pt}{4.818pt}}
\put(995,200){\makebox(0,0){\ {$6$}}}
\put(995.0,1730.0){\rule[-0.200pt]{0.400pt}{4.818pt}}
\put(1103.0,250.0){\rule[-0.200pt]{0.400pt}{4.818pt}}
\put(1103,200){\makebox(0,0){\ {$7$}}}
\put(1103.0,1730.0){\rule[-0.200pt]{0.400pt}{4.818pt}}
\put(1210.0,250.0){\rule[-0.200pt]{0.400pt}{4.818pt}}
\put(1210,200){\makebox(0,0){\ {$8$}}}
\put(1210.0,1730.0){\rule[-0.200pt]{0.400pt}{4.818pt}}
\put(1318.0,250.0){\rule[-0.200pt]{0.400pt}{4.818pt}}
\put(1318,200){\makebox(0,0){\ {$9$}}}
\put(1318.0,1730.0){\rule[-0.200pt]{0.400pt}{4.818pt}}
\put(1425.0,250.0){\rule[-0.200pt]{0.400pt}{4.818pt}}
\put(1425,200){\makebox(0,0){\ {$10$}}}
\put(1425.0,1730.0){\rule[-0.200pt]{0.400pt}{4.818pt}}
\put(350.0,250.0){\rule[-0.200pt]{258.967pt}{0.400pt}}
\put(1425.0,250.0){\rule[-0.200pt]{0.400pt}{361.350pt}}
\put(350.0,1750.0){\rule[-0.200pt]{258.967pt}{0.400pt}}
\put(150,1300){\makebox(0,0){\Large{$N(\rho)$}}}
\put(862,25){\makebox(0,0){\large{$\rho$}}}
\put(350.0,250.0){\rule[-0.200pt]{0.400pt}{361.350pt}}
\put(538,254){\makebox(0,0){$\times$}}
\put(592,342){\makebox(0,0){$\times$}}
\put(646,490){\makebox(0,0){$\times$}}
\put(699,326){\makebox(0,0){$\times$}}
\put(753,266){\makebox(0,0){$\times$}}
\put(807,266){\makebox(0,0){$\times$}}
\put(861,264){\makebox(0,0){$\times$}}
\put(914,266){\makebox(0,0){$\times$}}
\put(968,276){\makebox(0,0){$\times$}}
\put(1022,435){\makebox(0,0){$\times$}}
\put(1076,1438){\makebox(0,0){$\times$}}
\put(1129,1141){\makebox(0,0){$\times$}}
\put(1183,1602){\makebox(0,0){$\times$}}
\put(1237,335){\makebox(0,0){$\times$}}
\put(538,259){\circle*{18}}
\put(592,636){\circle*{18}}
\put(646,1118){\circle*{18}}
\put(699,411){\circle*{18}}
\put(753,255){\circle*{18}}
\put(592,268){\circle{18}}
\put(646,287){\circle{18}}
\put(699,271){\circle{18}}
\put(753,261){\circle{18}}
\put(807,257){\circle{18}}
\put(861,259){\circle{18}}
\put(914,254){\circle{18}}
\put(968,264){\circle{18}}
\put(1022,336){\circle{18}}
\put(1076,672){\circle{18}}
\put(1129,546){\circle{18}}
\put(1183,726){\circle{18}}
\put(1237,278){\circle{18}}
\end{picture}
\end    {center}
\vskip 0.15in
\caption{The number of instantons in 2000 lattice fields 
using only the largest positive and negative peaks in each 
field. All in the deconfined phase, at $T=T_c$, on a $16^3 5$ 
lattice in $SU(6)$. Separately for the lattice fields with
$Q\not= 0$ and for the fields with  $Q=0$, $\times$.
For the former we show separately 
sizes from peaks with the same sign as $Q$ ($\bullet$) and
with the opposite sign ($\circ$).}
\label{fig_rhodeconfn6}
\end    {figure}
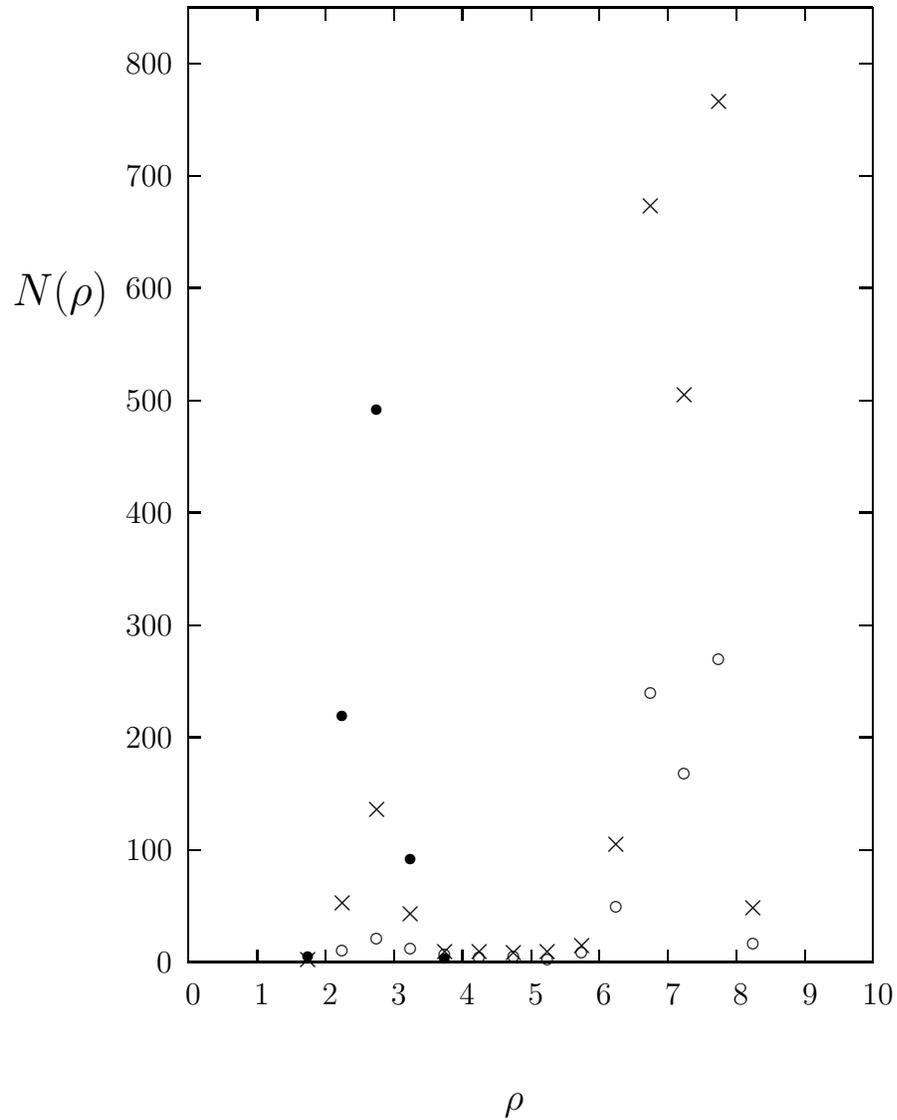

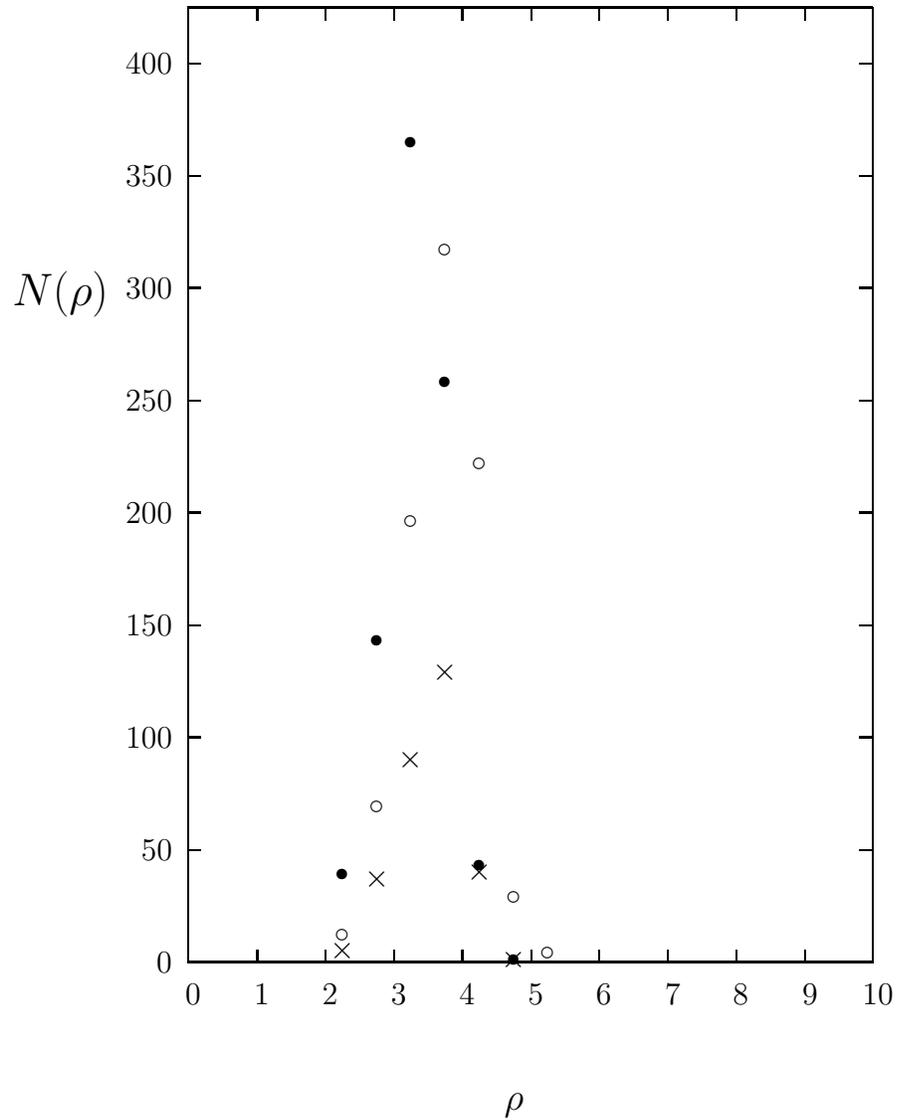
\begin  {figure}[p]
\begin  {center}
\leavevmode
\setlength{\unitlength}{0.240900pt}
\ifx\plotpoint\undefined\newsavebox{\plotpoint}\fi
\sbox{\plotpoint}{\rule[-0.200pt]{0.400pt}{0.400pt}}%
\begin{picture}(1500,1800)(0,0)
\font\gnuplot=cmr10 at 12pt
\gnuplot
\sbox{\plotpoint}{\rule[-0.200pt]{0.400pt}{0.400pt}}%
\put(350.0,250.0){\rule[-0.200pt]{4.818pt}{0.400pt}}
\put(325,250){\makebox(0,0)[r]{\ \ {$0$}}}
\put(1405.0,250.0){\rule[-0.200pt]{4.818pt}{0.400pt}}
\put(350.0,426.0){\rule[-0.200pt]{4.818pt}{0.400pt}}
\put(325,426){\makebox(0,0)[r]{\ \ {$50$}}}
\put(1405.0,426.0){\rule[-0.200pt]{4.818pt}{0.400pt}}
\put(350.0,603.0){\rule[-0.200pt]{4.818pt}{0.400pt}}
\put(325,603){\makebox(0,0)[r]{\ \ {$100$}}}
\put(1405.0,603.0){\rule[-0.200pt]{4.818pt}{0.400pt}}
\put(350.0,779.0){\rule[-0.200pt]{4.818pt}{0.400pt}}
\put(325,779){\makebox(0,0)[r]{\ \ {$150$}}}
\put(1405.0,779.0){\rule[-0.200pt]{4.818pt}{0.400pt}}
\put(350.0,956.0){\rule[-0.200pt]{4.818pt}{0.400pt}}
\put(325,956){\makebox(0,0)[r]{\ \ {$200$}}}
\put(1405.0,956.0){\rule[-0.200pt]{4.818pt}{0.400pt}}
\put(350.0,1132.0){\rule[-0.200pt]{4.818pt}{0.400pt}}
\put(325,1132){\makebox(0,0)[r]{\ \ {$250$}}}
\put(1405.0,1132.0){\rule[-0.200pt]{4.818pt}{0.400pt}}
\put(350.0,1309.0){\rule[-0.200pt]{4.818pt}{0.400pt}}
\put(325,1309){\makebox(0,0)[r]{\ \ {$300$}}}
\put(1405.0,1309.0){\rule[-0.200pt]{4.818pt}{0.400pt}}
\put(350.0,1485.0){\rule[-0.200pt]{4.818pt}{0.400pt}}
\put(325,1485){\makebox(0,0)[r]{\ \ {$350$}}}
\put(1405.0,1485.0){\rule[-0.200pt]{4.818pt}{0.400pt}}
\put(350.0,1662.0){\rule[-0.200pt]{4.818pt}{0.400pt}}
\put(325,1662){\makebox(0,0)[r]{\ \ {$400$}}}
\put(1405.0,1662.0){\rule[-0.200pt]{4.818pt}{0.400pt}}
\put(350.0,250.0){\rule[-0.200pt]{0.400pt}{4.818pt}}
\put(350,200){\makebox(0,0){\ {$0$}}}
\put(350.0,1730.0){\rule[-0.200pt]{0.400pt}{4.818pt}}
\put(458.0,250.0){\rule[-0.200pt]{0.400pt}{4.818pt}}
\put(458,200){\makebox(0,0){\ {$1$}}}
\put(458.0,1730.0){\rule[-0.200pt]{0.400pt}{4.818pt}}
\put(565.0,250.0){\rule[-0.200pt]{0.400pt}{4.818pt}}
\put(565,200){\makebox(0,0){\ {$2$}}}
\put(565.0,1730.0){\rule[-0.200pt]{0.400pt}{4.818pt}}
\put(673.0,250.0){\rule[-0.200pt]{0.400pt}{4.818pt}}
\put(673,200){\makebox(0,0){\ {$3$}}}
\put(673.0,1730.0){\rule[-0.200pt]{0.400pt}{4.818pt}}
\put(780.0,250.0){\rule[-0.200pt]{0.400pt}{4.818pt}}
\put(780,200){\makebox(0,0){\ {$4$}}}
\put(780.0,1730.0){\rule[-0.200pt]{0.400pt}{4.818pt}}
\put(888.0,250.0){\rule[-0.200pt]{0.400pt}{4.818pt}}
\put(888,200){\makebox(0,0){\ {$5$}}}
\put(888.0,1730.0){\rule[-0.200pt]{0.400pt}{4.818pt}}
\put(995.0,250.0){\rule[-0.200pt]{0.400pt}{4.818pt}}
\put(995,200){\makebox(0,0){\ {$6$}}}
\put(995.0,1730.0){\rule[-0.200pt]{0.400pt}{4.818pt}}
\put(1103.0,250.0){\rule[-0.200pt]{0.400pt}{4.818pt}}
\put(1103,200){\makebox(0,0){\ {$7$}}}
\put(1103.0,1730.0){\rule[-0.200pt]{0.400pt}{4.818pt}}
\put(1210.0,250.0){\rule[-0.200pt]{0.400pt}{4.818pt}}
\put(1210,200){\makebox(0,0){\ {$8$}}}
\put(1210.0,1730.0){\rule[-0.200pt]{0.400pt}{4.818pt}}
\put(1318.0,250.0){\rule[-0.200pt]{0.400pt}{4.818pt}}
\put(1318,200){\makebox(0,0){\ {$9$}}}
\put(1318.0,1730.0){\rule[-0.200pt]{0.400pt}{4.818pt}}
\put(1425.0,250.0){\rule[-0.200pt]{0.400pt}{4.818pt}}
\put(1425,200){\makebox(0,0){\ {$10$}}}
\put(1425.0,1730.0){\rule[-0.200pt]{0.400pt}{4.818pt}}
\put(350.0,250.0){\rule[-0.200pt]{258.967pt}{0.400pt}}
\put(1425.0,250.0){\rule[-0.200pt]{0.400pt}{361.350pt}}
\put(350.0,1750.0){\rule[-0.200pt]{258.967pt}{0.400pt}}
\put(150,1300){\makebox(0,0){\Large{$N(\rho)$}}}
\put(862,25){\makebox(0,0){\large{$\rho$}}}
\put(350.0,250.0){\rule[-0.200pt]{0.400pt}{361.350pt}}
\put(592,268){\makebox(0,0){$\times$}}
\put(646,381){\makebox(0,0){$\times$}}
\put(699,568){\makebox(0,0){$\times$}}
\put(753,705){\makebox(0,0){$\times$}}
\put(807,391){\makebox(0,0){$\times$}}
\put(861,254){\makebox(0,0){$\times$}}
\put(592,388){\circle*{18}}
\put(646,755){\circle*{18}}
\put(699,1538){\circle*{18}}
\put(753,1161){\circle*{18}}
\put(807,402){\circle*{18}}
\put(861,254){\circle*{18}}
\put(592,292){\circle{18}}
\put(646,494){\circle{18}}
\put(699,942){\circle{18}}
\put(753,1369){\circle{18}}
\put(807,1034){\circle{18}}
\put(861,352){\circle{18}}
\put(914,264){\circle{18}}
\end{picture}
\end    {center}
\vskip 0.15in
\caption{The number of instantons in 1000 lattice fields 
using only the largest positive and negative peaks in each 
field. All in the confined phase, at $T=T_c$, on a $16^3 5$ 
lattice in $SU(6)$. Separately for the   lattice fields with
$Q\not= 0$ and for the fields with  $Q=0$, $\times$.
For the former we show separately 
sizes from peaks with the same sign as $Q$ ($\bullet$) and
with the opposite sign ($\circ$).}
\label{fig_rhoconfn6}
\end    {figure}

\end{document}